\documentclass[12pt]{article}
\pdfoutput=1                     %%%%%%%%%%%%%%%%% FIGURE .EPS to .PDF %%%%%%%%%%%%%%
\usepackage{graphicx,amssymb,bbold}
\usepackage{amsmath,amsfonts,epsfig}
\usepackage{chessfss}
\usepackage{braket}
\usepackage{tikz}
\newcommand{\pslash}{p\kern-1ex /}
\newcommand{\qslash}{q\kern-1ex /}
\newcommand{\lslash}{l\kern-1ex /}
\newcommand{\sslash}{s\kern-1ex /}
\newcommand{\kaslash}{k_a\kern-2ex /}
\newcommand{\kbslash}{k_b\kern-2ex /}
\newcommand{\Dslash}{{\cal D}\kern-1.5ex /}

\newcommand{\bc}{\overline{c}}

\newcommand{\beqa}{\begin{eqnarray}}
\newcommand{\eeqa}{\end{eqnarray}}

\newcommand{\bpm}{\begin{pmatrix}}
\newcommand{\epm}{\end{pmatrix}}
\newcommand{\bbm}{\begin{bmatrix}}
\newcommand{\ebm}{\end{bmatrix}}

\newcommand{\white}{\tikz[black, baseline=-0.6ex]\draw(0,0) circle (1.0ex) ;}%
\newcommand{\black}{\tikz[black,fill=black, baseline=-0.6ex]\filldraw(0,0) circle (1.0ex) ;}%
\newcommand{\link}[1]{| #1 \rangle\!\rangle}
\makeatletter

\@addtoreset{equation}{section}
\makeatother
\usepackage{subfig}
\begin{document}

%\voffset -0.7 true cm
%\hoffset 1.1 true cm
%\topmargin 0.0in
%\evensidemargin 0.0in
%\oddsidemargin 0.0in
%\textheight 8.6in
%\textwidth 7.1in
%\parskip 10 pt

\voffset -0.7 true cm
\hoffset 1.5 true cm
\topmargin 0.0in
\evensidemargin 0.0in
\oddsidemargin 0.0in
\textheight 8.6in
\textwidth 5.4in
\parskip 9 pt
 
\def\Tr{\hbox{Tr}}
\newcommand{\be}{\begin{equation}}
\newcommand{\ee}{\end{equation}}
\newcommand{\bea}{\begin{eqnarray}}
\newcommand{\eea}{\end{eqnarray}}
\newcommand{\beas}{\begin{eqnarray*}}
\newcommand{\eeas}{\end{eqnarray*}}
\newcommand{\nn}{\nonumber}
\font\cmsss=cmss8
\def\C{{\hbox{\cmsss C}}}
\font\cmss=cmss10
\def\bigC{{\hbox{\cmss C}}}
\def\scriptlap{{\kern1pt\vbox{\hrule height 0.8pt\hbox{\vrule width 0.8pt
  \hskip2pt\vbox{\vskip 4pt}\hskip 2pt\vrule width 0.4pt}\hrule height 0.4pt}
  \kern1pt}}
\def\ba{{\bar{a}}}
\def\bb{{\bar{b}}}
\def\bc{{\bar{c}}}
\def\bphi{{\Phi}}
\def\Bigggl{\mathopen\Biggg}
\def\Bigggr{\mathclose\Biggg}
\def\Biggg#1{{\hbox{$\left#1\vbox to 25pt{}\right.\n@space$}}}
\def\n@space{\nulldelimiterspace=0pt \m@th}
\def\m@th{\mathsurround = 0pt}

\begin{titlepage}
\begin{flushright}
%\rightline{OU-HET-xxx}
{\small OU-HET-938} 
 \\
\end{flushright}

\begin{center}

\vspace{5mm}

{\Large \bf Time Evolution of Complexity } \\[3pt] 
\vspace{1mm}
{\Large \bf in Abelian Gauge Theories }  \\[3pt] 
\vspace{3mm}
{\it {- And Playing Quantum Othello Game -}}
\vspace{6mm}

\renewcommand\thefootnote{\mbox{$\fnsymbol{footnote}$}}
Koji Hashimoto\footnote{koji@phys.sci.osaka-u.ac.jp},  
Norihiro Iizuka\footnote{iizuka@phys.sci.osaka-u.ac.jp}    
and  
Sotaro Sugishita\footnote{sugishita@het.phys.sci.osaka-u.ac.jp}

\vspace{3mm}

%${}^\textsymking$
{\small \sl Department of Physics, Osaka University} \\ 
{\small \sl Toyonaka, Osaka 560-0043, JAPAN}

\end{center}

\vspace{3mm}

\noindent
Quantum complexity is conjectured to probe inside of black hole horizons (or wormhole) via gauge gravity correspondence. In order to have a better understanding of this correspondence, we study time evolutions of complexities for Abelian pure gauge theories.  For this purpose, we discretize $U(1)$ gauge group as $\mathbf{Z}_N$ and also continuum spacetime as lattice spacetime, and this enables us to define a universal gate set for these gauge theories, and evaluate time evolutions of the complexities explicitly. We find that 
for a generic class of diagonal Hamiltonians 
to achieve a large complexity $\sim \exp(\mbox{entropy})$, which is one of the conjectured criteria necessary to have a dual black hole, the Abelian gauge theory needs to be maximally nonlocal.

\end{titlepage}

\setcounter{footnote}{0}
\renewcommand\thefootnote{\mbox{\arabic{footnote}}}

\newpage

\setcounter{tocdepth}{2}  %To define the depth of the table of contents
\tableofcontents

\newpage

%%%%%%%%%%%%%%%%%%%%%%%
\section{Introduction\label{sect:intro}}
%%%%%%%%%%%%%%%%%%%%%%%

\begin{figure}[tbp]
\begin{center}
 \includegraphics[width=0.26\textwidth]{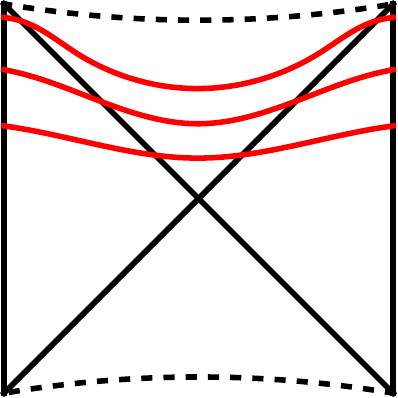} 
 \hspace{20mm}
 \includegraphics[width=0.26\textwidth]{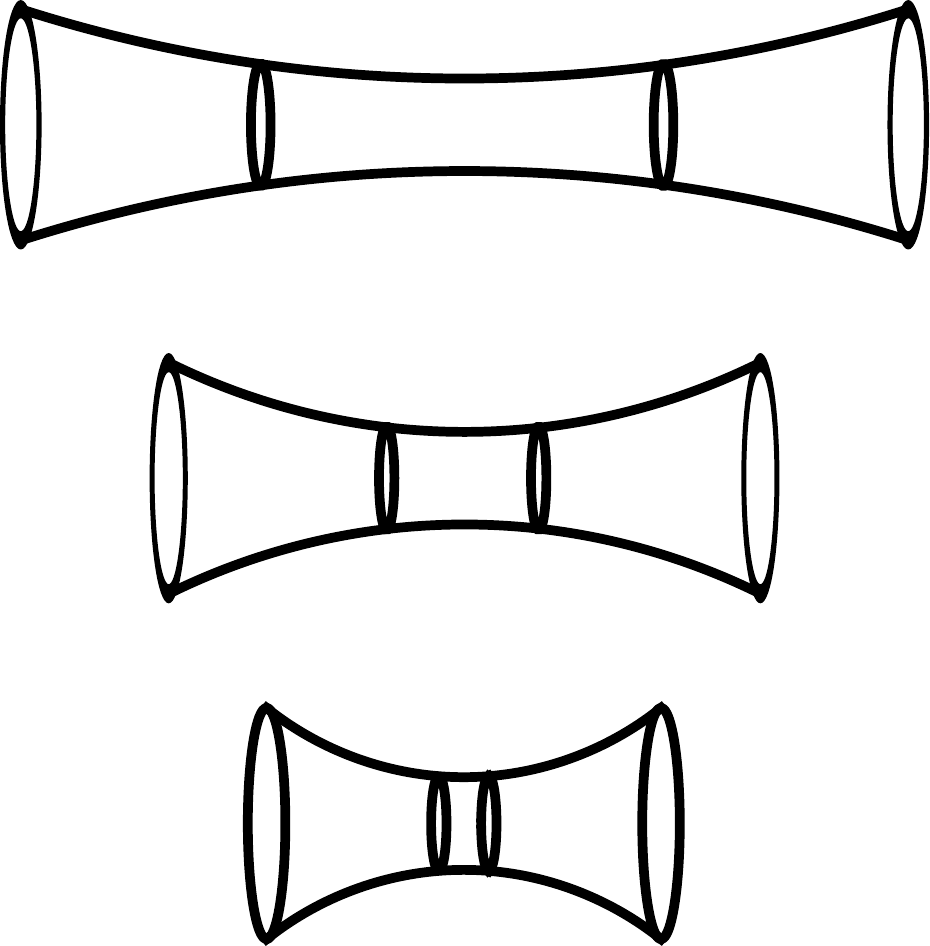} 
\caption{Eternal black hole with extremal surface anchored at the boundary time $t$ where $t$ runs upward in both CFTs (Left Figure). As time evolves upward, the wormhole inside black holes keeps growing linearly with respect to time $t$ (Right Figure). \label{introfigures}}  
\end{center}
\end{figure}

Understanding the inside of black hole horizons is a challenging problem in 
modern theoretical physics. 
The black hole firewall paradox \cite{Mathur:2009hf, Almheiri:2012rt} has sharpened the view that 
the black hole complementarity \cite{Susskind:1993if, Bigatti:1999dp} is not enough, and 
we need to modify our view of the inside of the horizon a bit more drastically once the black hole is entangled with 
Hawking radiations.  
One possible resolution of this firewall paradox is the 
ER = EPR conjecture \cite{Maldacena:2013xja}, namely that the Einstein-Rosen (ER) bridge 
wormhole connecting two boundaries is dual to the entangled (Einstein-Podolsky-Rosen (EPR)) boundary theories.  
A typical example of such ER = EPR is an eternal AdS black hole with two boundaries, which is dual to the 
thermo-field double (TFD) state \cite{Maldacena:2001kr}.  
By tracing out one boundary, one obtains a thermal ensemble for the other boundary, which corresponds to a black hole seen from the outside of the black hole horizon. 
On the other hand, by considering the TFD state (which is a pure state), one can probe the inside of the black hole horizon. 
This immediately leads to an intriguing puzzle;  
In the bulk, this ER bridge keeps growing linearly with respect to time $t$ forever at least classically. 
This linear growth can be seen by looking at the time evolution of an extremal surface 
anchored at the boundaries with a fixed time $t$, where $t$ runs forward in both CFTs \cite{Hartman:2013qma}
(see Fig.~\ref{introfigures}).
On the other hand, in quantum field theories, once the system is thermalized, 
it is unclear what kind of physical quantity keeps growing, since apparently the system ceases to grow
after the thermalization.
Recently, Susskind proposed that a quantum complexity is a key quantity to see the growth even 
after the system is thermalized \cite{Susskind:2014rva, Susskind:2014moa, Susskind:2016tae}. 

The quantum (computational) complexity $\mathcal{C}$, or complexity for short, 
is a notion used in quantum computation/information theory. 
In quantum mechanics,  the complexity is simply defined as 
how distant given the state is from a given ``reference'' state. 
Roughly speaking, given `gates' (a gate corresponds to `one step'), the complexity measures a
minimum amount of gates (which corresponds to `how many steps') one needs to move from the reference state 
to the given state. 
For a time-evolving state $\ket{\psi(t)}$, the time evolution of the 
complexity $\mathcal{C}(t)$ 
starts with $\mathcal{C}(0)=0$ and  
it typically increases linearly with respect to time $t$, and it reaches the maximum value at a time $t= t_{\rm max}$. 
The important point here is that there is a hierarchy for the time scale; entanglement entropy reaches its 
maximum value at the time scale of the power of the thermalization time $t_{\rm therm}$, which scales typically as powers of the entropy of the system.  
On the other hand, even after the system gets thermalized, 
the complexity keeps growing. 
In quantum mechanical systems,\footnote{In classical systems, the complexity reaches its maximum value  
at the time scale which typically scales as the entropy of the system.} 
complexity reaches its maximum value typically at the time which typically scales as the exponential of the entropy \cite{Susskind:2014moa}, 
\bea
t_{\rm max} \sim e^S \gg t_{\rm therm} \sim \left(S \right)^{p} \,,
\eea
where $p$ is generically some ${\cal{O}}(1)$ number. 
One of the goals in this paper is to have a better understanding of the time 
evolution of the complexity in gauge theories both qualitatively and quantitatively.  

It should be clear why we want to conduct analysis in gauge theories.  
Needless to say, gauge theories are a core of our modern understanding of physics describing not only all of the non-gravitational forces in our world but also 
they describe gravity too via holography.  
In order to apply the notion of the complexity, rather than spin systems, 
we have to deal with gauge theories. 
In this paper, as a first step toward understanding the time evolution of complexity in  generic gauge theories, 
we study the complexity in discrete Abelian gauge theories in 2+1 dimensions: namely 
$\mathbf{Z}_N$ gauge theories on a spatial two-dimensional lattice.

The reason why we consider  $\mathbf{Z}_N$ gauge theory is to discretize the continuous gauge group 
so that
we can handle it as if it is a qubit system.
The gauge group is recovered to $U(1)$ in the limit $N\to \infty$.  For the same reason, we 
adopt a lattice regularization for the two-dimensional space.\footnote{Generalization to higher dimensions, or to multiple $U(1)$ gauge group is straightforward.}  
Taking into account a gauge invariance, we may consider only physical operators for the universal gate sets, which 
we will explain later, and evaluate the complexity of the theory.   
Note that $\mathbf{Z}_2$ gauge theory is essentially the same as Kitaev's toric code \cite{Kitaev:1997wr}. 

By studying these complexity more in generic gauge groups, 
we would like to understand the following important questions; 
1) what kind of gauge theories really satisfy the criterion of the ${\cal{O}}(e^S)$ 
timescale for the growth of the complexity? and 2) what kind of gauge theories can  
consequently allow a gravity (and black hole) dual description?  
Of course, for the second question we assume that the complexity actually captures the growth of the inside of the wormhole geometry. Looking at the correspondence conjecture from the other way around,
it leads to a question of how the fastest computer can be realized by gauge theories. 

The organization of this paper is as follows. In Sec.~\ref{sec:2} 
we review the necessary ingredients; {\it i.e.,} gates, complexity, and $\mathbf{Z}_N$ lattice gauge theories.  
Sec.~\ref{sec:3} and Sec.~\ref{sec:quantum} are our main analysis, where we study the time evolution of the complexity in $\mathbf{Z}_N$ gauge theory both classically and quantum mechanically respectively. Sec.~\ref{sec:Summary} 
is for our summary and discussions.

Before closing the introduction, we comment on several closely related references. 
Two bulk duals of the boundary complexity have been proposed: 
One is the complexity = volume (CV) conjecture \cite{Stanford:2014jda, Susskind:2014jwa}, 
and the other is the complexity = action (CA) conjecture \cite{Brown:2015bva, Brown:2015lvg}. 
The CV conjecture states that the complexity at a time in the boundary is related to the maximal volume of a spatial slice in the dual bulk geometry,  
where the spatial slice is anchored to the boundary at the boundary time as in Fig.~\ref{introfigures}. 
The CA conjecture states that the complexity is given by the bulk action on the Wheeler-DeWitt patch, which is a region bounded by the future and the past null surface anchored also at the given boundary time. 
The qualitative behavior of the value of the action at late times is almost the same as the maximal volume. 
See also \cite{Chapman:2016hwi, Carmi:2016wjl, ccmms} for related works.

%%%%%%%%%%%%%%%%%%%%%%%%%%%%%%%%%%%%%%%%%%%%%%%%%%%%%%%%%%%%%%%%%%%%%%%%%%%%%%%%%%%%%%%%%%%%%%%%%%%%%%%%%%%%%%%%%%%%%%
\section{Gates and Complexity in $\mathbf{Z}_N$ Lattice Gauge Theories}
\label{sec:2}
%%%%%%%%%%%%%%%%%%%%%%%%%%%%%%%%%%%%%%%%%%%%%%%%%%%%%%%%%%%%%%%%%%%%%%%%%%%%%%%%%%%%%%%%%%%%%%%%%%%%%%%%%%%%%%%%%%%%%%
%%%%%%%%%%%%%%%%%%%%%%%%%%%%%%%%%%%%%%%%%%%%%%%%%%%%%%%%%%%%%%%%%%%%
\subsection{Overview of Gates and Complexity\label{sec_review_comp}} 
%%%%%%%%%%%%%%%%%%%%%%%%%%%%%%%%%%%%%%%%%%%%%%%%%%%%%%%%%%%%%%%%%%%%
In this subsection, we briefly review gates and complexity. 
For the reader who would like to know more in detail, see \cite{Nielsen, Aaronson:2016vto} for example.  

First we define a (quantum) gate for a given Hilbert space $\mathcal{H}$. 
A (quantum) gate is simply a unitary operator $G$ on $\mathcal{H}$. 
Let us consider some unitary operator $U$ on $\mathcal{H}$, and ask how to construct $U$ 
as a product of the gates taken  
from a set of gates $\{G_\alpha\}$. 
One might wonder what set $\{G_\alpha\}$ one should choose as a gate set, but 
there is no unique choice for that.

Given a gate set, in the context of quantum computation, there is a quantity 
characterizing how hard it is to construct $U$, and this quantity is the {\it complexity} of $U$. 
The complexity $\mathcal{C}(U)$ of a unitary operator $U$ is defined as the {\it minimum number} of gates necessary
to realize $U$ from her/his own given gate set.\footnote{Precisely speaking, such complexity is called {\it exact} complexity.} 
An approximate complexity $\mathcal{C}(U, \epsilon)$ can also be defined as 
the minimum number of gates to construct an operator $V$, satisfying $\|U-V\|<\epsilon$ with some norm  on the unitary operators $\mathcal{H}$ \cite{Nielsen:PRA73} where $\epsilon$ is a small positive number.

In this paper, 
we also consider the complexity $\mathcal{C}(\psi)$ (or the approximate complexity $\mathcal{C}(\psi, \epsilon)$) of a state $\ket{\psi} \in \mathcal{H}$, which is defined as the minimum number of gates to construct a 
unitary operator $U$ for $\ket{\psi} = U \ket{\psi_0}$ (or $\mid \ket{\psi} - U \ket{\psi_0}\mid<\epsilon$).  
Here $\ket{\psi_0}$ is a given initial state as a reference state \cite{Stanford:2014jda}  to evaluate the complexity for  
the state $\ket{\psi}$, 
and $U$ is written as a product of elements in the gate set.\footnote{\label{foot_state_op}
	Since the unitary operator $U$ satisfying $\ket{\psi}=U \ket{\psi_0}$ is not unique, 
	the complexity of state $\mathcal{C}(\psi)$ is slightly different from the complexity $\mathcal{C}(U)$. See our later discussions on the state dependence at Sec.~\ref{Z22-localanalysis}.} 
In particular, we consider the time evolution, $\ket{\psi(t)} = e^{- i H t} \ket{\psi(0)}$, and see the time-dependence of the complexity of $\ket{\psi(t)}$.

To make our argument more concrete, let us consider an $n$-qubit system $\mathcal{H}=(\mathbb{C}^2)^{\otimes n}$ as an example, 
where $\mathbb{C}^2$  is for two coefficients of $\ket{0}$ and $\ket{1}$, 
and the orthogonal bases can be chosen as 
$\ket{s_1 \ldots s_n}$  $(s_i=0$ or $1)$. 
Let us introduce two types of elementary gates in this system: 1) single qubit gates and 2) multiple qubit gates.
The single qubit gates are gates acting on a single qubit,
{\it i.e.}, a single qubit gate is a $2\times2$ unitary matrix.\footnote{
More precisely, a single qubit gate is a $2^n \times 2^n$ unitary matrix on $\mathcal{H}=(\mathbb{C}^2)^{\otimes n}$, which has a form 
$1\otimes \cdots \otimes U_{2 \times 2} \otimes \cdots \otimes 1$ where $1$ and $U_{2 \times 2}$ are unit and general $2\times2$ matrix respectively.
}    
Similarly the multiple qubit gates are gates acting on multiple qubits. 
For example, $k$-qubit gates are $2^k\times2^k$ unitary matrices acting on $k$ qubits simultaneously. 
An especially important multiple qubit gate exists at $k=2$, which is called a controlled-NOT (CNOT) gate, and we will explain it now.

A CNOT gate is a 2-qubit gate acting on a control qubit and a target qubit.  
It flips the target qubit if the control qubit is $\ket{1}$.  
In other words, a CNOT gate  acts on a control qubit $\ket{s_1}$ and a target qubit $\ket{s_2}$ as $\ket{s_1, s_2} \to \ket{s_1, s_2\oplus s_1}$ where 
$\oplus$ denotes addition modulo 2. 
In the computational basis $\{\ket{00},\ket{01},\ket{10},\ket{11}\}$, 
a CNOT gate is expressed as a simple matrix as 
\begin{align}
U^{\rm CNOT} = 
\begin{pmatrix}
1&0&0&0\\
0&1&0&0\\
0&0&0&1\\
0&0&1&0\\
\end{pmatrix}.
\label{CNOTdef}
\end{align}
A nontrivial but true fact is that if one has all single qubit gates and the CNOT gate 
for {\it every pair} of qubits, then one can construct any multiple $k$ qubit gate 
and furthermore, any unitary operator.

The gate set is called {\it universal} if any operator on $\mathcal{H}$ can be approximated with arbitrary accuracy by a product of elements of the gate set, 
and called exactly universal if any operator can be constructed exactly. 
It is known that a set of all the single qubit gates and the CNOT gates 
for every pair of qubits constitutes an exactly universal gate set for $n$-qubit systems (see, e.g., \cite{Nielsen}),
and we use this universal gate set in this paper.\footnote{
	For the universal gate set, one can use an imprimitive 2-qubit gate instead of the CNOT gate \cite{Brylinski,ZhangVala}. 
	Here,  an imprimitive gate is defined as a gate which entangles two qubits. 
	In quantum computation, a 2-qubit gate $V$ is called primitive if and only if it acts as $V\ket{s_1 s_2} = U \ket{s_1} \otimes U' \ket{s_2}$ 
	or $ U \ket{s_2} \otimes U' \ket{s_1}$ with some single qubit gates $U$ and $U'$.
	All the other 2-qubit gates are called imprimitive.}  
\footnote{
As an aside, let us note that 
we do not even need to include all single qubit gates in the gate set 
(although we do not take this choice in this paper). %because they can be constructed from the set given in eq.~\eqref{single_gateset}. 
For a single qubit system, 
overall phase shift operators, relative phase shift operators 
and rotation operators, which are respectively given by 
\begin{align}
\begin{pmatrix}
e^{i \alpha}&0\\
0&e^{i \alpha}\\
\end{pmatrix},
\quad
\begin{pmatrix}
e^{i \beta}&0\\
0&e^{-i \beta}\\
\end{pmatrix},
\quad
\begin{pmatrix}
\cos \gamma&-\sin \gamma\\
\sin \gamma&\cos \gamma\\
\end{pmatrix},
\label{single_gateset}
\end{align}
constitute an exactly universal gate set, 
because any $2\times 2$ unitary matrix $A$ can be 
decomposed as 
\begin{align}
A= 
\begin{pmatrix}
e^{i \alpha}&0\\
0&e^{i \alpha}\\
\end{pmatrix}
\begin{pmatrix}
e^{i \beta}&0\\
0&e^{-i \beta}\\
\end{pmatrix}
\begin{pmatrix}
\cos \gamma&-\sin \gamma\\
\sin \gamma&\cos \gamma\\
\end{pmatrix}
\begin{pmatrix}
e^{i \delta}&0\\
0&e^{-i \delta}\\
\end{pmatrix}.
\end{align}
Thus, if we take the above gate set \eqref{single_gateset}, 
the complexity for a single qubit is less than or equal to four. In this paper 
we 
%use just a single qubit gate in our universal gate set, 
include all single qubit gates into our universal gate set,
so the complexity for a single qubit is one.
}

So far we have considered qubits (2-level spins). 
The $N$-level spins are called qudits, and 
generalizing the above argument for $n$-qudit systems is straightforward. 
As in the qubit systems, we define single qudit gates as well as multiple qudit gates.
In particular, we can define the generalized CNOT gates for the qudit system as 
$U^{\rm CNOT}\ket{s_1, s_2} = \ket{s_1, s_2\oplus s_1}$ where $\oplus$ denotes addition modulo $N$  \cite{Brylinski}. 
A universal gate set in qudit system is 
a set of all the single qudit gates and the generalized CNOT gates for every pair of qudits, just as in the qubit systems.

%%%%%%%%%%%%%%%%%%%%%%%%%%%%%%%%%%%%%%%%%%%%%%%%%%%%%%%%%%%%%%%%%%%%
\subsection{$\mathbf{Z}_N$ Lattice Gauge Theory}
\label{Z2review}
%%%%%%%%%%%%%%%%%%%%%%%%%%%%%%%%%%%%%%%%%%%%%%%%%%%%%%%%%%%%%%%%%%%%
In this paper, we study the time evolution of the complexity of the state $\ket{\psi}$, $\mathcal{C}(\psi)$ in $\mathbf{Z}_N$ gauge theory. For that purpose, we also briefly review the $\mathbf{Z}_N$ gauge theory, especially its physical Hilbert space. For details, see Appendix \ref{AppA}. 

We consider two-dimensional space on the lattice as shown in Fig~\ref{2dlatticemain}.\footnote{
	We use the temporal gauge and consider a time-slice (see Appendix \ref{AppA}).} 
\begin{figure}[tbp]
\begin{center}
\includegraphics[height=5.0cm]{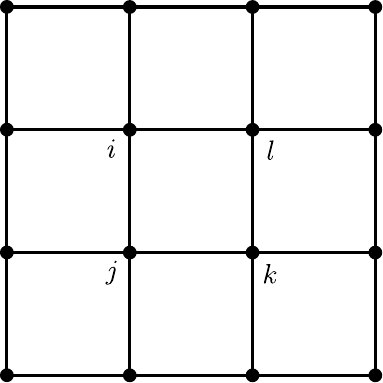}
\caption{
	Two-dimensional spacial lattice. Dynamical variables live on links.  
	\label{2dlatticemain}}  
%\label{2dlattice}
\end{center}
\end{figure} 
We use labels $i,j,\ldots$ for vertices and $i$-$j$ for links on the lattice. 
%We set $a$ as lattice spacing. 
In lattice gauge theory, 
we have an oriented link variable on each link $i$-$j$ as $L_{ij} = L_{ji}^{-1}=\exp(2\pi i\, n_{ij}/N) \in \mathbf{Z}_N$, where $n_{ij}$ is modulo $N$ integer, $n_{ij}=0,1,\cdots,N-1$. 
The link variable is roughly the exponential of the gauge field on the link. 
In pure $\mathbf{Z}_N$ gauge theory, link variables ({\it i.e.,} gauge fields) are the only dynamical degrees of freedom and 
the states are written as superpositions of the basis $\{\otimes_\text{all links} \link{n_{ij}}_{ij}\}$.\footnote{We use a double bracket $\link{\phantom{n}}$ to represent link-states.} 
%, where $n_{ij}$ is modulo $N$ integer.  

Gauge transformation $e^{2\pi i\, \delta_i/N}\in \mathbf{Z}_N$ on a vertex $i$ act on $L_{ij}$ as 
\be
L_{ij} \to e^{2\pi i\, \delta_i/N}L_{ij}=e^{2\pi i\,( n_{ij}+\delta_i)/N} \quad (\mbox{where} \,\,\, \delta_i =1, 2, \cdots , N-1)\,.
\ee
%For example, for $\mathbf{Z}_2$ case, $\theta_i \to \theta_i + \pi$. 
Thus, it results in shifting $n_{ij}$ on all links emanating from $i$ as 
\be
\label{ZNgaugetransformationmain}
n_{ij} \to n_{ij} + \delta_i \quad (\mbox{for all $j$ adjacent to $i$})\,.
\ee
%In the square lattice as Fig \ref{2dlatticemain}; this results in flipping all the link variables 
%which emanate from vertex $i$. 
In order to impose the gauge invariance, it is enough to consider transformations by the unit shift $e^{2\pi i/N}$ ({\it i.e.,} $\delta_i = 1$ case)
since other transformations $e^{2\pi i\, \delta_i/N}$, with $\delta_i = 2, \cdots, N-1 $, are generated from it. 
The operator $g_i$ corresponding to the gauge transformation $e^{2\pi i/N}$ at vertex $i$ is represented by a tensor product of shift operators $\tau_1^{(ij)}$ as 
\begin{align}
g_i = \underset{j \text{ (adjacent to } i)}{\otimes} \tau_1^{(ij)}, \quad 
\tau_1^{(ij)}\equiv
\begin{pmatrix}
0&0&\cdots&0&1\\
1&0&\cdots&0&0\\
0&1&\ddots&0&0\\ 
\vdots&&\ddots&&\vdots\\
0&0&&1&0
\end{pmatrix} \,. 
\label{gauge_trsf_ZN}
\end{align}
Here, we represent the shift operators $\tau_1^{(ij)}$ by a $N$ by $N$ matrix, which acts on $N$-dimensional vector space $\link{n_{ij}}_{ij}$. 
Eigenvalues of the shift operator $\tau_{ij}$ are $e^{2\pi i\, \beta_{ij}/N}$ with $\beta_{ij}=0,1,\cdots N-1$. We denote the corresponding eigenstates of the shift operator $\tau_{ij}$ by $\link{\beta_{ij}}_{ij}$, where we have 
$\beta_{ij}=-\beta_{ji} \pmod{N}$. 

To see the gauge invariant states, 
let us first introduce {\it extended} \footnote{We use the word ``extended" in the sense that we do NOT restrict states to gauge invariant ones.}  
Hilbert space ${\cal{H}^{\bf{ext}}}$ 
and then require the condition that physical states $\ket{\psi}$ must be gauge invariant, $g_i \ket{\psi}=\ket{\psi}$. 
Tensor product states $\{\otimes_\text{all links} \link{\beta_{ij}}_{ij}\}$ constitutes a basis of the extended space ${\cal{H}^{\bf{ext}}}$. 
Since transformation $g_i$ acts on each orthogonal state $\otimes_\text{all links} \link{\beta_{ij}}_{ij}$ as 
\begin{align}
g_i \, \left( \otimes_\text{all links} \link{\beta_{ij}}_{ij} \right) 
=\left(\prod_{j \text{ (adjacent to } i)} \!\!\!\!\!\! e^{i \frac{2\pi}{N} \beta_{ij}} \right)\,\left( \otimes_\text{all links} \link{\beta_{ij}}_{ij} \right),  
\end{align}
$\beta_{ij}$ must satisfy 
\begin{align}
\label{sum0condition}
\sum_{j \text{ adjacent to } i} \beta_{ij} =0 \pmod{N} 
\quad (\mbox{for each vertex  $i$})\,, 
\end{align}
for the gauge invariance.   
A basis of the physical Hilbert space $\mathcal{H}^\text{phys}$ is thus $\{\otimes_\text{all links} \link{\beta_{ij}}_{ij}\}$ $(\beta_{ij}=0,1,\cdots N-1)$ with the condition \eqref{sum0condition}. 
Be careful for the directions of the links in the sum in \eqref{sum0condition}; they
are all from $i$ to the adjacent vertices $j$.

\begin{figure}[htbp]
\begin{center}
\includegraphics[height=5.0cm]{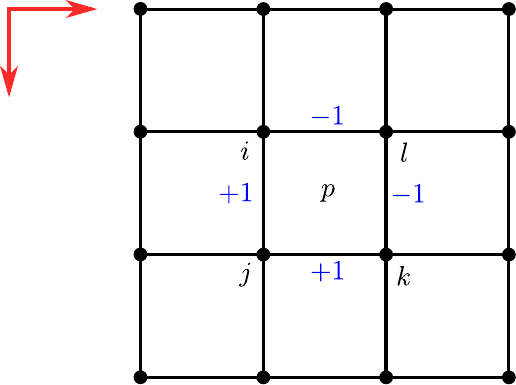}
\caption{Magnetic flux operator with unit strength on plaquette $p$. 
	A flux loop has positive strength $+1$ if we see it along the direction $i\to j \to k \to l \to i$. \label{loopmain}}  
\end{center}
\end{figure}

The gauge invariance condition \eqref{sum0condition} implies an electric flux conservation (the net flux out of any vertex vanishes).  
Let us define the direction in the square lattice as; from left to right and from up to bottom as the positive directions, see Fig.~\ref{loopmain}. The state with 
all $\beta_{ij} = 0$ has no electric flux. Consider a generic state for which we have some links with $\beta_{ij} \neq 0$. Then, any physical state, a state which satisfies the condition \eqref{sum0condition}, should look like, on the $\beta_{ij} = 0$ background, a flux (whose location is defined by $\beta_{ij} \neq 0$) flowing on links
while satisfying 
\eqref{sum0condition}.  

To describe the loops and the physical states, 
let us define a 
gauge invariant loop operator creating a flux loop flowing along a minimal plaquette, as shown in Fig.~\ref{loopmain}.  In 
$\mathbf{Z}_N$ gauge theory, there are $N$ species of loops associated with its flux strength $0, 1, 2, \cdots N-1$. (Fig.~\ref{loopmain} corresponds to the
unit strength.) 
We call the loop operators \textit{magnetic} flux operators.  
Each $N-1$ different magnetic flux operator acting on a minimal plaquette is creating the $\mathbf{Z}_N$ flux loop along the box edges. 
Acting a magnetic flux operator with strength $m$ is the same as acting the 
unit strength operator $m$ times. 
If one acts the same strength operator on two neighboring plaquettes, then there is a cancellation of the flux 
on the link-state shared by the two plaquettes, which 
results in a bigger flux loop. In this way, each state in a basis of the physical space $\mathcal{H}^\text{phys}$ is specified by how many magnetic flux operators with the unit strength are acting on each of the
$L \times L$ plaquettes on the no-flux state $\otimes_\text{all links} \link{\beta_{ij} = 0}_{ij}$.\footnote{
	On lattices with periodic boundary conditions, there are other physical states which has globally winding fluxes (see Appendix \ref{AppA}). They cannot be obtained from the no-flux state $\otimes_\text{all links} \link{\beta_{ij} = 0}_{ij}$ by acing local magnetic operators. However, since they are in different super-selection sectors, we do not consider these topologically non-trivial states in the following analysis.  
}

\begin{figure}[htbp]
	\begin{center}
	\includegraphics{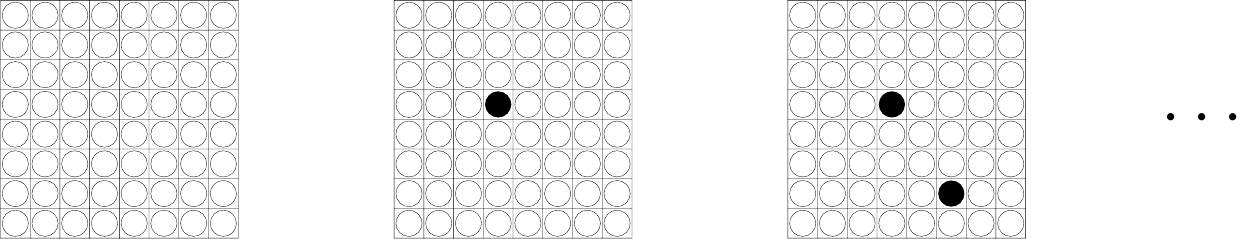}
		\caption{Othello game representation of a basis of the physical Hilbert space in $\mathbf{Z}_2$ gauge theory. 
		General physical states are given by superpositions of these Othello  configurations.
			{\label{reversi}}}
	\end{center}
\end{figure}

Each state in the basis actually 
looks like a configuration of a board game, as follows. 
For the $\mathbf{Z}_2$ case, this setting reminds us of a board game `Reversi' (known more as `Othello' game in Japan), see Fig.~\ref{reversi}. Starting with all white ($ \ket{\white}$) on each plaquette, which corresponds to the state $\otimes_\text{all links}\link{\beta_{ij} = 0}_{ij}$,  we flip the white one over to the
back (= black, $\ket{\black}$) on each plaquette.  
The flipping on a plaquette corresponds to acting the 
unit strength magnetic flux on the plaquette. 
Note that flipping twice makes it back to white due to $\mathbf{Z}_2$. 

We now move to the $\mathbf{Z}_N$ case, and introduce an $N$-colored Othello game. % from the monochrome game for generalization to $\mathbf{Z}_N$.  
The basis of the physical space $\mathcal{H}^\text{phys}$ in $\mathbf{Z}_N$ is obtained as follows:
\begin{enumerate}
\item Start with  the no flux state $\otimes_\text{all links} \link{\beta_{ij} = 0}_{ij}$, which we call all-white state, as in the first figure of Fig.~\ref{reversi}.
We represent this state as $\otimes_{\text{plaquettes } p}\ket{0}_{p}$. 
\item Then we add on each ``box'' a colored disk with a variation of $N-1$ different colors, as shown in the second and third figures in Fig.~\ref{reversi} for $N=2$ case. (See also Fig.~\ref{evolve_random_magnet}.)
The colors represent the strength of the magnetic flux on a plaquette. 
We denote each state corresponding to an $N$-colored Othello configuration by 
$\otimes_{\text{plaquettes } p}\ket{m_p}_{p}\,$, where $m_p$ is a modulo $N$ integer $(m_p=0,1,\cdots,N-1)$ and it  represents colors or strength of the magnetic flux on the plaquette $p$. Here $p$ is the label specifying the position of the ``box''. 
%\item Special rule is that acting on the {\it same} color operators on all of the box is regarded as ``trivial'' action. In other words, acting on whatever color operations at all the boxes simultaneously the state must be regarded as the same state. This is sort of ``gauge equivalence'' of this Othello game. 
%\item Then $\otimes_\text{all links} \ket{\beta_{ij} = 0}_{ij}$ state is equivalent to $\otimes_\text{all links} \ket{\beta_{ij} = k}_{ij}$ state where $k = 1, 2, \cdots, N-1$. In other words, our starting state is simply just the fixed color state. 
\end{enumerate}
Therefore, a basis of the physical space is $\{\otimes_{\text{plaquettes } p}\ket{m_p}_{p}\}$, 
and general physical states are obtained as the superpositions. 
This plaquette-state expression is a dual to the link-state description. 

Note that if we have a lattice space without boundary, such as a periodic lattice on $T^2$, then there is a global identification (global gauge symmetry); 
\be
\otimes_p \ket{m_p}_p  =  \otimes_p \ket{m_p  \oplus \delta}_p \,,
\label{globalsymbyboundarycondition}
\ee
for a plaquette-independent constant $\delta = 1, 2, \cdots, N-1$, where $\oplus$ denotes addition modulo $N$.

%%%%%%%%%%%%%%%%%%%%%%%%%%%%%%%%%%%%%%%%%%%%%%%
%%%%%%%%%%%%%%%%%%%%%%%%%%%%%%%%%%%%%%%%%%%%%%%
\subsection{Gate sets and locality in $\mathbf{Z}_N$ gauge theory}
\label{gate_gauge_th}
%%%%%%%%%%%%%%%%%%%%%%%%%%%%%%%%%%%%%%%%%%%%%%%
%%%%%%%%%%%%%%%%%%%%%%%%%%%%%%%%%%%%%%%%%%%%%%%

In this subsection we define universal gate sets in the lattice $\mathbf{Z}_N$ gauge theories. 
As we have seen, all of the orthogonal physical bases are obtained by adding magnetic fluxes at plaquettes to the reference physical state. Here we choose our reference state as the state 
$\otimes_{\rm plaquette} \ket{m_p = 0}_p$. 
If a magnetic flux is added at, say, plaquette $q$, then it becomes  $[\otimes_{{\rm plaquette} \, p (\neq q)} \ket{m_{p} = 0}_p ]\otimes \ket{m_q =1}_q$, etc.

Let us first consider a $\mathbf{Z}_2$ gauge theory on a $L\times L$ lattice 
(with or without periodic boundary conditions).  Then, physical state bases are $\otimes_{p} \ket{m_p}$ with $m_p=0$ or 1, which can be regarded as a state of a system with $L^2$ qubits on the plaquettes of the lattice. 
Thus, this physical Hilbert space is the same as an $L^2$-qubit system.\footnote{%
If we have periodic boundary conditions on all of the boundary, then we have the global gauge symmetry,  
$\otimes_{p} \ket{m_p}_p = \otimes_{p} \ket{m_p\oplus 1}_p$, which results in 
a physical Hilbert space $\sim (L^2-1)$-qubit system.} 
As we saw in Sec.~\ref{sec_review_comp}, the single qubit gates and the 
CNOT gates constitute a universal gate set of the qubit system.\footnote{Since the identification of states by the global gauge symmetry \eqref{globalsymbyboundarycondition} does not matter in the proof of the universality, they also constitute a universal gate set of $\mathbf{Z}_2$ gauge theory.} 
Note that all of the gates are unitary operators on gauge invariant physical states. For example, 
a single qubit gate acts on a plaquette $p$, which superposes $\ket{0}_p$ and $\ket{1}_p$. 

Extension to the $\mathbf{Z}_N$ gauge theory is straightforward. 
Ignoring the boundary condition and therefore the global identification \eqref{globalsymbyboundarycondition}, physical states in $\mathbf{Z}_N$ gauge theory are labeled as $\otimes_{p} \ket{m_p}_p$ with $m_p=0, \ldots, N-1$. 
Thus, the physical Hilbert space of the $\mathbf{Z}_N$ gauge theory is the same as that of the $L^2$-qudit system. 
As we also saw in Sec.~\ref{sec_review_comp}, 
the set of all the single qudit gates $U_p$ ($U(N)$ matrix acting on the plaquette $p$) 
and the generalized CNOT gates 
$U^{\rm CNOT}_{(p,q)}$ (acting on the control qudit at the plaquette $p$, and the target qudit 
at the plaquette $q$)  is universal in the qudit system, and  thus, 
they form a universal gate set $\mathcal{U}$ in $\mathbf{Z}_N$ gauge theory:
%
%We need the definition of the universal gate set ${\cal U}$. Noting that the Hilbert space of
%the $\mathbf{Z}_2$ gauge theory is equivalent to $L^2$ qubits, we use the universal gate
%set of the qubits: All single qubit gates and neighboring CNOT gates, as described in section ??.
%The single qubit gate is denoted by $U_{a,b}$, which is a $U(2)$ matrix and so has four real parameters
%at every qubit $(a,b)$.
%It satisfies 
%trivially the commutativity
%$U_{a,b}U_{c,d}=U_{c,d}U_{a,b}$.
%As for the CNOT gates, we allow them with arbitrary pair of the sites (the control and the target), 
%$(a,b)$ and $(a',b')$. Then the universal gate set is
\begin{align}
{\cal U} \equiv \{U_p, \;  U^{\rm CNOT}_{(p,q)}\; | \;  p,q\mbox{: plaquettes} \} \, .
\label{Uni}
\end{align}
In this paper, for calculating the complexity, we use this ${\cal U}$ unless otherwise stated.

Our goal is to study the complexity in quantum field theory. 
Thus, it is natural to choose a universal gate set such that it respects the spatial {\it locality}.   
Since the degrees of freedom live on links or equivalently plaquettes for lattice systems, 
there is a notion of neighboring plaquettes.  
We define \textit{neighboring} multiple qubit (or qudit) gates as gates acting on only the multiple plaquettes
which are next to each other.\footnote{%
In literatures (see e.g. \cite{Brylinski}), a gate acting on $k$-qubits simultaneously is called local 
(or $k$-local \cite{Brown:2017jil}) if $k$ is much less than the total number of qubits $L^2$, {\it i.e.}, $k \ll L^2$.  
This notion of the $k$-locality will be used in Sec.~\ref{Sec:klocal} to classify Hamiltonians in this paper too. 
Our terminology of {\it neighboring} is different from the notion of the $k$-locality. 
}  
Note that quantum field theories 
can be regarded as an IR limit of a spin system. Therefore any multiple qubit (qudit) 
gates acting on finite distance-away plaquettes behave as local interactions in the continuum limit,  which is $L \to \infty$ and plaquette size goes to zero limit. 
Only multiple qubit (qudit) gates acting on the plaquettes whose distance scale as ${\cal{O}}(L)$ (the size of the system) are spatially nonlocal in the continuum limit.   However just as usual spin system such as the Ising model, restricting to only neighboring multiple gates is specially interesting with respect to see the effects of locality for discrete systems.

In fact, for a gate set to be universal, it is enough to have only the neighboring gate sets: 
Any (generalized) CNOT gate can be constructed by a product of neighboring (generalized) CNOT gates, as follows.
For any (generalized) CNOT gate $U^{\rm CNOT}_{(p,q)}$, we have a relation
\begin{align}
U^{\rm CNOT}_{(p,q)}=U^{\rm CNOT}_{(r,q)} \,\left[ U^{\rm CNOT}_{(p,r)}\right]^{N-1} \, 
U^{\rm CNOT}_{(r,q)} \, U^{\rm CNOT}_{(p,r)} \, \left[U^{\rm CNOT}_{(r,q)}\right]^{N-2} \,,
\label{local_cnot_z2}
\end{align}
in the $N$-level qudit system.
Suppose the pair $(p,q)$ is not a neighboring plaquette pair, while $(p,r)$ and $(r,q)$ 
are neighboring ones, respectively. 
Then \eqref{local_cnot_z2} shows that 
the (generalized) CNOT gate $U^{\rm CNOT}_{(p,q)}$ acting on the non-neighboring plaquettes $(p,q)$ 
can be constructed from the neighboring (generalized) CNOT gates.
Thus, combining only the neighboring (generalized) CNOT gates, 
we can construct any (generalized) CNOT gate. 
We can define the neighboring universal gate set ${\cal U}_{\rm neighbor} $
on the $L\times L$ lattice as
\begin{align}
{\cal U}_{\rm neighbor} 
\equiv \{U_{a,b}, \;  U^{\rm CNOT}_{((a,b),(a+1,b))}, \; 
U^{\rm CNOT}_{((a,b),(a,b+1))}\; | \;  a,b=1,2,\cdots,L \} \, .
\label{Ulocal}
\end{align}
As we mention, if one is interested in understanding the effects of locality in evaluating the 
complexity in discrete systems, we shall use this ${\cal U}_{\rm neighbor}$
instead of ${\cal U}$.
Note that %the number of necessary CNOT gates, and as a result, 
complexity generically increases
by using $\mathcal{U}_{\rm neighbor}$ compared to $\mathcal{U}$.

%%%%%%%%%%%%%%%%%%%%%%%%%%%%%%%%%%%%%%%%%%%%%%%
%%%%%%%%%%%%%%%%%%%%%%%%%%%%%%%%%%%%%%%%%%%%%%%
\section{``Classical'' complexity in gauge theory}
\label{sec:3}
%%%%%%%%%%%%%%%%%%%%%%%%%%%%%%%%%%%%%%%%%%%%%%%
%%%%%%%%%%%%%%%%%%%%%%%%%%%%%%%%%%%%%%%%%%%%%%%
In this section, as a warm-up for the full quantum treatment in the next section, 
we consider a classical analogue of the complexity. We use 
toy models of $\mathbf{Z}_N$ gauge theories in which quantum superposition of states never appear,
which enables us to 
compute the complexity easily. We call the models \textit{random flux models}. 
The time evolution in the model is at random as a random quantum circuit in \cite{Brown:2016wib}.

Using the models, we will calculate the time evolution of the complexity and find the followings:
\vspace{-7pt}
\begin{itemize}
\item The complexity follows the typical expected time evolution: it starts from zero, 
grows at first linearly in time (= steps), and then reaches the maximum ${\cal C}_{\rm max}$, 
and stays there for a long time.
\item Introducing a nonlocality in the time evolution (which we call ``Othello rule'') accelerates
the growth of the complexity.
\item The growth speed is evaluated 
as a function of the theory parameters ($N$ of $\mathbf{Z}_N$ and spatial size $L$). 
\end{itemize}
First in section Sec.~\ref{sec:Rfm} we define the random flux models, and then 
in Sec.~\ref{sec:Crfm} we numerically calculate the time evolution of the complexity.
We present various numerical data for the evolution, and in particular
introduce the Othello rule to find how the nonlocality enhances the
growth rate of the complexity nontrivially.
%We will see the typical behavior of the complexity and how it depends on parameters ($N$ of $\mathbf{Z}_N$ and spatial size $L$). 

%In this section, we consider toy models of $\mathbf{Z}_N$ gauge theories where we can compute the complexity easily. We call the models \textit{random flux models}. 
%The time evolution in the model is at random as a random quantum circuit in \cite{Brown:2016wib}. We will see the typical behavior of the complexity and how it depends on parameters ($N$ of $\mathbf{Z}_N$ and spatial size $L$). 

%%%%%%%%%%%%%%%%%%%%%%%%%%%%%%%%%%%%%%%%%%%%%%%
\subsection{Random flux models}
\label{sec:Rfm}
%%%%%%%%%%%%%%%%%%%%%%%%%%%%%%%%%%%%%%%%%%%%%%%

\begin{figure}[htbp]
	\begin{center}
		\includegraphics{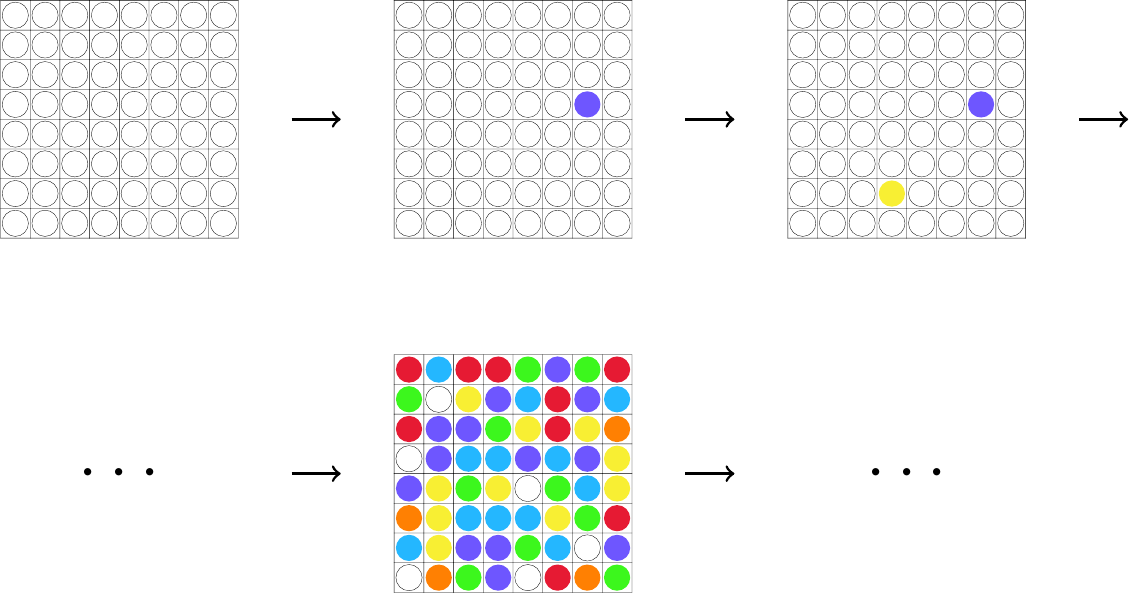}
		\caption{A time evolution of a state in the random flux model. The state at a step $t$ is obtained from the state at $(t-1)$ by acting a single qudit gate randomly. The white disk represents that the magnetic flux in the plaquette is zero, and other color disks do that the magnetic fluxes are nonzero. }
		\label{evolve_random_magnet}
	\end{center}
\end{figure}

Let us define the random flux model for $\mathbf{Z}_N$ gauge theory on a $L \times L$ periodic spatial lattice. 
The model uses a discretized time evolution, and 
a state evolves step by step randomly by simply adding a random magnetic flux as in Fig.~\ref{evolve_random_magnet}. 
The rule of the time evolution is as follows. 
Suppose that we have a state $\ket{m_{(1,1)}}_{(1,1)}\otimes \cdots \otimes \ket{m_{(L,L)}}_{(L,L)}$ at a step $t$, where $(a,b)$ are labels for plaquettes representing $a$-th row and $b$-th column plaquettes. Here $m_{(a,b)}$, which 
represents the strength of the magnetic flux at the plaquette $(a,b)$, is a modulo $N$ integer.    
%Then, the state at the next step $(t+1)$ is given by $\ket{m_{(1,1)}}_{(1,1)}\otimes \cdots \otimes \ket{m_{(a_p,b_p)} \oplus n_{(a_p,b_p)}}_{(a_p,b_p)} \otimes \cdots \otimes \ket{m_{(L,L)}}_{(L,L)}$, 
%where a magnetic flux with the strength $n_{(a_p,b_p)}$ is added at the plaquette $(a_p,b_p)$, where the plaquette position $(a_p,b_p)$ for the each step is chosen randomly. The strength $n_{(a_p,b_p)}$ is selected from $\{1, \cdots, N-1\}$ uniformly at random.  
Then, the state at the next step $(t+1)$ is given by $\ket{m_{(1,1)}}_{(1,1)}\otimes \cdots \otimes \ket{m_{(a,b)} \oplus n_{(a,b)}}_{(a,b)} \otimes \cdots \otimes \ket{m_{(L,L)}}_{(L,L)}$, 
where a magnetic flux with the strength $n_{(a,b)}$ is added at the plaquette $(a,b)$, where the plaquette position $(a,b)$ for the each step is chosen randomly. The strength $n_{(a,b)}$ is selected from $\{1, \cdots, N-1\}$ uniformly at random.  
Since in this model
any superposition of such states or entangled states do not appear in the time evolution, 
the notion of the complexity here is purely %may not be really of a quantum nature but closer to a 
\textit{classical}. 
In fact, for a $\mathbf{Z}_2$ gauge group, 
this model is exactly the same as the classical coins considered in \cite{Susskind:2014moa}. 
%{\it i.e.} 
%\footnote{The state at each step is described by classical bits. Only minor difference is the identification under the flip of all the bits due to the periodic lattice nature \eqref{globalsymbyboundarycondition}.} %such as $(110010...1)=(001101...0)$.}  
That is, the time evolution at each interval is 
given by acting a single qudit 
gate at a plaquette randomly.\footnote{This is similar to a random quantum circuit in \cite{Brown:2016wib}.} 
Since fluxes are added randomly, we call these classical toy models the random flux models. 

We take the state with no magnetic flux, $\otimes_{(a,b)} \ket{0}_{(a,b)}$, as the initial state and also the reference state  to compute the complexity. 
In this classical problem, any state evolving from the initial state can be prepared (from the reference state) 
by a sequence of 
actions of only {\it single} qudit gates in our universal gate set $\mathcal{U}$ in \eqref{Uni}.  
Thus, the complexity at each step is the minimum number of single qudit gates to prepare the state. 

From now on, we often use $p$ to specify a plaquette, {\it i.e.,} $p = {(a,b)}$. In this notation, the reference state is written as $\otimes_{p} \ket{0}_{p}$.  

%%%%%%%%%%%%%%%%%%%%%%%%%%%%%%%%%%%%%%%%%%%%%%%
\subsection{Complexity in random flux models}
\label{sec:Crfm}
%%%%%%%%%%%%%%%%%%%%%%%%%%%%%%%%%%%%%%%%%%%%%%%
%%%%%%%%%%%%%%%%%%%%%%%%%%%%%%%%%%%%%%%%%%%%%%%
\subsubsection{Counting of the complexity in random flux models}
%%%%%%%%%%%%%%%%%%%%%%%%%%%%%%%%%%%%%%%%%%%%%%%

We first consider $\mathbf{Z}_2$ gauge theories. 
In the random flux model with the $\mathbf{Z}_2$ gauge symmetry, 
a state $\ket{\psi(t)}$ at step $t$ has a form $\otimes_{p} \ket{m_p}_p$ with $m_p=0$ or $1$. 
Let's compute the complexity of the state $\ket{\psi(t)}$. 
Let the number of plaquettes with flux $m$ be $n_m$. 
Since there are $L^2$ plaquettes on a $L\times L$ lattice, 
$n_0 + n_1=L^2$ holds. 
The state $\ket{\psi(t)}$ can be obtained from the reference state $\otimes_{p} \ket{0}_p$ by adding $n_1$ magnetic fluxes. 
However, since the state $\otimes_{p} \ket{m_p\oplus 1}_p$ is the same as $\otimes_{p} \ket{m_p}_p$  due to the identification explained in section \ref{gate_gauge_th}, 
$\ket{\psi(t)}$ can also be obtained by adding $n_0$ magnetic fluxes into $\otimes_{p} \ket{1}_p =\otimes_{p} \ket{0}_p$. Thus, the complexity of $\ket{\psi(t)}$ is $\min\{n_0,n_1\}$. 
Since $n_0 + n_1=L^2$ should be satisfied, the maximal value of $\min\{n_0,n_1\}$ is 
$\lfloor L^2/2 \rfloor$, 
{\it i.e.}, 
the maximum complexity for the model is $\mathcal{C}_{\rm max} = \lfloor L^2/2 \rfloor$.

\begin{figure}[htbp]
	\begin{center} 
		\includegraphics[scale=0.8]{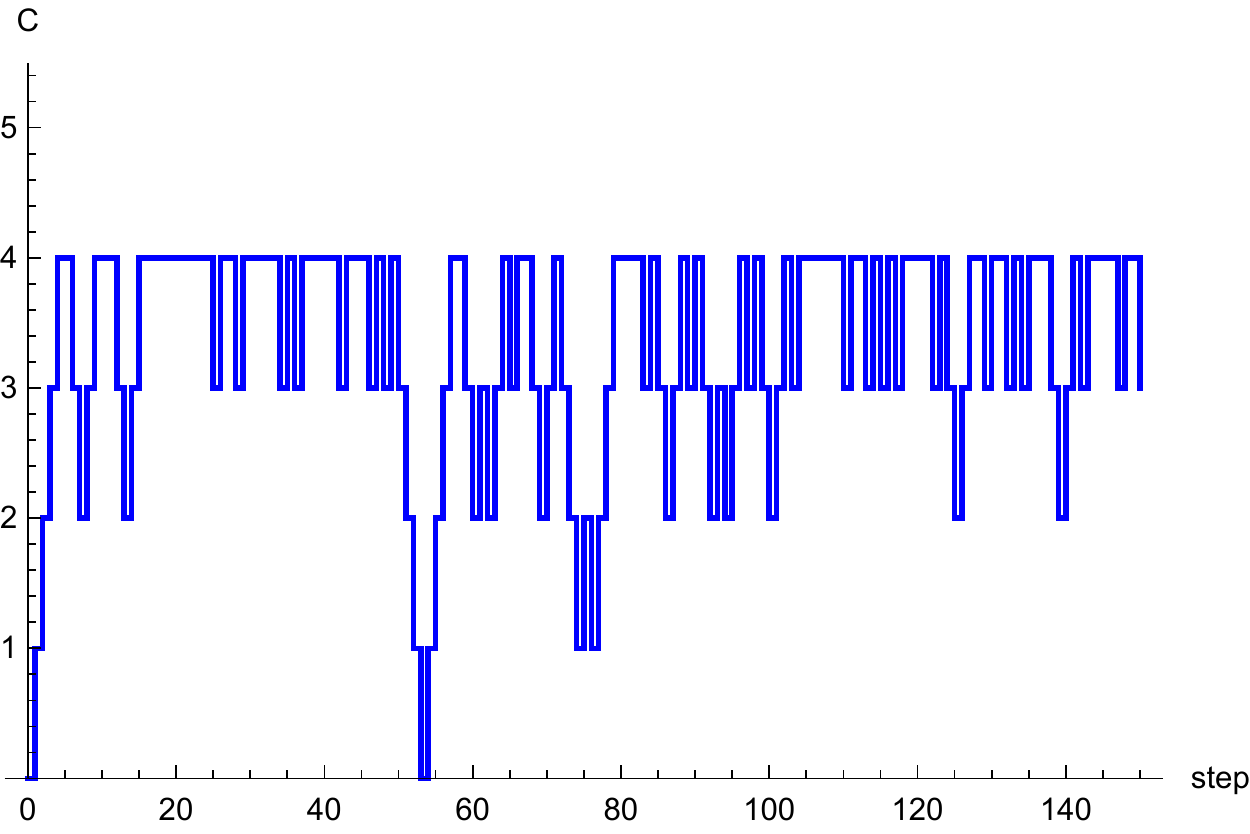}
		\caption{A time evolution of complexity in the random flux model with $\mathbf{Z}_{2}$ gauge symmetry on a lattice with the spatial size $3\times 3$. The maximum value of the complexity is $\mathcal{C}_{\rm max}=4$. Although the complexity typically fluctuates just below the maximum value, it can take zero at a finite time, {\it i.e.,} the state can return to the initial state by the Poincar\'e recurrence. 
			\label{z2_comlexity_typical}}
	\end{center}
\end{figure}

Fig.~\ref{z2_comlexity_typical} shows a time evolution of the complexity on a $3\times 3$ lattice. The complexity typically grows at early times linearly with time (= step number) and then fluctuates below the maximum value $\mathcal{C}_{\rm max}$. The state can be close to the initial state (the Poincar\'e recurrence) and the complexity can take small value at late times, although the probability is very small for a lattice with the large size.   

Next, we count the complexity of a state $\ket{\psi}$ in $\mathbf{Z}_N$ theory. 
Let $n_m$ be the number of plaquettes with magnetic flux $m$ $(m=0,1,\ldots N-1)$ in the state $\ket{\psi}$ where we have $\sum_{m}n_m=L^2$. 
We can construct $\ket{\psi}$ from the reference state $\otimes_{p} \ket{0}_p$ by acting single qudit
gates with non-zero magnetic flux $n_1, n_2, \cdots, n_{N-1}$. 
The number of used gates is $\sum_{m\neq 0} n_m = L^2 -n_0$. 
In general, however, it is not the minimal number of gates to construct $\ket{\psi}$. 
Since we have $\otimes_{p} \ket{0}_p = \otimes_{p} \ket{m}_p$ due to the identification rule  explained in Sec.~\ref{gate_gauge_th}, 
the state $\ket{\psi}$ is also obtained by using single magnet gates with flux $m'\neq m$.  
The number of used gates is then $\sum_{m'\neq m} n_{m'} = L^2 -n_m$. 
Thus, the complexity (the minimal value of gates to construct $\ket{\psi}$) is  
$\mathcal{C}=L^2-\max\{n_0,n_1,\ldots,n_{N-1}\}$. 
Since we have $\sum_{m}n_m=L^2$, 
the maximum complexity $\mathcal{C}_{\rm max}$ is $\lfloor L^2(1-1/N) \rfloor$.

%%%%%%%%%%%%%%%%%%%%%%%%%%%%%%%%%%%%%%%%%%%%%%%
\subsubsection{Extension to nonlocal interactions}
%%%%%%%%%%%%%%%%%%%%%%%%%%%%%%%%%%%%%%%%%%%%%%%
We extend the random flux models by allowing nonlocal interactions,  
and see how it changes the time evolution of complexity. 

In the previous model, only one plaquette is changed at each step by adding a magnetic flux.   
We can generalize it so that $q$ plaquettes are changed simultaneously at each step. 
At each step, $q$ plaquettes are chosen randomly, and a magnetic flux with random strength is added at each chosen plaquette. 
However, this generalization is equivalent to simply changing the time-scale of the $q=1$ model.

\begin{figure}[htbp]
	\begin{center}
		\includegraphics{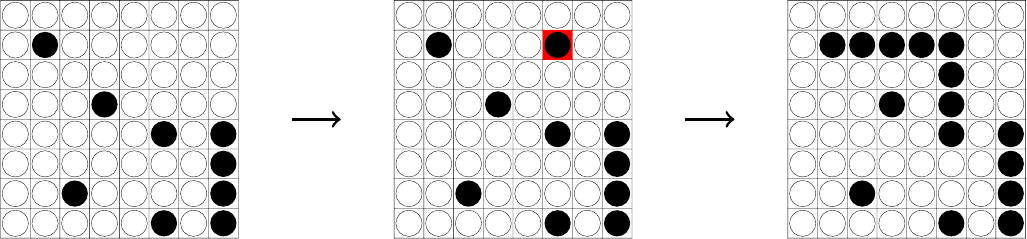}
		\caption{A time evolution with the Othello rule in a $\mathbf{Z}_2$ gauge theory. 
			A plaquette which is chosen at random in the left figure is highlighted by red color in the middle figure, and the color of the disk changes from white to black (step (i)). 
			Since there are white disks which are sandwiched by the highlighted black disk and  other black disks, their colors are also changed into black in the right figure (step (ii)). 
			The change from the left to the right is a combined single unit of the time evolution. }
		\label{othello_rule}
	\end{center}
\end{figure}

We may introduce another type of the random evolution, which we call ``Othello rule''.  
See Fig.~\ref{othello_rule}. 
%%%%%%
%%%%%%%
The Othello rule consists of the two succeeding actions: (i) The random flux, and (ii) Othello flips. 
For $\mathbf{Z}_2$ gauge theories, we represent the strength of magnetic flux 0 or 1 in the plaquette by white or black disk. 
In the random flux model with $q=1$, a plaquette is chosen at random and the color of the disk at the plaquette is changed at each step of time evolution. 
%This step is the same for the Othello rule, and 
The left figure in Fig.~\ref{othello_rule} changes to the middle one where the chosen plaquette is highlighted by red color. This is the process (i).  
Suppose that the new color of the disk is black (white).  
Then, if there are white (black) disks which are sandwiched between the chosen disk and other black (white) ones in the horizontal or vertical line,\footnote{Note that 
in this rule we do not consider the periodic boundary condition.} their colors are also changed into black (white). The sandwich rule is similar to the rule in board game `Reversi' (known more as `Othello' game in Japan). In Fig.~\ref{othello_rule}, the middle figure
is changed to the right one. This is the process (ii). Whole change by action (i) and (ii) is a combined single unit of the time evolution. 

Since disks separated by a large distance $\mathcal{O}(L)$ can change at a single time-step 
of the time evolution,  
we can say that the model is nonlocal. Whether the sandwich rule occurs in a time step depends on 
the state at the time. The Othello rule resembles to the CNOT gate because whether the 
flip of the target qubit occurs depends on the state of the control qubit. 
We shall see the similarity in more detail in Appendix \ref{sec:QO}.

Similarly, we may apply the Othello rule to the $\mathbf{Z}_N$ gauge theories. 
We can also represent the strength of the magnetic flux by disks with $N$ colors as shown in Fig.~\ref{evolve_random_magnet}. 
In the Othello rule, if a new color disk at the randomly chosen plaquette sandwiches the different color disks with the same color disk, the colors of the sandwiched disks changes to the same colors as that at the chosen plaquette. 

%%%%%%%%%%%%%%%%%%%%%%%%%%%%%%%%%%%%%%%%%%%%%%%
\subsubsection{Typical time evolution of complexity in random flux models}
%%%%%%%%%%%%%%%%%%%%%%%%%%%%%%%%%%%%%%%%%%%%%%%

\begin{figure}[tbp]
	\begin{center}
		\subfloat[$L=8$]{
			\includegraphics[scale=0.6]{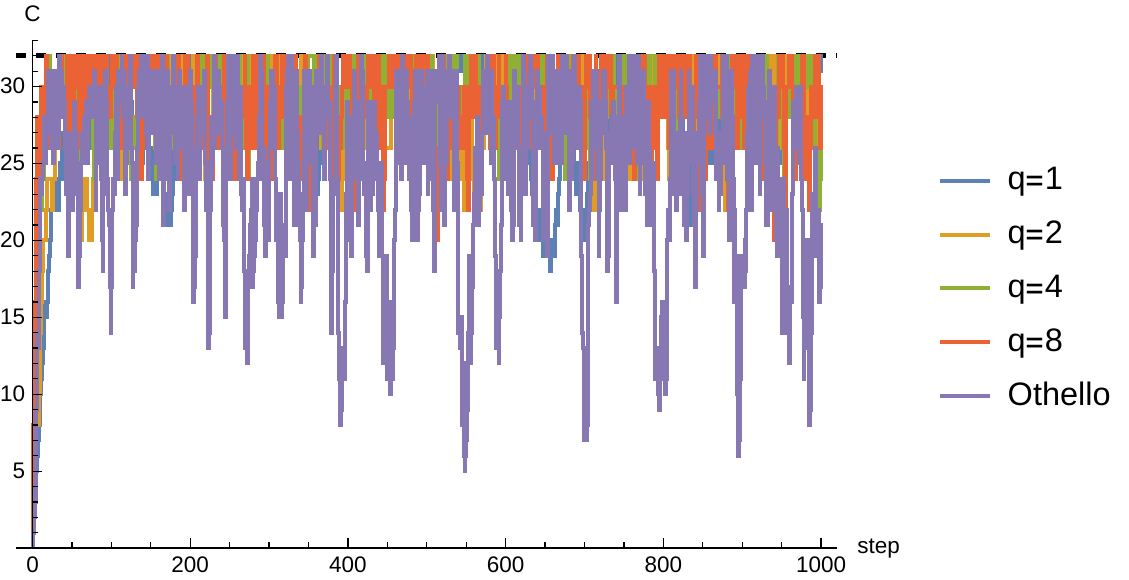}
		}
		\subfloat[$L=8$ (average)]{
			\includegraphics[scale=0.6]{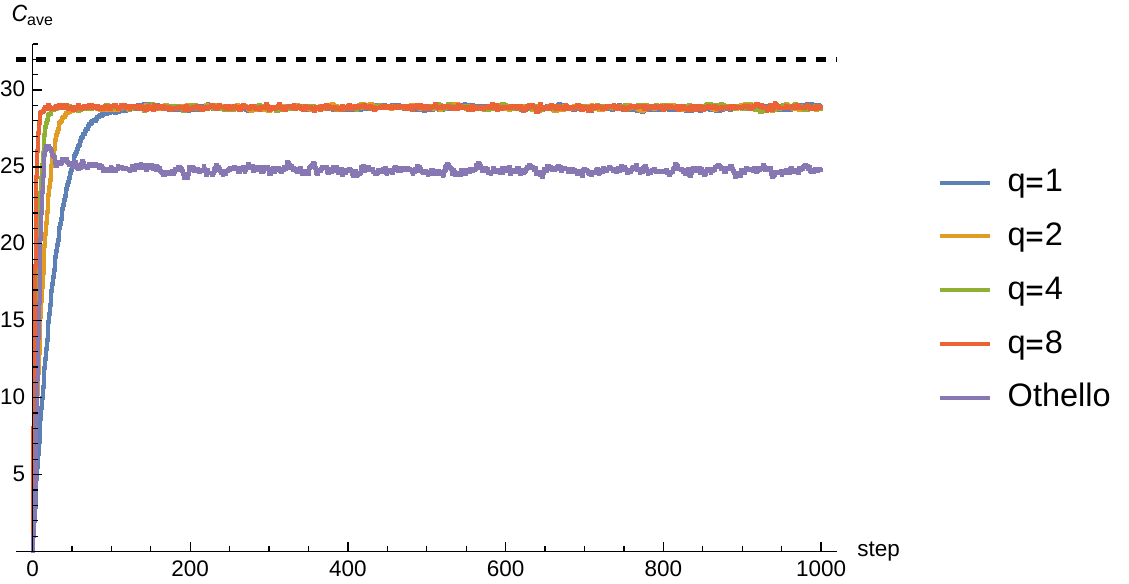}
		}\newline
		\subfloat[$L=16$]{
			\includegraphics[scale=0.6]{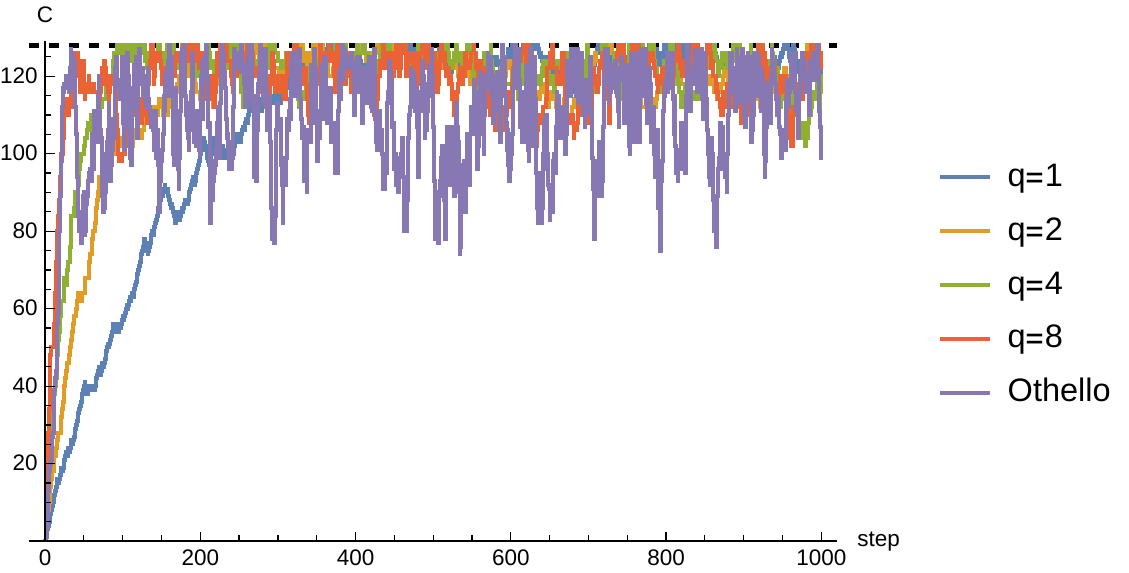}
		}
		\subfloat[$L=16$ (average)]{
			\includegraphics[scale=0.6]{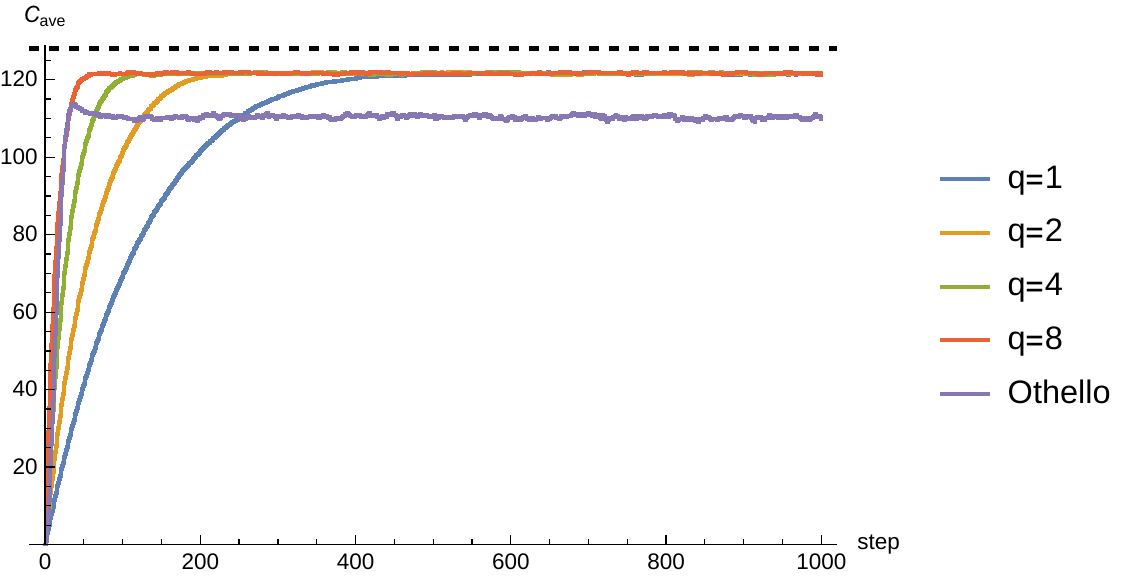}
		}\newline
		\subfloat[$L=32$]{
			\includegraphics[scale=0.6]{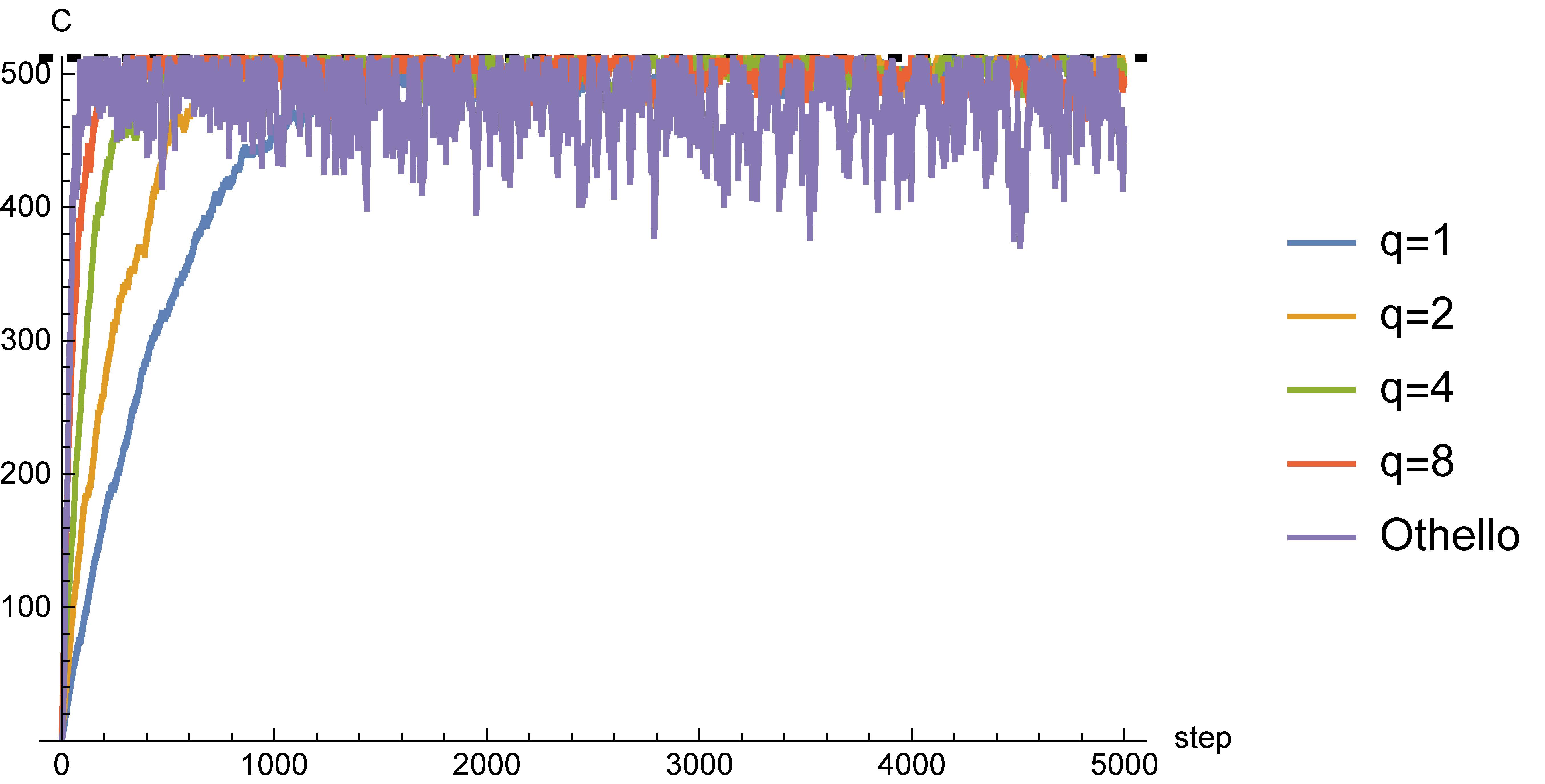}
		}
		\subfloat[$L=32$ (average)]{
			\includegraphics[scale=0.6]{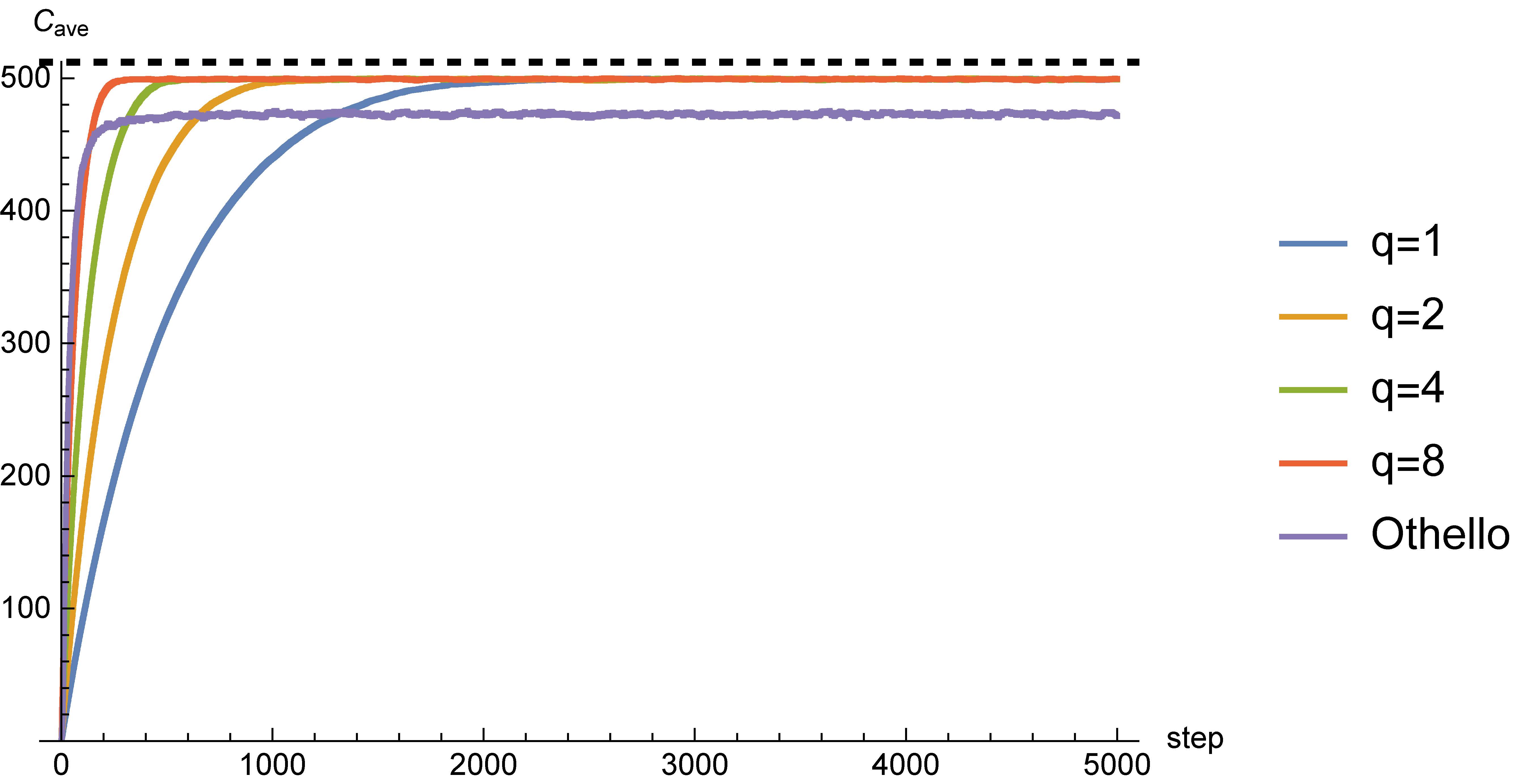}
		}
		\caption{Time evolution of the complexity in $\mathbf{Z}_{2}$ gauge theory on a lattice with the width $L=8, 16, 32$. Figures in the left side show the evolutions of complexities for $q=1, 2, 4, 8$ models and the Othello rule. Right figures show the averages over  $10^3$ samples. The dotted black line represents the possible maximum value $\mathcal{C}_{\rm max}=\lfloor L^2/2 \rfloor$. 
			\label{classic_complexities_z2}}
	\end{center}    
\end{figure}

We show, in Fig.~\ref{classic_complexities_z2}, the time evolution of the complexity for random flux models with $q=1,2,4,8$ interactions and the Othello rule for the $\mathbf{Z}_2$ gauge group on $L\times L$ spacial lattices ($L=8,16,32$). 
Each plot in the left figures is a sample, and one in the right figures shows an average over 1000 samples. 
All of them show the following typical shape of the time evolution:
\begin{itemize}
\item The complexity grows at early times.
\item Then it almost reaches the maximum $\mathcal{C}_{\rm max}$.
\item It fluctuates just below $\mathcal{C}_{\rm max}$.
\end{itemize}

\begin{figure}[tbp]
	\begin{center}
		\subfloat[$L=8$]{
			\includegraphics[scale=0.6]{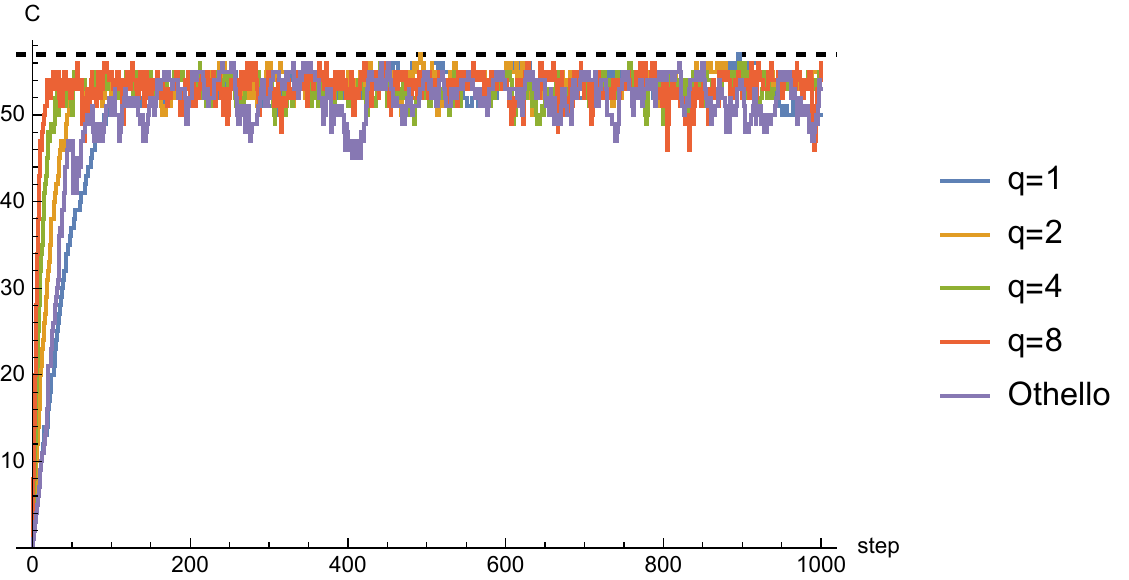}
		}
		\subfloat[$L=8$ (average)]{
			\includegraphics[scale=0.6]{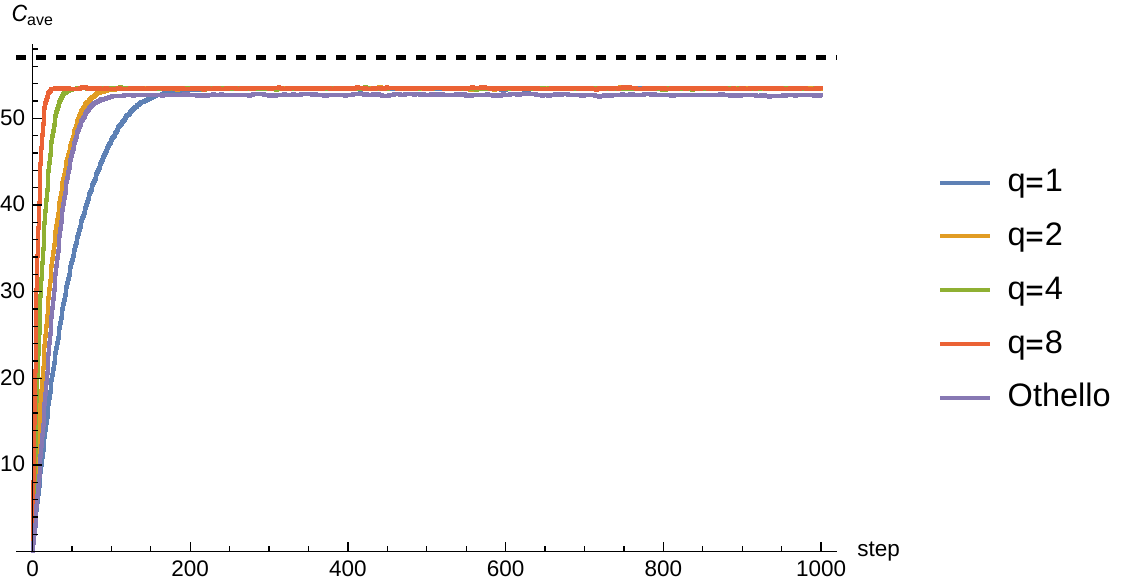}
		}\newline
		\subfloat[$L=16$]{
			\includegraphics[scale=0.6]{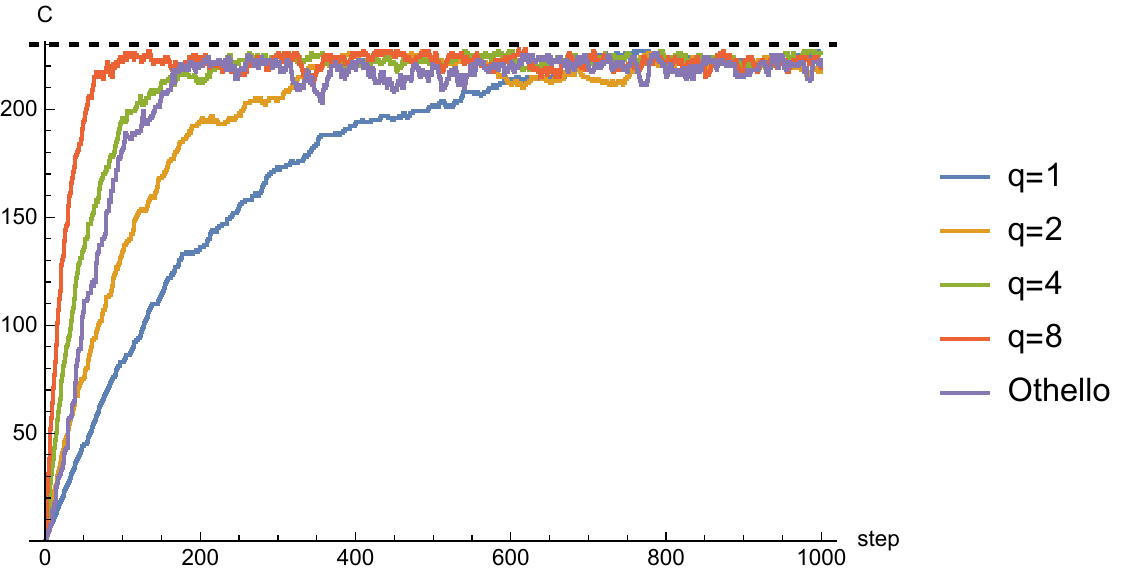}
		}
		\subfloat[$L=16$ (average)]{
			\includegraphics[scale=0.6]{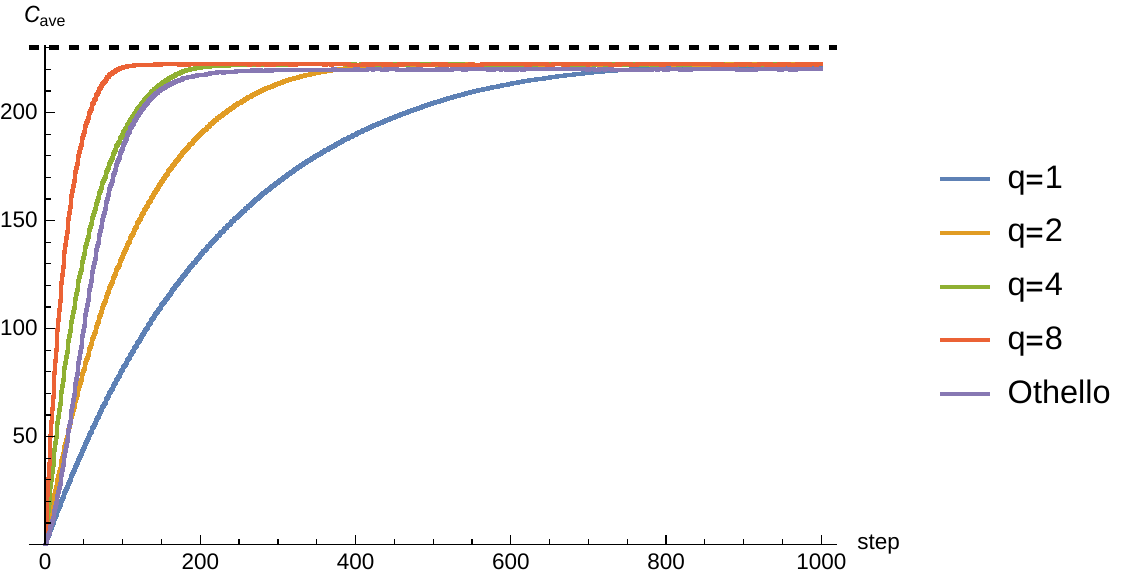}
		}\newline
		\subfloat[$L=32$]{
			\includegraphics[scale=0.6]{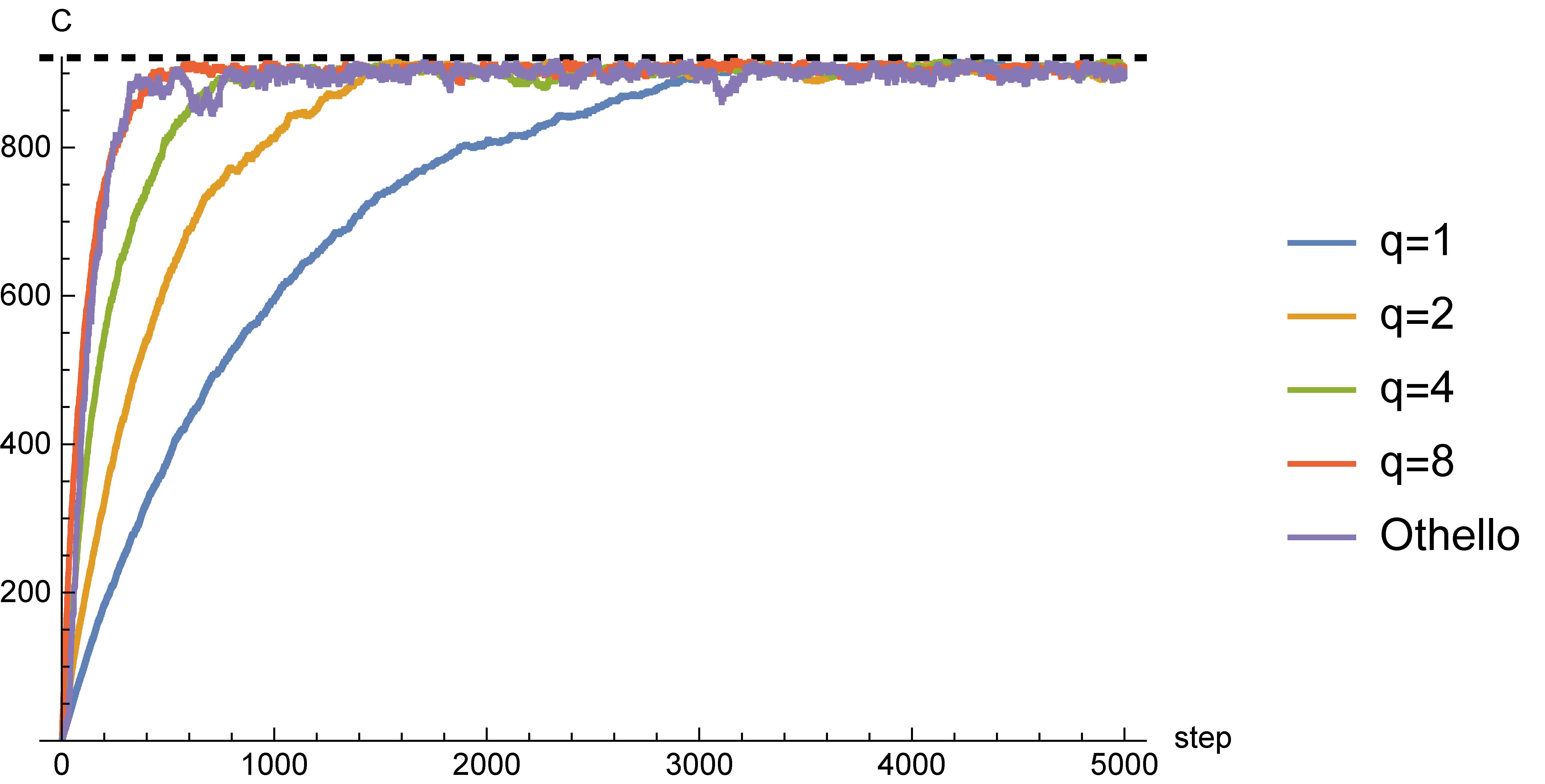}
		}
		\subfloat[$L=32$ (average)]{
			\includegraphics[scale=0.6]{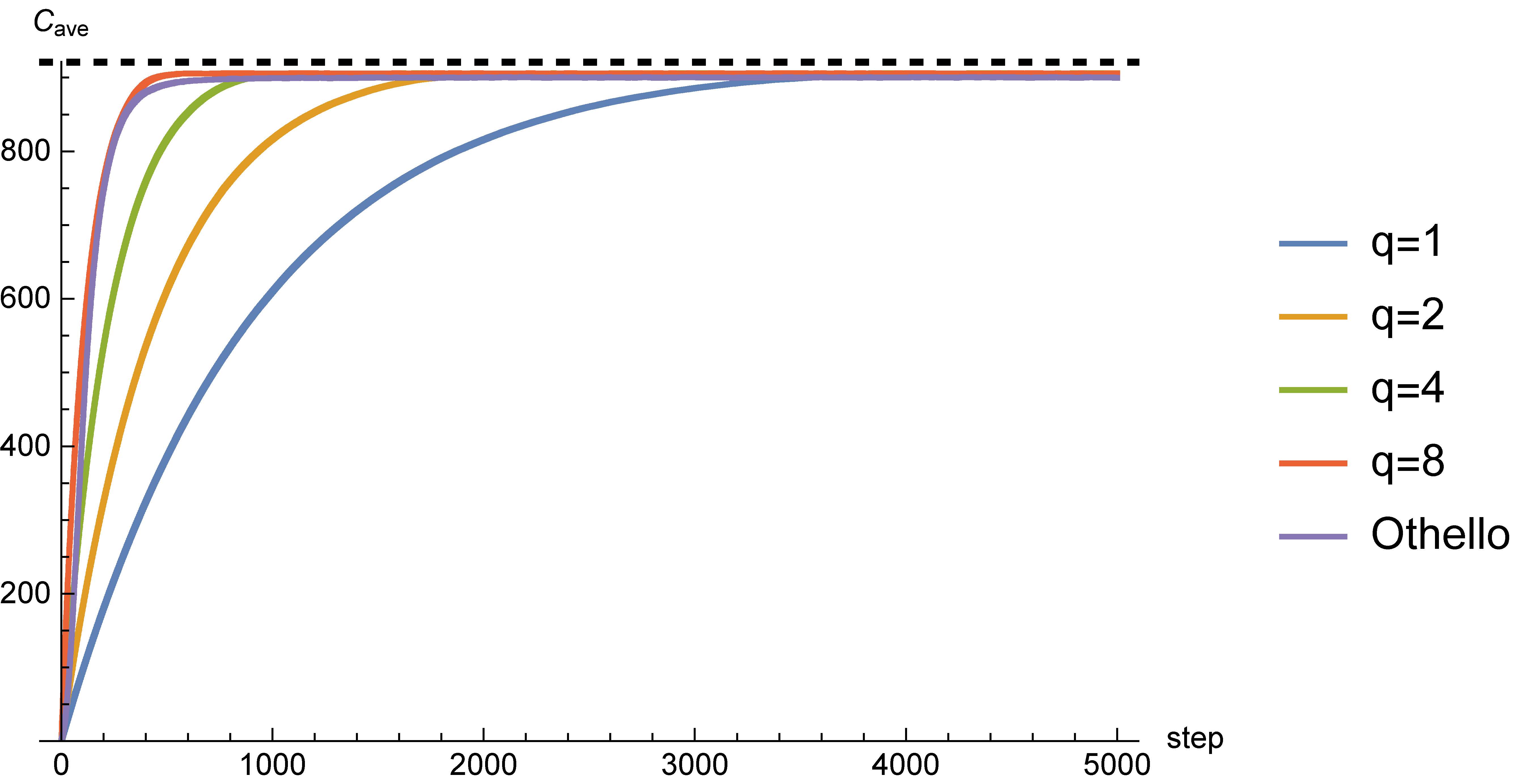}
		}
		\caption{Time evolution of the complexity in $\mathbf{Z}_{10}$ gauge theory. 
			Typical behaviors are almost the same as $\mathbf{Z}_{2}$ theory, though the fluctuation in the Othello rules is smaller compared to $\mathbf{Z}_{2}$ theory. The possible maximum value of the complexity is $\mathcal{C}_{\rm max}=\lfloor (9/10)L^2 \rfloor$.
			\label{classic_complexities_z10}}
	\end{center}    
\end{figure}

We also show the time evolution of $\mathbf{Z}_{10}$ gauge theories 
in Fig.~\ref{classic_complexities_z10}. 
They share the same properties with the $\mathbf{Z}_2$ gauge theories.\footnote{
As for the fluctuation after the complexity reaches around the maximum, 
the $\mathbf{Z}_{10}$ gauge theories have smaller fluctuations compared to the
$\mathbf{Z}_{2}$ gauge theories.
The Othello rule gives a larger fluctuation compared to that of the $q=1,2,4,8$ interactions, so 
the averaged complexity is smaller at late times than those for the others.
}

Looking at the detailed difference among the models, we notice that 
the complexity in the Othello rule grows the fastest when the lattice size is large. 
This shows the important fact that the nonlocality of the system rules 
makes the growth rate bigger.

%They show that fluctuations are smaller than those in $\mathbf{Z}_{2}$ theory. 
%They implies that However, since the fluctuation in the Othello rule is large, the averaged complexity for the rule is smaller at late times than those for the others.
%%%%

\begin{figure}[tbp]
	\begin{center}
		\includegraphics[scale=0.9]{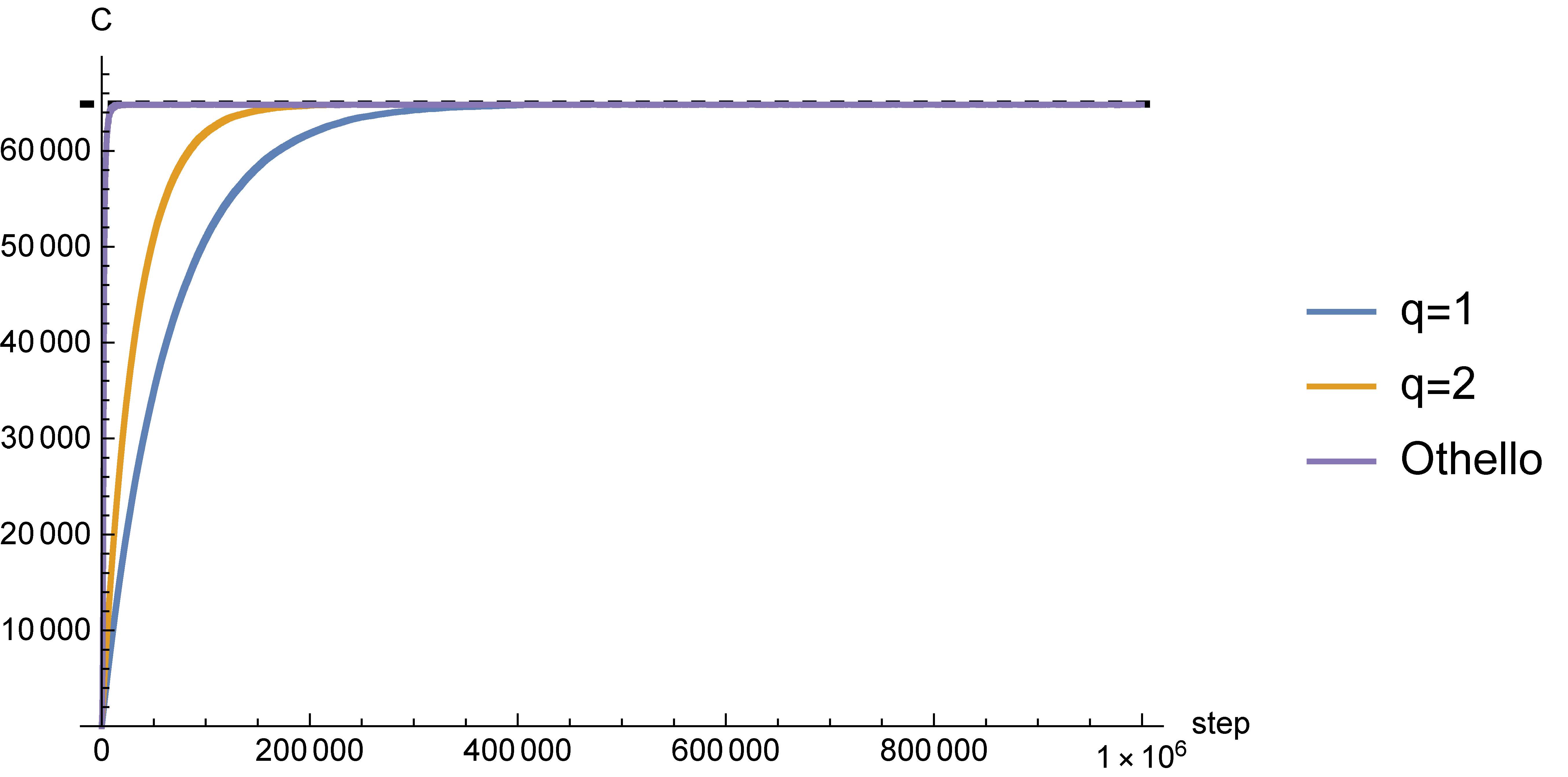}\\
		\vspace{10mm}
		\includegraphics[scale=0.9]{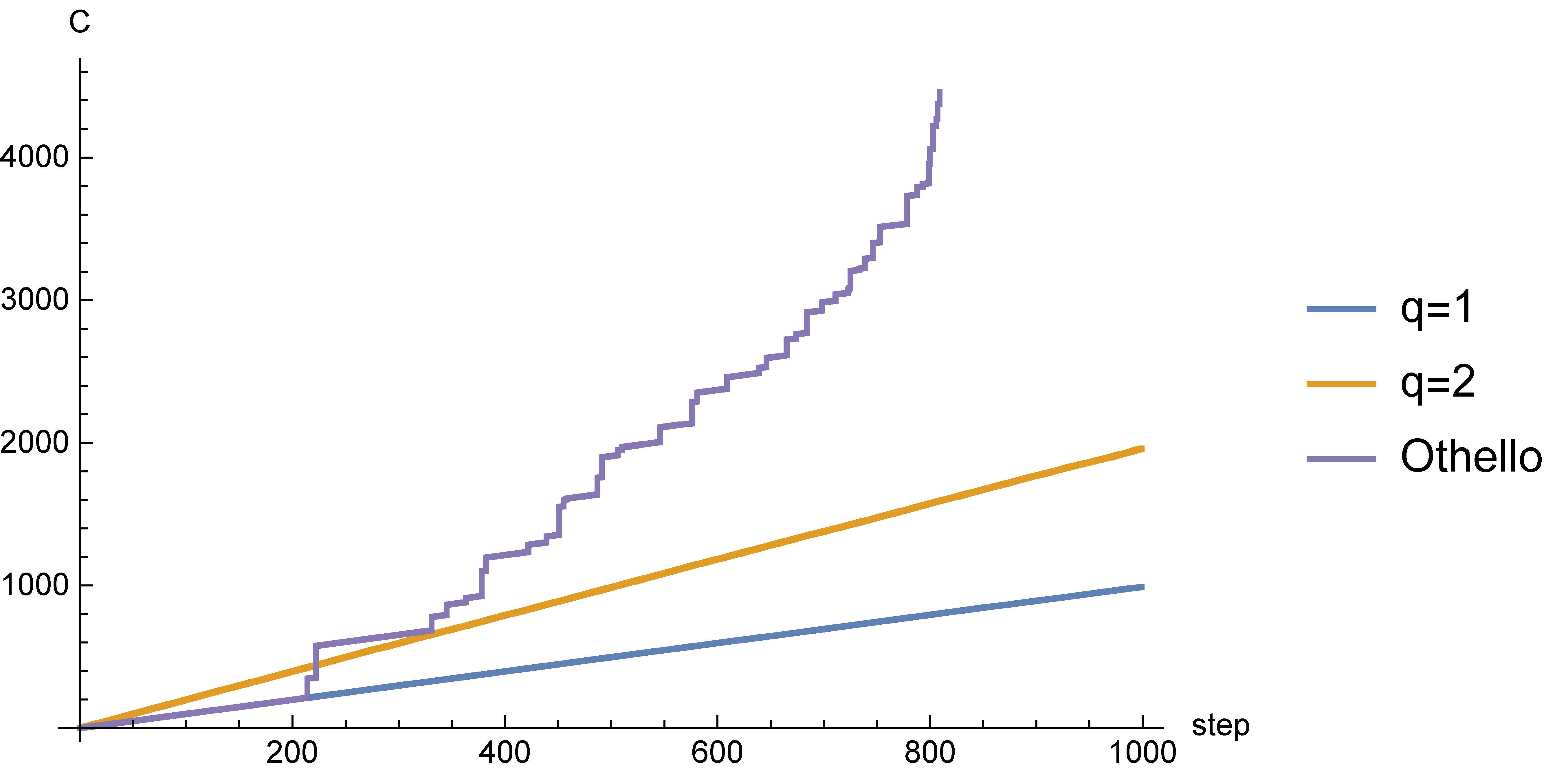}
		\caption{Time evolution of the complexity in $\mathbf{Z}_{100}$ gauge theory. The spatial lattice size is $2^8\times 2^8$. The upper figure shows the evolution for $0 \leq t \leq 10^6$ for $q=1$ (blue), $q=2$ (yellow) and Othello (purple). The dashed horizontal line represents the possible maximum value $\lfloor 2^{16} (99/100) \rfloor =64880$. Note that each plot is just a random sample, not the average of samples. 
			The lower figure is the zoom-in for $0\leq t \leq 10^3$. It represents the linear growth of the complexity for $q=1, 2$ at early times as $\mathcal{C} \sim q \, t$. \label{fig_Z100_L=256}}
	\end{center}    
\end{figure}

To find how the nonlocal Othello rule goes much faster, we consider the 
random flux models with $q=1,2$ and the Othello rule for $\mathbf{Z}_{100}$ gauge theory on a $2^8 \times 2^8$ spacial lattice. 
Fig.~\ref{fig_Z100_L=256} shows the growth of the complexity.
It exhibits a linear growth of the complexity for $q=1, 2$ models at early times, {\it i.e.} 
the complexity grows as $\mathcal{C} \sim q \, t$. For the Othello rule, the complexity grows in the same way as $q=1$ at very early times, but the growth suddenly becomes much faster, 
which is qualitatively different
from the $q=1,2$ models.
So, we conclude that the nonlocal Othello rule accelerates the growth rate drastically. 

Let us evaluate the speed of the growth numerically, to distinguish the nonolocal Othello rule and
the local rules quantitatively,
by looking at the
parameter dependence of the complexity in random flux models. 

\begin{figure}[tbp]
	\begin{center}
		\includegraphics[scale=0.9]{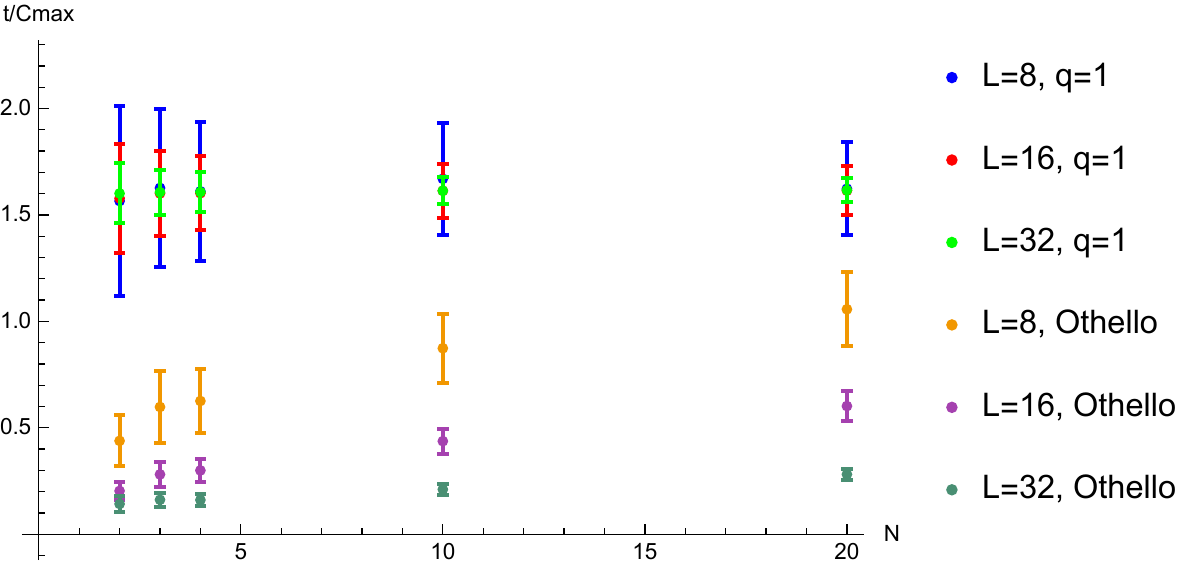}
		\includegraphics[scale=0.9]{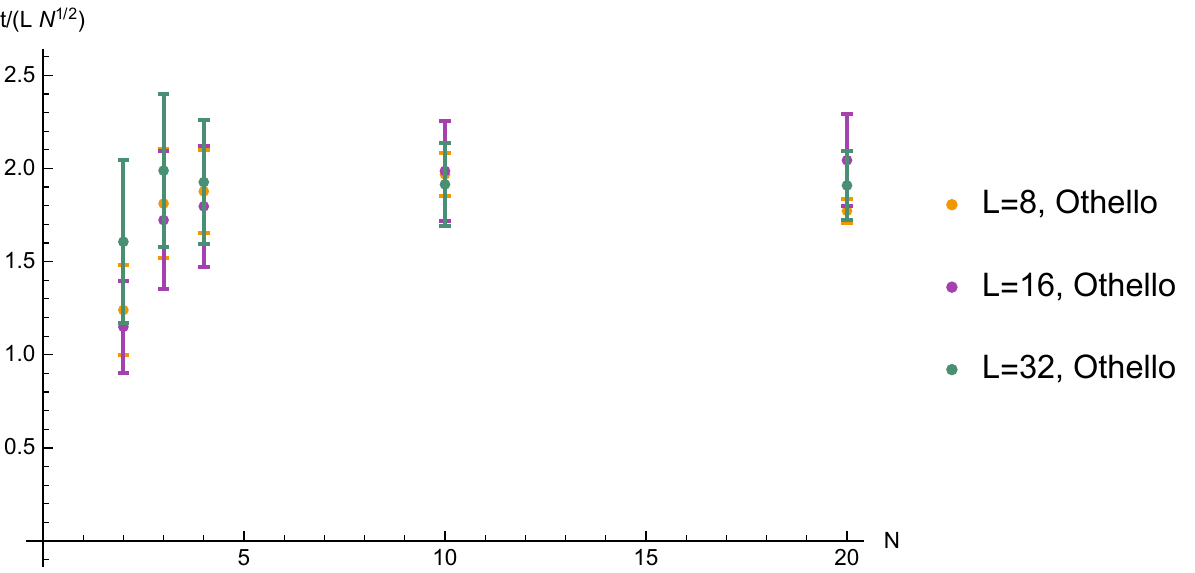}
		\caption{Plot of the time that the complexity first reaches the 80\% of the possible maximum value $\mathcal{C}_{\rm max}=\lfloor L^2 (N-1)/N \rfloor$ for $\mathbf{Z}_N$ gauge theories. We consider $\mathbf{Z}_N$ gauge theories $(N=2,3,4,10)$ on $L \times L$ lattices $(L=8,16,32)$ for $q=1$ and the Othello rule. In the upper figure, each dot represents the average of the time, which is rescaled by $\mathcal{C}_{\rm max}$, over 1000 samples, and the associated error bar represents the standard deviation. In the lower figure, the vertical scale is changed to $t/(L N^{1/2})$. \label{average_80percent}}
	\end{center}    
\end{figure}

Fig.~\ref{average_80percent} shows the averaged time to reach the 80\% of the possible maximum value $\mathcal{C}_{\rm max}=\lfloor L^2 (N-1)/N \rfloor$ for the 
$\mathbf{Z}_N$ gauge theories.\footnote{It is very rare that the complexity exactly reaches $\mathcal{C}_{\rm max}$. This is the reason why we consider the time to reach the 80\% of $\mathcal{C}_{\rm max}$.} 
The numerical data show 
that the time is proportional to $L^2 (N-1)/N$ for the random flux model 
with $q=1$. The data also show that the time is roughly proportional 
to $L \sqrt{N}$ for the model with the Othello rule.  
Therefore, we find that, for large $L$ with a fixed $N$, the Othello rule is parameterically
faster. This is the importance of the nonlocality to have a faster growth of the complexity.

The time dependence of the complexity in the random flux models resembles the behavior of the complexity in the random circuit model \cite{Brown:2016wib}. 
In the both models the complexity grows linearly in the time at early times and then fluctuates near the maximum value. However, the maximum values $\mathcal{C}_{\rm max}$ are different. 
In our models $\mathcal{C}_{\rm max}$ is proportional to $L^2$. For the $\mathbf{Z}_N$ gauge theories, $L^2$ corresponds to the number of qudits. 
The time to get maximum complexity is proportional to $L^2$ (excepted for the one with 
the Othello rule). The results agree with the general arguments in \cite{Susskind:2014moa}. 
On the other hand, $\mathcal{C}_{\rm max}$ in the random circuit model \cite{Brown:2016wib} is expected to be the power of the number of qubits, {\it i.e.}, proportional to $e^{L^2}$. The difference is due to the fact that our random flux models in this section are \textit{classical}. 
In the random flux models, the superposition of the states does not appear in the time evolution. 
Although it makes easier to calculate the complexity due to no entanglement, 
the model never achieve the large complexity of order ${\cal O}(e^S)$. 
In order to realize the large complexity which could be responsible for having a gravity dual, we need to consider more quantum models. 
In the next section, we fully treat the quantum $\mathbf{Z}_N$ gauge theories and will find
the large complexity.

%%%%%%%%%%%%%%%%%%%%%%%%%%%%%%%%%%%%%%%%%%%%%%%

%%%%%%%%%%%%%%%%%%%%%%%%%%%%%%%%%%%%%%%%%%%%%%%
%%%%%%%%%%%%%%%%%%%%%%%%%%%%%%%%%%%%%%%%%%%%%%%
\section{Complexity in quantum gauge theory}
\label{sec:quantum}
%%%%%%%%%%%%%%%%%%%%%%%%%%%%%%%%%%%%%%%%%%%%%%%
%%%%%%%%%%%%%%%%%%%%%%%%%%%%%%%%%%%%%%%%%%%%%%%

We would like to calculate quantum complexity ${\cal C}(t)$ 
for any given Hamiltonian, 
a reference state (a state at $t=0$)
and a universal gate set. However, it is a hard task to evaluate the complexity for general Hamiltonians including the dependence of the reference state. Thus, in this paper, we consider the complexity of time evolution operator for arbitrary diagonal Hamiltonians: 
%In this section, we describe and conduct the calculation of ${\cal C}(t)$ %and compare the properties 
%for arbitrary Hamiltonians: 
we will see especially the model dependence of the maximum complexity $\mathcal{C}_{\rm max}$ and 
the growth speed of the complexity. 
We will find that 
both the quantities increase drastically
with the nonlocality encoded in the Hamiltonians. In particular, to achieve $\mathcal{C}_{\rm max}
\sim e^S$, where $S$ is entropy of the system, the diagonal Hamiltonian needs to be maximally nonlocal.

First in Sec.~\ref{sec:Z2} we describe the 
$\mathbf{Z}_2$ gauge theories in 2 spatial dimensions for simplicity. 
The extension to the 
$\mathbf{Z}_N$ gauge theories and the $N\to\infty$ limit is straightforward, and will
be described in Sec.~\ref{sec:ZN}.
The results for $\mathcal{C}_{\rm max}$ and the growth speed are summarized 
in Table \ref{tab1} and Table \ref{tab2}, for the $\mathbf{Z}_2$ and $\mathbf{Z}_N$ gauge theories
respectively.

%%%%%%%%%%%%%%%%%%%%%%%%%%%%
\subsection{Complexity in $\mathbf{Z}_2$ gauge theory}
\label{sec:Z2}

Time evolution of the $\mathbf{Z}_2$ gauge theory is
provided by the unitary operator $e^{-iHt}$ for a given Hamiltonian $H$.
Then, %with the time-evolved state 
%\be
%\ket{\psi(t)} = e^{- i H t} \ket{\psi(0)} \,,
%\ee 
the quantum complexity 
${\cal C}(t)$ of the unitary operator $e^{-iHt}$ is defined as a minimum number of the quantum gates in 
the universal gate set ${\cal U}$ \eqref{Uni} which satisfies\footnote{As mentioned in footnote \ref{foot_state_op}, this complexity of operator is different from the complexity of state $\ket{\psi(t)} = e^{- i H t} \ket{\psi(0)}$ with the reference state $\ket{\psi(0)}$.} 
\begin{align}
\Bigm|\Bigm| \, e^{- i H t}- \prod_i U_{i}  
\Bigm|\Bigm| \, \, < \epsilon
\label{epdef}
\end{align}
where $|| \cdot ||$ is defined as 
\be
\Bigm|\Bigm| \, U  - V   
\Bigm|\Bigm|^2  \equiv \frac{1}{\Tr (1)} \Tr \Bigl[ (U  - V)^\dagger (U  - V) \Bigr] \,. 
\ee
$\epsilon$ ($\ll 1$) is a cut-off parameter for the effectiveness of the complexity,
and the quantum gates $U_i$ provide a minimum set 
to satisfy \eqref{epdef}.

%%%%%%%%%%%%%%%%%%%%%%%%%%%%%%%

\subsubsection{Classification of Hamiltonians by $k$-locality}
\label{Sec:klocal}

Let us consider a $\mathbf{Z}_2$ gauge theory on a $L\times L$ lattice.
For our later purpose, we classify all possible Hamiltonians by introducing the notion of
$k$-local Hamiltonians.

As described earlier, gauge-invariant states are given by a superposition of 
the basis vectors $\otimes_{(a,b)}  \ket{m_{(a,b)}}$
where $m=0,1$. Here $(a,b)$ specifies the position of the plaquette; $1\leq a,b\leq L$.
The dimension of the Hilbert space\footnote{In this section, we do not impose 
the global gauge symmetry \eqref{globalsymbyboundarycondition}
since we regard the $L\times L$ lattice as a part of a bigger lattice.}
 is $2^{L^2}$, so 
the gauge-invariant Hamiltonian is an element of $u(2^{L^2})$ which is an arbitrary 
$2^{L^2}\times 2^{L^2}$ Hermitian matrix.
It is written by an arbitrary combination of the plaquettes: the magnetic flux
operators penetrating the plaquettes of the lattice.

The standard lattice gauge theories employ a plaquette action 
which  is given by a sum of all the plaquettes  \cite{Kogut:1974ag}.
Among those, an example Hamiltonian consisting just of spatial plaquettes is 
\begin{align}
H_0 = J \sum_{(a,b)=(1,1)}^{(L,L)} M_{(a,b)} \,,
\label{Hamil-lat}
\end{align}
where $M_{(a,b)}$ is the magnetic flux operator acting on the plaquette $(a,b)$.%
\footnote{This Hamiltonian does not contain electric flux terms of the standard one-plaquette lattice
Hamiltonian. 
}
The real constant $J$ is the overall strength of the Hamiltonian, and determines the energy scale of it.
This magnetic flux operator $M_{(a,b)}$ acts on gauge-invariant states as
$M_{(a,b)} \ket{0}_{(a,b)} = \ket{1}_{(a,b)}$, and $M_{(a,b)} \ket{1}_{(a,b)} = \ket{0}_{(a,b)}$.
The gauge-invariant states are given by qubits assigned to the plaquettes, and the
magnetic flux operator is represented by $\sigma_1$ matrix acting on the qubit.
In this regard, the plaquette Hamiltonian \eqref{Hamil-lat} is written as
\begin{align}
H_0 = J \sum_{(a,b)} \left[
\mathbf{1}\otimes
\cdots \otimes \mathbf{1}\otimes
\underbrace{\sigma_1}_{(a,b)} \otimes \mathbf{1}\otimes
\cdots
\right]
\label{H0ab}
\end{align}
where the $\sigma_1$ entry is inserted at the plaquette location $(a,b)$.
For example, if we pick up a subsector consisting of a $1\times 2$ lattice, we have simply
$H_0 = J \left[
\mathbf{1}\otimes
\sigma_1 + \sigma_1 \otimes \mathbf{1}
\right]$.

We can generalize the Hamiltonian \eqref{Hamil-lat} such that each plaquette takes different 
weights:
%The complexity takes only two values, $0$ and $L^2$. This is simply because
%the Hamiltonian \eqref{Hamil-lat} treats all the plaquettes with the same weight.
With arbitrary real coefficients $J_{(a,b)}$,
\begin{align}
H_0 = \sum_{(a,b)} J_{(a,b)} M_{(a,b)} \, .
\label{HJM}
\end{align}
This Hamiltonian feels randomly distributed magnetic fluxes penetrating the 2-dimensional
surface.
%Since the distribution is random, $J_{(a,b)}$ is generically irrational
%with each other.  

The Hamiltonian $H_0$ of \eqref{Hamil-lat} or \eqref{HJM} is the simplest one, since the effect of each element is restricted to acting on 
only one plaquette. 
This Hamiltonian contains only magnetic flux operators. 
However one can introduce more
interacting Hamiltonians where several different qubits are entangled. 
Generically, one can consider a Hamiltonian, whose elements are given as 
\begin{align}
\left(\mathbf{1} \; \mbox{or} \; \sigma_{i_{(1,1)}}\right)\otimes
\left(\mathbf{1} \; \mbox{or} \; \sigma_{i_{(1,2)}}\right)\otimes
\left(\mathbf{1} \; \mbox{or} \; \sigma_{i_{(1,1)}}\right)\otimes
\cdots \otimes 
\left(\mathbf{1} \; \mbox{or} \; \sigma_{i_{(L,L)}}\right)
\,,
\label{0x0x}
\end{align}
where $i_{(1,1)}, i_{(1,2)}, i_{(1,3)}, \cdots , i_{(L,L)} $= 1, 2, 3;  
there are $L^2$ qubits and on each qubit, we have a choice among $\mathbf{1}$, $\sigma_1, \sigma_2, \sigma_3$. 
%\if0
%In order to classify the properly of all the Hamiltonians,
%we would like to focus on the eigenvalues, since 
%the time evolution via the Hamiltonian is characterized by them. 
%In other words, we consider the diagonalized Hamiltonian in order to see the time evolution. 
%This does not loose any generality; since if the Hamiltonian is not diagonalized and written as \eqref{0x0x} for example, one can 
%always choose another basis and make the Hamiltonian diagonalized. 
%%In the unperturbed Hamiltonian $H_0$ given by \eqref{H0ab}, the magnetic flux operator is
%%inserted at each face of the lattice, with the same weight $J$. We introduce the interacting Hamiltonian
%%$H_I$ and study the maximum complexity, and seek for the fastest Hamiltonian. See Table \ref{tab1}
%%for the summary of our study.
%%
%%
%%
%%According to the guidelines for the calculation of the complexity, 
%%we are interested in the eigenvalues of the Hamiltonian.
%%In general, the eigenvalues are given by all the elements of the maximal
%%torus of $u(2^{L^2})$. 
%%The dimension of the maximal torus is $2^{L^2}-1$. 
%%
%\fi
If we add the trace part, then a
generic Hamiltonian can be written as\footnote{
This generic Hamiltonian of course includes electric flux terms
as well as the magnetic terms, see Appendix \ref{AppA}. 
}
\begin{align}
H = \sum_{I_{(1,1)}, I_{(1,2)}, \cdots , I_{(L, L)}} a_{I_{(1,1)} I_{(1,2)} \cdots I_{(L, L)}}  \,
\sigma_{I_{(1,1)}} \otimes \sigma_{I_{(1,2)}} \otimes \cdots \otimes \sigma_{I_{(L,L)}} \, ,
\label{0303}
\end{align} 
where $I_{(a,b)}=0, 1, 2, 3$ and we define $\sigma_0\equiv \mathbf{1}$. 
The trace part is given by 
$\mathbf{1}\otimes\mathbf{1}\otimes \cdots \otimes \mathbf{1}$.

To classify the Hamiltonians, 
let us introduce the notion of $k$-local Hamiltonians \cite{Brown:2017jil}. 
The Hamiltonian is called $k$-local if the maximum number of $\sigma_i$ ($i$ = 1, 2, 3) in the terms 
of the Hamiltonian \eqref{0303} is $k$. For example, the one-plaquette Hamiltonian $H_0$ of
\eqref{Hamil-lat} or \eqref{HJM} is 1-local, since $M_{(a,b)}$ act only on one plaquette at $(a,b)$ 
as is seen explicitly in \eqref{H0ab}.  
Since we have $L^2$ sites, $k$ can go up to $L^2$. 
We shall call $k$-local Hamiltonian ``nonlocal Hamiltonian", if $k = {\cal{O}}(L^2)$. This is because it 
involves the operation acting on all of the  lattice scale plaquettes simultaneously. 

We will see that the integer $k$ for the $k$-local Hamiltonians {\it dictates how large the maximum 
complexity $\mathcal{C}_{\rm max}$ is}, and {\it how fast the complexity grows in the time evolution}.

%%%%%%%%
\subsubsection{Complexity in one-plaquette Hamiltonian system}
\label{sec:one-p}

%We are ready for calculating the complexity for the $\mathbf{Z}_2$ gauge theory.

\begin{figure}
\begin{center}
\includegraphics[scale=0.5]{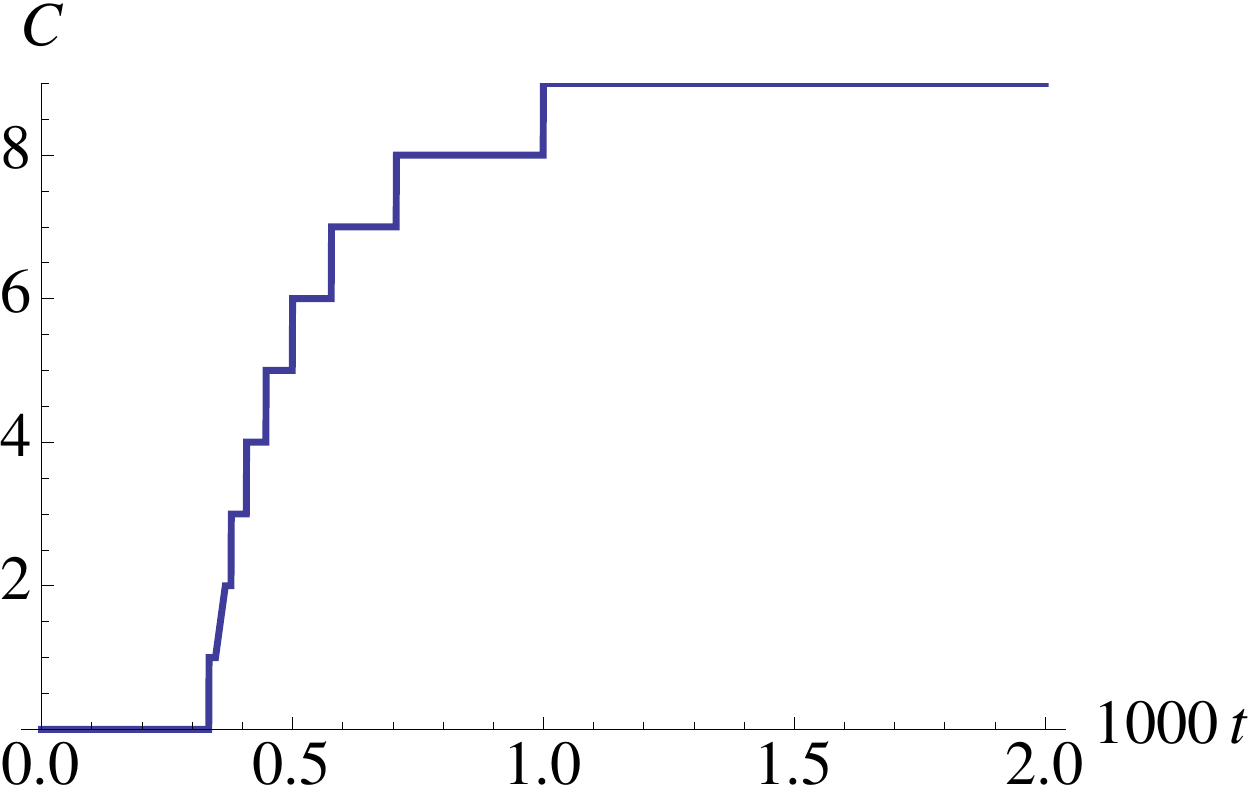}
\hspace{5mm}
\includegraphics[scale=0.52]{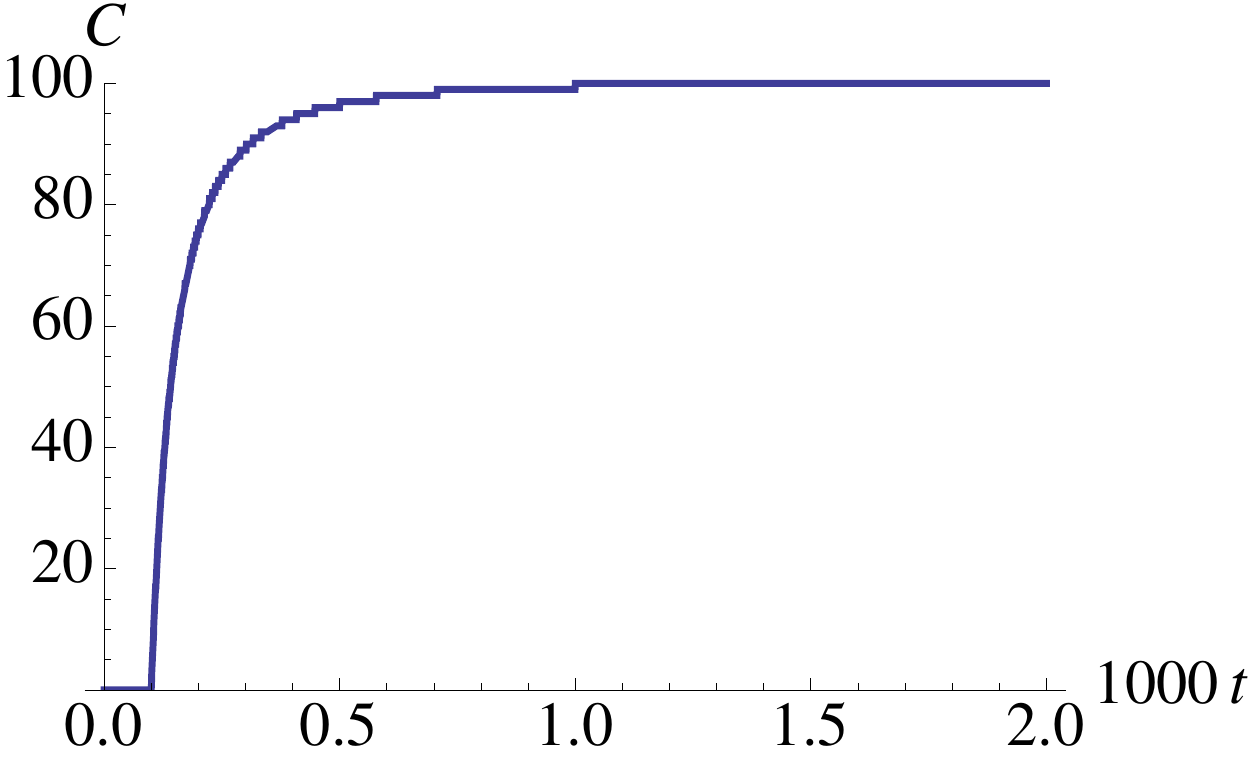}
\end{center}
%\vspace{-13mm}
\caption{The time evolution of the quantum complexity ${\cal C}(t)$ given by \eqref{Crand}. 
For an illustration we have chosen $\epsilon=10^{-3}$ and $J=1$.
We chose $L=3$ (Left) and $L=10$ (Right). }
\label{fig:Crand}
\end{figure}

%First, let us calculate the time evolution of the complexity for the system given by the
%lattice Hamiltonian 
Let us calculate the time evolution of the complexity for 
the simplest one-plaquette Hamiltonian \eqref{Hamil-lat}, or equivalently \eqref{H0ab}, which is 1-local.
The time evolution operator is written as 
%\footnote{Here we maintain the non-diagonalized Hamiltonian of the form \eqref{H0ab}, instead of the diagonalized form Hamiltonian \eqref{0303} since the calculation is straightforward.}
\begin{align}
e^{-iHt} = \prod_{(a,b)} X_{(a,b)}(t) \, , \quad 
X_{(a,b)}(t) \equiv \exp[-i J \sigma_1 t]\, .
\end{align}
%the time-evolved state is given by
%$|\psi(t)\rangle =\prod_{a,b} X_{(a,b)}(t) |\psi(0)\rangle$.
%which is already of the form of the definition of the complexity, \eqref{epdef}.
Then the complexity is defined as the minimum number of single qubit gates $U_i$ which are necessary to
satisfy the following equation,
\begin{align}
\Bigm|\Bigm| \, \prod_{(a,b)} X_{(a,b)} - \prod_{i} U_i 
\Bigm|\Bigm| \, < \epsilon\, .
\label{Caccept}
\end{align}
Initially at $t=0$, the complexity is zero, ${\cal C}=0$. Then, at time $t=t^{(1)}$
which is a solution of the equation 
\begin{align}
2\left(1-(\cos(Jt^{(1)}))^{L^2}\right) = \epsilon^2 \, ,
\end{align}
one needs at least one gate $U$ to satisfy the inequality \eqref{Caccept},
and the complexity grows to ${\cal C}=1$. Then, next, at time $t=t^{(2)}>t^{(1)}$,
one needs at least two gates $U$ to satisfy \eqref{Caccept}. The time $t^{(2)}$ 
is given by
a solution of the equation
\begin{align}
2\left(1-(\cos(Jt^{(2)}))^{L^2-1}\right) = \epsilon^2 \, .
\end{align}
This procedure continues until the complexity reaches its maximum value, 
\begin{align}
{\cal C}_{\rm max} = L^2 \, .
\end{align}
The timing of the growth of the complexity is solved for $\epsilon\ll 1$ as
\begin{align}
t^{(n)} = \frac{\epsilon}{J\sqrt{L^2+1-n} } \, .
\label{tn}
\end{align}
Thus, we obtain the complexity at early times as
\begin{align}
{\cal C}(t)=\sum_{n=1}^{L^2} \theta(t-t^{(n)}) \, .
\label{Crand}
\end{align}
Note that the complexity is apparently a periodic function of time, with the period $2\pi/J$.
The expression \eqref{Crand} is valid only for the early times, $t \leq t^{(L^2)} = {\cal O}(\epsilon/J)$ 
for $\epsilon \ll 1$. See Fig.\ref{fig:Crand}.

Setting ${\cal C}_{\rm max} = L^2$, and ${\cal C}(t) = n$,   
\eqref{tn} implies  
\be
t J\sqrt{{\cal C}_{\rm max} + 1 -{\cal C}(t)}=\epsilon \,,  
\ee
and this gives the complexity density as a function of time as 
\begin{align}
\label{timeevoCforZ21local}
{{\cal C}}(t) = L^2
\left[1 + \frac{1}{L^2}- \left(\frac{\epsilon}{JL }\right)^2 \frac{1}{t^2}\right]\,, 
\end{align}
for the range $t^{(1)}\leq t \leq t^{(L^2)}$, which is, $\epsilon/JL \leq  t \leq \epsilon/J$. 

In summary, the time evolution of the complexity is as follows; For $0\leq t \leq t^{(1)} = \epsilon/JL$,
the complexity is zero. At $t=t^{(1)}$, the complexity density 
starts to grow and obey  \eqref{timeevoCforZ21local} and 
it asymptotically approaches
the maximum complexity density $L^2$ at $t \ge t^{(L^2)} = \epsilon/J$.

Next, let us consider slightly generalized Hamiltonian \eqref{HJM}, which is again 1-local.
The initial growth of the complexity is calculated as the same expression \eqref{Crand}, but now with
$t^{(n)}$ defined by
\begin{align}
t^{(n)}\equiv \frac{\epsilon}{\sqrt{\sum_{k=1}^{L^2+1-n}(J^{(k)})^2}}   \,.
\end{align}
Here we aligned the coefficients 
$J_{(a,b)}$ to form $\{J^{(1)}, J^{(2)}, \cdots, J^{(L^2)} \}$
such that $|J^{(1)}| < |J^{(2)}|<  \cdots < |J^{(L^2)}|$. We have assumed $\epsilon \ll 1$ again. 
%To derive the result \eqref{Crand}, we noticed that every component of the Hamiltonian
%$J_{(a,b)} M_{(a,b)}$ contributes as a rotating phase between $|0\rangle$ and $|1\rangle$.
%So every sector grows at very early times, and the difference of the growing speed, $J_{(a,b)}$,
%provides the steps in \eqref{Crand}.
The complexity reaches maximum at $t = t_{\rm max}$, where 
\begin{align}
\label{tmaxforcom}
t_{\rm max} = t^{(L^2)} = \frac{\epsilon}{J^{(1)}} = \epsilon \times \mbox{max}
\left\{ (J_{(a,b)})^{-1} \right\} \, .
\end{align}
If we define the the speed $v$ to reach the maximum complexity as 
\be
v\equiv \frac{{\cal C}_{\rm max}}{t_{\rm max}} \,,
\ee  
then, 
\begin{align}
v = \frac{L^2}{\epsilon} \mbox{min} \left\{ J_{(a,b)}\right\} \, .
\end{align}

After the initial growth, the complexity reaches
the maximum value $L^2$ and keeps the value close to $L^2$.  
Occasionally the complexity decreases  by a small integer but it goes back soon to $L^2$. 
At a very few occasions it can go down to zero. This occasion
is at the time when all multiples of $2\pi/J_{(a,b)}$ coincide with each other 
within the accuracy given by $\epsilon/J_{(a,b)}$. This can be understood as follows:
every plaquette has a rotating relative phase between $|0\rangle$ and $|1\rangle$,
and the complexity goes down to zero only at the instance when all the phases vanish
simultaneously within a given accuracy.

Note that the maximum complexity is ${\cal O}{(L^2)}$, far smaller than generic cases where we expect the maximum complexity is ${\cal O}(2^{L^2})$. The reason is that the Hamiltonian \eqref{Hamil-lat}, or \eqref{HJM} 
is just 1-local, {\it i.e.,} it consists of just the single qubit gates, and each qubit never interact 
with each other, as a result, they never create entanglement.  
Due to this independent structure of the gates, and thus the maximum complexity is just equal to the number of sites which is ${\cal O}{(L^2)}$. 

We are more interested in generic cases and how large the maximum complexity can become, 
and how fast the growth of the complexity can be, depending on the choice of interacting Hamiltonians. In the following, we will study more generic $k$-local ($k > 1)$ situations.

%%%%%%%%
\subsubsection{Complexity in 2-local case} %2-site case}
\label{Z22-localanalysis}

Let us introduce ``interactions'' to the one-plaquette Hamiltonian \eqref{Hamil-lat} of the $\mathbf{Z}_2$ 
gauge theory. We will see how these interactions increase the maximum complexity.  

%To study generic cases later, 
Let us start with the simplest situation in which we introduce an entanglement
between two neighboring qubits. 
Consider a sub-sector involving only the two qubits, resulting in $2^2$ dimensional Hilbert space.
%The Hamiltonian is given by the unperturbed Hamiltonian $H_0$ given by 
%\eqref{H0ab} and an interaction Hamiltonian $H_I$,
%\begin{align}
%H = H_0 + H_I \,  .
%\end{align}
The Hamiltonian is an arbitrary Hermitian $4\times 4$ matrix, expanded as
%. So the Hamiltonian  can be generically written as
\begin{align}
H = a_{00} \; (\mathbf{1}\otimes \mathbf{1})
+ \sum_{i=1}^3 a_{0i}\; (\mathbf{1}\otimes \sigma_i)
+ \sum_{i=1}^3 a_{i0}\; (\sigma_i \otimes \mathbf{1})
+ \sum_{i,j}a_{ij} \; (\sigma_i \otimes \sigma_j)
\end{align}
with real coefficients.
The first $a_{00}$ term is boring since it commutes with everything, therefore we set $a_{00}=0$ for simplicity.  
$a_{i0}$ and $a_{0i}$ terms
are essentially the same as \eqref{H0ab} once we diagonalize the Hamiltonian by appropriate basis choice. 
So we call them $H_0$, since it acts on the single qubits independently. In other words, it is 1-local.
The interaction Hamiltonian is given by the last term which we call $H_I$. 
This term entangles the two qubits, therefore, this $H_I$ is 2-local. 
The generalization to the $L\times L$ lattice will be studied later.
%In addition, we assume an exchange symmetry among the two qubits, which leads to $a_{ij}=a_{ji}$.
%So, we are led to the generic interaction with a symmetric real coefficient $a_{ij}$,
%H_I = \sum_{i,j}a_{ij} \; (\sigma_i \otimes \sigma_j) \, .
%\label{HI}
%\end{align}
%The physical difference between the original Hamiltonian $H_0$ and the interaction Hamiltonian $H_I$
%is that $H_0$ generates no entanglement while $H_I$ generates entanglement among the two qubits.
%In fact, one can show that time evolution by $H_0$ does not produce entanglement entropy for
%product state at $t=0$.

We are interested in how the complexity ${\cal C}(t)$ grows in time. In general, it is a difficult
problem to decompose the unitary transformation for time evolution, $\exp[-iHt]$, into a product of
minimal number of gates. Therefore in this paper, 
we focus on generic {\it diagonal} Hamiltonians.
The diagonal Hamiltonian $\Lambda$ can be regarded as a full set of
possible electric flux terms in view of ordinary lattice gauge theories. As shown in
Appendix \ref{AppA}, the electric flux operator, acting on each link, 
provides just a phase factor.
So any electric flux operator is generated by a diagonal Hamiltonian.\footnote{
While restricting to only either electric or magnetic Hamiltonians prevents us from taking a Lorentz-invariant
continuum limit, quantum mechanically the Hamiltonian itself is a consistent theory.
}
For a generic diagonal Hamiltonian $\Lambda$, 
we count the minimal number of gates which are used for reconstructing 
$\exp[-i\Lambda t]$. 

% where $\Lambda$ is a diagonalized Hamiltonian.
As a side note,
assuming that $H$ is a generic Hamiltonian and $\Lambda$ is the diagonalization of $H$,
we can argue that $\Lambda$ captures a state-independent part of the complexity. %This can be seen as follows: 
Suppose we diagonalize the Hamiltonian by a constant unitary matrix $U_0$ as 
\begin{align}
U_0 HU_0^\dagger = \Lambda\, , \quad \Lambda \equiv 
\mbox{diag} \{e_1,e_2,e_3,\cdots, e_{2^{L^2}}\} \, .
\end{align}
The real number $e_i$'s are the eigenvalues of the Hamiltonian. By our choice of $a_{00} =0$, we have $\sum_i e_i=0$. Using this expression, we find that the $U_0$ part does not
change in the time evolution, while the time evolution is encoded completely in the eigenvalue part,
\begin{align}
\exp[-i H t] = U_0^\dagger \exp [-i\Lambda t] U_0 \, . 
\label{decomposition}
\end{align}
The complexity depends on the reference state at $t=0$, and this dependence is expected to
be encoded in $U_0$. %The complexity of $U_0$ does not evolve with time, therefore, 
%Since $U_0$ is time-independent and also state-dependent, 
%the time dependence of the state-independent part of the 
%complexity, which is expected to be 
%universal among all reference states, is captured by $\Lambda$.
Due to the fact 
${\cal{C}}(\exp[-i H t]) \neq {\cal{C}}(U_0^\dagger) + {\cal{C}}(\exp [-i\Lambda t])+ {\cal{C}}(U_0)$ generically, in order to 
see the state-dependent part of the complexity, we need to evaluate ${\cal{C}}(\exp[-i H t])$ directly. In this paper we treat only the complexity of
$\exp [-i\Lambda t]$, and leave the evaluation of the complexity of $(\exp[-i H t])$
for a future problem.

It is easy to decompose the unitary operator $\exp[-i\Lambda t]$  for the time evolution with
$\Lambda \equiv \mbox{diag} \{e_1,e_2,e_3, e_4\}$ into a product of
the gates. We prepare three gates (which are commutative with each other)
\begin{align}
 U_1(a) &\equiv \exp[-i at  (\sigma_3 \otimes \mathbf{1})]\,, \\
 U_2(b) &\equiv \exp[-i bt  (\mathbf{1}\otimes \sigma_3)]\, ,  \\
 U_{\rm ent}(c) 
&\equiv \exp[-i ct  (\sigma_3 \otimes \sigma_3)] = U^{\rm CNOT}_{12} U_2(c)U^{\rm CNOT}_{12}\, . 
\label{u3c}
\end{align}
Here $a$, $b$ and $c$ are real parameters.\footnote{In our convention, $\sigma_3 \otimes \mathbf{1} = \rm{Diag} (1, 1, -1, -1)$.} The gates $U_1(a)$ and $U_2(b)$ are single-qubit gates,
acting on the first or the second qubit respectively. The last one $U_{\rm ent}(c)$ is entangling the two
qubits, which can be constructed by using two CNOT gates and a single qubit gate, as shown above.
So, all of these are independent and are constructed by the gates in the universal gate set.
Using these unitary operators, arbitrary eigenvalues can be reproduced as
\begin{align}
\exp[-i\Lambda t] = U_1(a)U_2(b) U_{\rm ent}(c) \,,
\end{align}
where 
\begin{align}
&e_1 = a + b + c \, ,\,\, \,\, \,  \quad e_2  = a-b-c\, , \nn \\\quad
&e_3  = -a+b-c \, ,\quad e_4  = -a -b+c \, .
\label{lambda4}
\end{align}
This uniquely determines $(a,b,c)$ for any given Hamiltonian.\footnote{Some 
other choice of the order of the eigenvalues can be taken into account by 
exchanging $\{a,b,c\}$ and changing the signs of them. As we will see, the complexity
is determined simply the set $\{ |a|,|b|,|c|\}$.} 
%The constant unitary matrix $U_0$
%can be reconstructed by a product of constant gates, so we finally obtain a formula
%\begin{align}
%\exp[-i H t] = U_0^\dagger U_1(a)U_2(b) U_{\rm ent}(c)U_0 \, 
%\end{align}
%which fully decomposes the time evolution unitary operator into a product off the gates,
%where the time dependence is explicit.

Now, we count the number of gates to evaluate the complexity. For example, 
for the case where $|a|,|b|,|c|$ are parameterically different from each other,
ignoring the state-dependent
constant complexity of $U_0$,  the time evolution of the  complexity is
\begin{align}
{\cal C}(t) = \theta(|\sin at| -\epsilon)+ \theta(|\sin bt| -\epsilon)+ 3\theta(|\sin ct| -\epsilon) \, .
\end{align}

The maximum complexity is given by ${\cal C}_{\rm max} = 5$.
The time when the system reaches the maximum complexity,
denoted by $t_{\rm max}$, is given by
\begin{align}
t_{\rm max} = \epsilon \times \mbox{max}\{|a|^{-1}, |b|^{-1}, |c|^{-1} \} \, ,
\label{tmax}
\end{align}
for $\epsilon \ll 1$. So the speed $v\equiv {\cal C}_{\rm max}/t_{\rm max}$ to reach the maximum complexity
is
\begin{align}
v
= \frac{5}{\epsilon} \mbox{min}\{|a|, |b|, |c| \}  \, .
\end{align}

Let us ask one of the main questions in this paper: 
{\it what kind of Hamiltonian gives the fastest growth of the complexity?}
If we look at the formula \eqref{tmax}, the growth rate can be arbitrarily fast, if we make $a,b,c \to \infty$.
However, this does not make sense, since the eigenvalues of the Hamiltonian become $\pm \infty$; the theory 
does not have well-defined spectrum. To evaluate a class of well-defined Hamiltonians with overall time-scale fixed, we need to fix a variance 
$\sigma$ of the eigenvalues of the Hamiltonian \cite{Brown:2017jil},
\begin{align}
\sigma^2 = \frac14 \sum_{k=1}^4 e_k^2 \, .
\end{align}
%in order to have a well-defined spectrum. 
The proper question is; {\it given a fixed variance $\sigma$ of the eigenvalues, 
what kind of Hamiltonian 
gives the fastest growth of the complexity?} 
In order to answer this question, notice the following equality
\begin{align}
\frac14 \sum_k e_k^2 = a^2+b^2+c^2 \, .
\label{labc}
\end{align}
This means that fixing the variance $\sigma$ means fixing $a^2+b^2+c^2$. Under this constraint,
we want $t_{\rm max}$, given by \eqref{tmax}, to take the minimal value. It is easy to find that $t_{\rm max}$
is minimized when
\begin{align}
|a|=|b|=|c| =\frac{\sigma}{\sqrt{3}}\, .
\label{abc}
\end{align}
Using the relations \eqref{lambda4}, this set of gates with $|a|=|b|=|c|$ is shown to be equivalent to 
the following relation among the eigenvalues\footnote{All the other solutions
are given just by permutation of the solution \eqref{soll}.}
\begin{align}
e_1 = -3 e_2 = -3e_3 = -3e_4 = \pm {\sqrt{3}}{\sigma}\, .
\label{soll}
\end{align}
This is the set of the the eigenvalues which reaches the maximum complexity at the fastest speed:
\begin{align}
v_{\rm fastest}= \frac{\sigma}{\epsilon}\frac{5}{\sqrt{3}}  \, .
\end{align}

The fastest Hamiltonian
which 
has the eigenvalues satisfying \eqref{soll} is found to be
\begin{align}
H=
U_0^\dagger
\left[
J(\sigma_3\otimes\mathbf{1})+J(\mathbf{1}\otimes\sigma_3)
\pm J (\sigma_3\otimes \sigma_3)
\right]
U_0 \, . \label{fastest}
\end{align}
The coefficient \eqref{fastest} shows that the strength of the last term, which we call interaction, 
needs to be equal to that of $H_0$. Therefore, 
if one regards 1-local Hamiltonian as ``kinetic terms'' and 2-local Hamiltonian as ``interactions'', then 
the fastest Hamiltonian can be achieved when the strength of the interaction is as large as kinetic terms.  
%XXX This is not $\lambda \to \infty$ type, rather it is $\lambda \to O(1)$ type.  

The three terms of the Hamiltonian 
\eqref{fastest} correspond to $U_1, U_2$ and $U_{\rm ent}$, respectively. They commute with
each other and span the maximal torus of the Hamiltonian space $su(4)$.  
%So, the essential requirement about 
The fastest Hamiltonian is summarized as follows:
\begin{itemize}
\item The Hamiltonian contains all possible components of the maximal torus of the Hamiltonian space.
\item The parameters in the Hamiltonian are equal to each other. 
\end{itemize}
As is clearly seen from \eqref{abc} (see also \eqref{tmax} and \eqref{tmaxforcom}), given the fixed variance $\sigma$, 
in order to achieve the fastest evolution of the total complexity, we need to arrange 
all of the parameters equal. %We call this parametric equality {\it communism}.    
We call this equal distribution of parameters {\it communism}.% 
\footnote{For the reader curious about this ideology, see for example, \cite{Marx:1848}.}

The gate $U_{\rm ent}$ using the CNOT gates is important for having larger complexity. As
shown in Appendix \ref{sec:QO}, the CNOT gate is interpreted as a quantum version of the Othello rule.
Thus the quantum Othello rule is responsible for a large complexity and also for the fast time evolution of the complexity.

%%%%%%%%
\subsubsection{Complexity in general Hamiltonian system}

Let us proceed to consider the general case of $L\times L$ lattice on which our $\mathbf{Z}_2$ gauge theory
lives. From now on, we consider a large $L$ limit, $L \to \infty$,  
and study how the maximum complexity and the fastest Hamiltonian behaves there. 
The generic Hamiltonian is given by \eqref{0303}.  However, 
we will focus on counting the minimal number of gates which are used for reconstructing 
$\exp[-i\Lambda t]$, where $\Lambda$ is a diagonalized Hamiltonian.
This is because it captures the universal part of the complexity as we have seen in Sec.~\ref{Z22-localanalysis}. 
This means that we analyze the generic Hamiltonian 
\begin{align}
H = \sum_{I_{(1,1)}, I_{(1,2)}, \cdots , I_{(L, L)}} a_{I_{(1,1)} I_{(1,2)} \cdots I_{(L, L)}}  \,
\sigma_{I_{(1,1)}} \otimes \sigma_{I_{(1,2)}} \otimes \cdots \otimes \sigma_{I_{(L,L)}} \, ,
\label{genericdiagonalHamiltonian}
\end{align} 
where $I_{(a,b)}$ runs $0$ and $3$ only (remember we define $\sigma_0\equiv \mathbf{1}$), 
instead of \eqref{0303} where $I_{(a, b)}$ runs $0, 1, 2, 3$. 
In the following, we calculate the maximum complexity of the $k$-local Hamiltonian of the type of \eqref{genericdiagonalHamiltonian}
and investigate the fastest Hamiltonian for any given $k$.

\noindent
\underline{Maximum complexity}
\vspace{-3mm}
\begin{description}
\item[1-local Hamiltonian.]
The $k=1$ (1-local) Hamiltonian is the unperturbed Hamiltonian, which is equivalent to \eqref{H0ab} 
or \eqref{HJM} by the change of the basis, and we have already studied that in Sec.~\ref{sec:one-p}.  
There, we found that the maximum complexity is 
\begin{align}
{\cal C}_{\rm max}=L^2 \,.
\label{1C}
\end{align}
The value is identical to the maximum complexity for the classical case. Therefore, when all the
terms in the Hamiltonian are not entangled, we find that the maximum complexity is the same order 
as the classical case.

\item[2-local Hamiltonian.]
Next, let us consider the 2-local Hamiltonian. In addition to the terms of the 1-local Hamiltonian,
we have entangling terms which relate two sites among $L^2$. The number of
possible kinds of terms is $L^2 (L^2-1)/2$, and for each term we need 3 complexity
(the reason
was explained earlier for the 2-site case. Any entangling term which relates two sites
should use two CNOT gates and one qubit gate, so in total we have 3). Therefore, together with 
the 1-local contribution \eqref{1C}, we obtain
\begin{align}
{\cal C}_{\rm max}=L^2 + 3\frac{L^2(L^2-1)}{2!} =  \frac{3}{2} L^4 + {\cal O}(L^2) \,.
\label{2C}
\end{align}

\item[$k$-local Hamiltonian ($k \ll L^2$).]
Let us consider a 3-local case first as an example.
To reproduce a term entangling three qubits in the Hamiltonian, we need 5 gates.
Suppose the term is labeled as $(p,q,r)$ among $L^2$. Then 
$U^{\rm CNOT}_{(p,q)}U^{\rm CNOT}_{(q,r)}U(r)U^{\rm CNOT}_{(q,r)}U^{\rm CNOT}_{(p,q)}$ 
with a single qubit gate $U(r)$ can do the job.
Therefore the total complexity is
\begin{align}
{\cal C}_{\rm max}=L^2 + 3\frac{L^2(L^2\!-\!1)}{2!} + 5 \frac{L^2(L^2\!-\!1)(L^2\!-\!2)}{3!}
=  \frac{5}{3!} L^6 + {\cal O}(L^4) \,.
\label{3C}
\end{align}
From this analysis, it is obvious that for $k$-local Hamiltonian with $k\ll L^2$,
the maximum complexity behaves as
\begin{align}
{\cal C}_{\rm max}=  \frac{(2k-1)}{k!} L^{2k} + {\cal O}(L^{2k-2}) \,.
\label{kC}
\end{align}

\item[Non-local Hamiltonian.] If we allow arbitrary entanglement of the qubits in the Hamiltonian,
we have the $L^2$-local Hamiltonian, which is maximally nonlocal.
It has all nonlocal interactions, 
for example, $\sigma_3 \otimes \sigma_3\otimes \cdots \otimes \sigma_3$
which consists of the magnetic flux at all plaquettes of the lattice. This term is the most nonlocal term
among possible interaction terms, since it entangles all the qubits at the same time.
As for the maximum complexity, 
including all the contributions of $k$-local terms, we obtain
\begin{align}
{\cal C}_{\rm max}=  \sum_{k=1}^{L^2}
(2k-1) %{}_{L^2}C_k 
{ L^2 \choose k} = 2^{L^2}(L^2-1)+1 \sim  2^{L^2}L^2 \, .
\label{L2C}
\end{align}
which scales as ${\cal{O}}(e^S)$. 
\end{description}

Let us compare the cases. A summary is given in Table \ref{tab1}.
The expression \eqref{L2C} for the maximally nonlocal Hamiltonian
is distinct from the $k$-local ($k\ll L^2$) cases: the maximum complexity
grows as $2^{L^2}$, not a power law in $L$.
As described in the introduction, theories with gravity dual allowing (eternal) black holes
are expected to have the maximum complexity of an exponential order in the number of qubits.
We conclude here that among ${\mathbf Z}_2$ gauge theories 
with diagonal Hamiltonians 
in 2 spatial dimensions,
theories which %show the maximum complexity is ${\cal{O}}(e^S)$ and therefore 
might have a gravity dual, are only the ones with maximally nonlocal Hamiltonians. 
These nonlocal Hamiltonians allow all possible number of entangling qubits. 
Physical implication of such theories is discussed in Sec.~\ref{sec:Summary}.  
Our conclusion is probably due to the restriction to diagonal Hamiltonians, which enables us to evaluate the complexity. For general non-diagonal Hamiltonians, it is expected that the maximal complexity reaches the order of the exponential of $L^2$.

\renewcommand{\arraystretch}{2}
\begin{table}[t]
\begin{center}
  \begin{tabular}{c||c|c|c|c}
  Hamiltonian & 1-local &2-local & $k$-local & \shortstack{maximally\\nonlocal}
  \\
  \hline\hline
%Maximum complexity ${\cal C}_{\rm max}$ & $L^2$ & $\displaystyle\frac32 L^4$ 
%& $\displaystyle\frac56 L^6$  & $2^{L^2}L^2$
Maximum complexity ${\cal C}_{\rm max}$ & $L^2$ & $\displaystyle  L^4$ 
& $\displaystyle  L^{2k}$  & $2^{L^2}L^2$
    \\
  \hline
%Fastest speed $v$ & $\displaystyle\frac{JL^2}{\epsilon}$ &
%$\displaystyle\frac{\sigma}{\epsilon}\frac{3}{\sqrt{2}}L^2$
Fastest speed $v$ & $\displaystyle\frac{JL^2}{\epsilon}$ &
$\displaystyle\frac{\sigma}{\epsilon} L^2$
 &
%$\displaystyle\frac{\sigma}{\epsilon}\frac{5}{\sqrt{6}}L^3$
%  & $\displaystyle\frac{\sigma}{\epsilon}2^{L^2/2}L^2$
$\displaystyle\frac{\sigma}{\epsilon} L^k$
  & $\displaystyle\frac{\sigma}{\epsilon}\, 2^{L^2/2}L^2$
  \end{tabular}
  \caption{A summary table of the maximum complexity and the fastest speed to reach the maximum, for 
  classes of Hamiltonians of ${\mathbf Z}_2$ gauge theories. Here $k$ for $k$-local is $k \ll L^2$. We list only the leading order terms
  for $L\gg 1$ and we omit ${\cal{O}}(1)$ coefficients.}
  \label{tab1}
  \end{center}
\end{table}
\renewcommand{\arraystretch}{1.0}

\noindent
\underline{Fastest Hamiltonian}

For the eigenvalues $e_k$ of the Hamiltonian $(k=1,2,\cdots,2^{L^2})$, we can prove the following equation for the energy variance $\sigma$, 
\begin{align}
\sigma^2
\left(= \frac{1}{2^{L^2}}
\sum_{k=1}^{2^{L^2}} e_k^2 
\right)
= \sum_{I_{(1,1)} I_{(1,2)} \cdots I_{(L, L)} }   (a_{I_{(1,1)} I_{(1,2)} \cdots I_{(L, L)}} )^2 \,.  
\end{align}
where the sum is for all $I_{(a, b)} = 0 , 3$. %, each runs $0$ and $3$ only. 
This equation is analogous to \eqref{labc}. The left hand side is the variance of the eigenvalues of 
the total Hamiltonian, which we fix to be some constant. Then, following the same reasoning as that
of the 2-site case, we come to a conclusion that the fastest Hamiltonian should have all equal
$|a_{I_{(1,1)} I_{(1,2)} \cdots I_{(L, L)}} |$.

Let us calculate the speed of the growth of the complexity. As we have studied, we define the
speed $v$ as ${\cal C}_{\rm max}/t_{\rm max}$ where $t_{\max}$ is the time when the system
reaches the maximum complexity. Following our study for the 2-site case, we find
\begin{align}
t_{\rm max} = \epsilon \times \mbox{max} \left\{ | a_{I_{(1,1)} I_{(1,2)} \cdots I_{(L, L)}}  |^{-1} \right\} . 
\end{align}
For all equal $|a_{I_{(1,1)} I_{(1,2)} \cdots I_{(L, L)}} |$, we just distribute the energy variance to
each $|a_{I_{(1,1)} I_{(1,2)} \cdots I_{(L, L)}} |$ equally. For $k$-local Hamiltonians, 
the number of independent terms in the diagonalized Hamiltonian is $n_k \equiv 
\sum_{s=1}^{k} {{L^2} \choose s}$. Then we find that the fastest Hamiltonian has
\begin{align}
t_{\rm max} = \frac{\epsilon \sqrt{n_k}}{\sigma} \, .
\end{align}
Therefore, the speed $v$ of the growth of the complexity is calculated as
\begin{align}
v = \left\{ 
\begin{array}{ll}
\displaystyle
\frac{\sigma}{\epsilon}\frac{(2k-1)}{\sqrt{k!}} L^k + \mbox{lower order in $L$} & \quad (k\ll L^2)\\[5mm]
\displaystyle
\frac{\sigma}{\epsilon}\frac{2^{L^2}(L^2-1)+1 }{2^{L^2/2}} 
\sim 
\frac{\sigma}{\epsilon} L^2 2^{L^2/2}
& \quad (k=L^2) \,.
\end{array}
\right.
\end{align}
The summary is given in Table \ref{tab1}.

%%%%%%%%
\subsubsection{Locality in Hamiltonians and gates}

So far we have seen that maximum complexity is much smaller than ${\cal{O}}(e^S)$ for local Hamiltonians, {\it i.e.,} $k$-local Hamiltonians where $k \ll L^2$. 
One might ask the following question; if we use only the neighboring universal gate sets which we discussed at the end of Sec.~\ref{gate_gauge_th}, 
then even with local Hamiltonian, can we achieve ${\cal{O}}(e^S)$ maximal complexity? 
The answer of this question is no. We can never achieve ${\cal{O}}(e^S)$ complexity even if we restrict our choices of the gate sets to neighboring ones, and we will show this in this subsection. 

The $k$-local Hamiltonians used so far is not spatially local on the lattice.
The spatial locality of the theory is guaranteed if the distance of the 
entanglement of the qubits in the Hamiltonian
does not grow as ${\cal O}(L)$, in the large $L$ limit. 

To introduce spatially local Hamiltonians, 
we define ``adjacent $k$-local" Hamiltonians as those consisting of terms of 
entanglement of at most $k$ {\it neighboring} qubits. For example, an adjacent
$2$-local Hamiltonian is given in the diagonalized form by
\begin{align}
H =  
H_0 &+  \sum_{(a,b)} J^{(2)}_{(a,b)}\left[
\mathbf{1}\otimes
\cdots \otimes \mathbf{1}\otimes
\underbrace{\sigma_3}_{(a,b)} \otimes \mathbf{1}\otimes
\cdots \otimes 
\mathbf{1}\otimes
\underbrace{\sigma_3}_{(a+1,b)} \otimes \mathbf{1}\otimes
\cdots
\right]
\nonumber \\
&+ 
\sum_{(a,b)} J^{(3)}_{(a,b)}\left[
\mathbf{1}\otimes
\cdots \otimes \mathbf{1}\otimes
\underbrace{\sigma_3}_{(a,b)} \otimes \mathbf{1}\otimes
\cdots \otimes 
\mathbf{1}\otimes
\underbrace{\sigma_3}_{(a,b+1)} \otimes \mathbf{1}\otimes
\cdots
\right]\, , 
\label{a2}
\end{align}
where $H_0$ is defined in \eqref{H0ab}.
%This Hamiltonian is spatially local, because it consists of only terms with 
%neighboring qubits, in the large $L$ limit.
The maximum complexity of the Hamiltonian \eqref{a2} is
\begin{align}
{\cal C}_{\rm max}^{\rm \;(adj. \; 2-local)}= L^2 + 2L(L-1)\times 3 = 7L^2-6L\, . 
\end{align}
Here $2 L (L -1)$ is the number of neighboring pairs in the $L \times L$ lattice. 
So, compared to the previous $2$-local case in \eqref{2C}, the maximum complexity
decreases and behaves as if it is $1$-local. 
Similarly for adjacent 3-local Hamiltonians, we can compute
\begin{align}
{\cal C}_{\rm max}^{\rm \;(adj. \; 3-local)}
& = {\cal C}_{\rm max}^{\rm \;(adj. \; 2-local)}
+\left(2L(L-2)+4(L-1)^2\right)\times 5 
\nonumber \\
&
= 37L^2 -66L+20 
\, . 
\end{align}
Again, the maximum complexity grows only as $L^2$.
%, which is the same as the 1-local Hamiltonian. 
Therefore, the maximum complexity of 
any adjacent $k$-local Hamiltonian ($k\ll L^2$) grows as $L^2$, and the system behaves
similar to that of a $1$-local Hamiltonian. 
%\footnote{Of course, adjacent $L^2$-local
%is identical to the $L^2$-local, because for $k=L^2$ all the qubits are entangled
%and thus automatically neighbored.}.

Then, how much ``nonlocal" the generic 2-local system is, compared to 
the adjacent 2-local system?
To answer this question, we can use the other universal gate set 
${\cal U}_{\rm neighbor}$ defined in \eqref{Ulocal}. Using
it, we can estimate more vividly the nonlocality of the 2-local
Hamiltonians. In fact, the maximum complexity of the 2-local Hamiltonian 
measured in ${\cal U}_{\rm neighbor}$ is bounded as\footnote{
A 2-local operator $V_{(p,p+d)}$ acting on qubits at $p$ and $p+d$ such as 
\begin{align*}
V_{(p,p+d)}=\mathbf{1}\otimes
\cdots \otimes \mathbf{1}\otimes
\underbrace{\sigma_3}_{p} \otimes \mathbf{1}\otimes
\cdots \otimes 
\mathbf{1}\otimes
\underbrace{\sigma_3}_{p+d} \otimes \mathbf{1}\otimes \cdots,
\end{align*}
can be decomposed as 
$V_{(p,p+d)}=U V_{p+d} U$,  
where $U$ is a product of neighboring CNOT gates 
\begin{align*}
U= \left(\prod_{i=1}^{d} U^{\rm CNOT}_{(p+d-i,p+d-i+1)}\right)\,\left(\prod_{i=1}^{d-1} U^{\rm CNOT}_{(p+i,p+i+1)}\right), 
\end{align*}
and $V_{p+d}$ is a single-qubit gate acting on the qubit at $p+d$ defined as 
\begin{align*}
V_{p+d}=\mathbf{1}\otimes
\cdots \otimes \mathbf{1}\otimes
\underbrace{\sigma_3}_{p+d} \otimes \mathbf{1}\otimes
\cdots \,. 
\end{align*} 
The number of neighboring CNOT gates used in the above $U$ is $2d-1$. 
Thus, the complexity of 2-local operator $V_{(p,p+d)}$ should be less than $4d-1$, although there might be a more efficient decomposition.  
}
\begin{align}
{\cal C}_{\rm max} \leq L^2 + \sum_{\{(a,b)\neq(a',b')\}}
\!\!\left[\,4(|a-a'|+|b-b'|) -1\,
\right]\,.
\label{2ab}
\end{align}
It is easy to see that the right-hand side behaves as $\mathcal{O}(L^5)$ for a large $L$. Remember that,
if it were measured by the previous ${\cal U}$, we would obtain ${\cal C}_{\rm max}\sim L^4$ 
(see Table \ref{tab1}). 
The reason why we obtain larger complexity for the gate set ${\cal U}_{\rm neighbor}$ is that the number of neighboring CNOT gates to create a distant entanglement 
grows proportionally to the distance.
However, although the complexity measured by ${\cal U}_{\rm neighbor}$ is larger than that by ${\cal U}$, we do not obtain the exponentially large complexity for the 2-local Hamiltonian.

As is seen in this example, the notion of spatial locality could depend on what gate set we use.
The original ${\cal U}$ defined in \eqref{Uni} is useful for explicitly evaluating the complexity.
On the other hand, 
to respect the spatial locality, it would be better to use ${\cal U}_{\rm neighbor}$
defined in \eqref{Ulocal}.
% --- Using ${\cal U}_{\rm neighbor}$, even the 
%2-local Hamiltonian has an exponentially
%large complexity.
However, the exponential behavior of the complexity $e^{S}\sim e^{L^2}$, 
which is one of the conjectured criterion for having a gravity dual \cite{Susskind:2014moa}, 
is never achieved in both universal gate sets except for the maximally nonlocal Hamiltonian. 
The difference of the universal gate sets may be just
the difference of the regularizations, in view of the criterion.

%%%%%%%%%%%%%%%%%%%%%%%%%%%%%%%%%%%%%%

\subsection{Complexity in ${\mathbf Z}_N$ gauge theory}
\label{sec:ZN}

In order to study the $U(1)$ gauge theory, we generalize the results of the previous section
to the ${\mathbf Z}_N$ gauge theory and take the $N\to\infty$ limit.
The evaluation of the complexity in ${\mathbf Z}_N$ gauge theory goes in a similar manner.
The only difference  is, this time we use the single qudit gates, and there are $N-1$ 
independent operations of the CNOT: $(U^{\rm CNOT})^i$, with  $i=1,\cdots,N-1$.
To see the difference, it is instructive to consider the 2-site case first.
After the analyses of the 2-site case, we summarize the results of the complexity
in the general cases, and finally study the $N\to\infty$ limit and the continuum limit 
($L \to \infty$ and plaquette size going to zero limit).

\subsubsection{Complexity in 2-site ${\mathbf Z}_N$ gauge theory}
 
Let us consider a Hamiltonian of a ${\mathbf Z}_N$ gauge theory on a 2-site.
We are interested in a diagonalized eigenvalues of the Hamiltonian. The Hilbert space is
$N^2$-dimensional, so in total we have $N^2$ eigenvalues, $\{ e_1,e_2, \cdots, e_{N^2} \}$.
For a single qudit gate, diagonal eigenvalues are reproduced from the diagonal
generators of $u(N)$, $\lambda_I$ $(I=0,1,\cdots,N-1)$.
We normalize the generators as ${\rm tr}\lambda_I^2=N$ (with no summation over $I$).
(For the case of the ${\mathbf Z}_2$ gauge theories, $\lambda_0 = {\bf 1}$, $\lambda_1 = \sigma_3$.)  
A generic diagonalized Hamiltonian of the 2-site ${\mathbf Z}_N$
gauge theory is written as
\begin{align}
\Lambda = \frac{1}{N}\sum a_{I,J}\lambda_I \otimes \lambda_J \, .
\end{align}
We prepare the following gates,
\begin{align}
&U_1(a) \equiv \exp\left[- \frac{i}{N} \sum_{i=1}^{N-1}a_i \lambda_i \otimes \lambda_0 \right] \,, \quad
U_2(b) \equiv \exp\left[- \frac{i}{N} \sum_{i=1}^{N-1} \lambda_0 \otimes b_i\lambda_i \right]\, ,  \\
&U_{\rm ent}^{(s)}(c^{(s)}) \equiv  (U^{\rm CNOT})^s U_2(c^{(s)})(U^{\rm CNOT})^{N-s} 
\qquad (s=1,2,\cdots, N-1)\,.
\label{u3sc}
\end{align}
Here, $U_1$ and $U_2$ are single-qudit gates. The last one is the entangling gate,
corresponding to \eqref{u3c}. As opposed to the case of the ${\mathbf Z}_2$ gauge theories,
this time we have $(N-1)$ species of the entangling gates.

This set of the gates suffices our purpose of reconstructing all the eigenvalues of the Hamiltonian.
Noticing that the complexity of \eqref{u3sc} is $N+1$, the total complexity is calculated as
a sum of the complexity contributed from $U_1(a), U_2(b), U_{\rm ent}^{(s)}(c)$,
\begin{align}
{\cal C}_{\rm max} = 1+1+ \sum_{s=1}^{N-1} (N+1) = N^2+1 \, .
\end{align}
%and the matrix for diagonalizing the Hamiltonian, respectively.

Let us study what is the fastest Hamiltonian on the 2 sites.
It is possible to show the following relation
\begin{align}
\sigma^2 \equiv \frac{1}{N^2}\sum_{i=1}^{N^2}(e_i)^2
=\sum_{i,j=0}^{N-1} (a_{i,j})^2
= \sum_{i=1}^{N-1} (a_i)^2 +  \sum_{i=1}^{N-1} (b_i)^2 + 
\sum_{s=1}^{N-1} \sum_{i=1}^{N-1} (c^{(s)}_i)^2. 
\end{align}
This equation is analogous to \eqref{labc} in the ${\mathbf Z}_2$ gauge theories.
Noticing that the first term $\sum_{i=1}^{N-1} (a_i)^2$ is responsible for the first qudit gate
and the second term $\sum_{i=1}^{N-1} (b_i)^2$ is for the second qudit gate,
we conclude that the fastest Hamiltonian should satisfy
\begin{align}
\sum_{i=1}^{N-1} (a_i)^2 =  \sum_{i=1}^{N-1} (b_i)^2 =
 \sum_{i=1}^{N-1} (c^{(s)}_i)^2 = \frac{\sigma^2}{N+1} \, .
 \label{faN}
\end{align}
The fastest speed to reach the maximum complexity is
\begin{align}
v_{\rm fastest}= \frac{\sigma}{\epsilon}\frac{N^2+1}{\sqrt{N+1}}  \,. 
\end{align}

In summary, we can calculate the complexity in a manner that is the same as that for the ${\mathbf Z}_2$
gauge theories. The only difference in calculations comes in the independent combination of the
CNOT gates. The total complexity and the fastest speed depends nontrivially on $N$.
In the fastest Hamiltonian \eqref{faN}, the energy variance $\sigma$ is equally distributed
to the first and the second qudit gates ($a_i$ and $b_i$), as well as to the entangling gates
($c_i^{(s)}$). Since there are $N-1$ entangling gates ($s=1,2,\cdots,N-1$), the major part
among the distribution of $\sigma$ is the entangling term $(c_i^{(s)})$. In the $N\to\infty$
limit, the entangling terms completely dominate.
This means that 
the fastest Hamiltonian dominantly consists of entangling terms, and in this sense, the
fastest Hamiltonian is strongly coupled.

%%%%%%%%%%%%%%%%%%%%%%%%%%%%%%%%%%%%%%

\subsubsection{Continuum limit of the complexity}

The extension to the case of 
of the $L\times L$ lattice is straightforward, and we present our results in Table \ref{tab2}.
Obviously, if we take $N=2$, it reduces back to the previous Table \ref{tab1}.
Now we have the complete $N$ dependence in the complexity of the ${\mathbf Z}_N$ gauge
theory.

In Table \ref{tab2}, we list also the results for the adjacent 2-local Hamiltonians.
Again, we find that the adjacent 2-local Hamiltonian has the maximum complexity proportional
to $L^2$.

As the continuum limit\footnote{We call the $L\to\infty$ limit
 a continuum limit, as we introduce the lattice spacing $a$ and keep $La =$ fixed, where 
$a (\to 0)$ plays the role of a UV cut-off.} of the ${\mathbf Z}_N$ gauge theory $L\to\infty$, it is natural to consider
a complexity density. Suppose the original $L\times L$ lattice has a 2-dimensional volume $V$. 
Then, using the lattice spacing $a$, the volume is written as $V=L^2 a^2$. The complexity 
density $\hat{\cal C}$ is defined as ${\cal C}/V = {\cal C}/(L^2a^2)$.
Using this complexity density, we find
\begin{align}
\hat{\cal C} = \left\{
\begin{array}{ll}
1/a^2 & \mbox{1-local}\\
2N^2/a^2 & \mbox{adj.~2-local}\\
N^{L^2}/a^2 & \mbox{nonlocal}\,.
\end{array}
\right.
\end{align}
Therefore, taking the limit $L\to \infty$, the complexity density is independent of $L$ for
spatially local $\mathbf{Z}_N$ gauge theories, 
while it diverges exponentially for the maximally nonlocal $\mathbf{Z}_N$ gauge theories.

Since $N^{L^2}$ is the dimension of the Hilbert space, it could be written as $e^S$ where
$S$ is the entropy. Then, for the maximally nonlocal ${\mathbf Z}_N$ gauge theory,
\begin{align}
\hat{\cal C} = \frac1{a^2} e^S \, .
\end{align}
Whether the maximum complexity is ${\cal{O}}(e^S)$ corresponds to the criterion of having a dual black hole, 
then we conclude that 
\textit{the criterion of having a dual black hole is
satisfied only by the maximally nonlocal theory, among all possible ${\mathbf Z}_N$ (or $U(1)$)
gauge theories with diagonal Hamiltonians.}

\renewcommand{\arraystretch}{2}
\begin{table}[t]
\begin{center}
  \begin{tabular}{c||c|c|c|c}
  Hamiltonian & 1-local &2-local & adj.~2-local 
  & \shortstack{maximally\\nonlocal}
  \\
  \hline\hline
%Maximum complexity ${\cal C}_{\rm max}$ & $L^2$  & $\displaystyle\frac{N^2}2 L^4$ 
%& $2N^2L^2$  
%& $N^{L^2}L^2$
Maximum complexity ${\cal C}_{\rm max}$ & $L^2$  & $\displaystyle {N^2} L^4$ 
& $N^2L^2$  
& $N^{L^2}L^2$
    \\
  \hline
Fastest speed $v$ & $\displaystyle\frac{JL^2}{\epsilon}$ &
%$\displaystyle\frac{\sigma}{\epsilon}\frac{N^{3/2}}{\sqrt{2}}L^2$
% & $\displaystyle\frac{\sigma}{\epsilon}\sqrt{2N}L$
%  & $\displaystyle\frac{\sigma}{\epsilon}N^{(L^2-1)/2}$
$\displaystyle\frac{\sigma}{\epsilon} {N^{3/2}} L^2$
 & $\displaystyle\frac{\sigma}{\epsilon}{N^{1/2}}L$
  & $\displaystyle\frac{\sigma}{\epsilon}N^{(L^2-1)/2}$
  \end{tabular}
  \caption{A summary table of the maximum complexity and the fastest speed to reach the maximum, for 
  classes of Hamiltonians of ${\mathbf Z}_N$ gauge theories. We list only the leading terms for
  $L\gg 1$ and $N\gg 1$ and we omit $O(1)$ coefficients.}
  \label{tab2}
  \end{center}
\end{table}
\renewcommand{\arraystretch}{1.0}

%
%\begin{align}
%\end{align}
%\begin{align}
%\end{align}
%\begin{align}
%\end{align}
%
%
%%
%%\begin{align}
%%\end{align}
%%
%%\begin{align}
%%\end{align}
%%
%%\begin{align}
%%\end{align}
%%
%%\begin{align}
%%\end{align}
%%
%%\begin{align}
%%\end{align}
%%
%%\begin{align}
%%\end{align}
%%
%

%%%%%%%%%%%%%%%%%%%%%%%%%%%%%%%%%%%%%%%%%%%%%%%%%%%%%%%%%%%%%%%%%%%%
\section{Summary and discussions}
\label{sec:Summary}
%%%%%%%%%%%%%%%%%%%%%%%%%%%%%%%%%%%%%%%%%%%%%%%%%

In this paper, we have evaluated the time evolution of the complexity in the $\mathbf{Z}_N$ gauge theories
on the 2-dimensional $L\times L$ spatial lattice. 
One of the motivations to study complexity in gauge theories is the conjecture that 
it gives the criterion of having a dual gravity black hole if  
the maximum complexity behaves as $\mathcal{C}_{\rm max} =  {\cal{O}}(e^S)$ where $S$ is the entropy \cite{Susskind:2014moa}.  
In order to evaluate the complexity, we considered only diagonal Hamiltonians, which contain either electric or magnetic flux operators. 
Our results of the complexity for $k$-local diagonal Hamiltonians  
are summarized in Table \ref{tab1} for $\mathbf{Z}_2$ gauge theories 
and Table \ref{tab2} for the $\mathbf{Z}_N$ gauge theories. 
These results show that for $\mathbf{Z}_N$ gauge theories
with diagonal Hamiltonians, 
only nonlocal Hamiltonians can satisfy the criterion of $\mathcal{C}_{\rm max} =  {\cal{O}}(e^S)$. 
%the nonlocal Hamiltonian consists of terms entangling arbitrary number of qudits where the qudits live on the plaquettes of the lattice and define the Hilbert space of the $\mathbf{Z}_N$ gauge theories. 

Why can we expect that only the nonlocal $\mathbf{Z}_N$ Hamiltonians could have dual black hole description? 
Here is a possible argument: we know that 2+1 dimensional maximally supersymmetric 
$SU(N_c)$ Yang-Mills theories in the large $N_c$ and strong coupling limit allows a gravity dual \cite{Itzhaki:1998dd}. 
Now, if we want to understand these $SU(N_c)$ theory within the scheme of our $U(1)$ gauge theories, we need to path-integrate out all of the off-diagonal elements of the
gauge fields in such a way that the non-Abelian gauge group reduces to several $U(1)$'s. 
Since gauge fields are massless, this path-integration generically induces maximally nonlocal interactions 
of a long range. From this view point, it is natural that for the $U(1)$ gauge theories to satisfy the criterion of having a gravity dual, long-range nonlocal interactions are necessary.  
This viewpoint also motivates us to study the complexity in local non-Abelian theories, since 
it will clarify the importance of locality and non-Abelian nature of gauge theories 
to satisfy the criterion of having a gravity dual.

In this paper, we look at how the complexity evolves in time. In particular we found, both in
the classical (Sec.3) and quantum (Sec.4) examples, that the complexity grows first %linearly 
and then saturates at $\mathcal{C}_{\rm max}$. 
This property is expected in 
\cite{Brown:2016wib, Brown:2017jil} as the second law of complexity.\footnote{It depends on the Hamiltonian whether the complexity fluctuates below the maximal value for a long time. Actually, for an integrable system, the complexity is a periodic function where the period is the inverse of the typical energy scale.}  
The evaluated speed grows when the time evolution
involves more nonlocality. For the classical case we studied with the Othello rule which is an example 
exhibiting the nonlocality,
and for the quantum case we worked with 
generic entangling Hamiltonians (using the CNOT gate which is a quantum 
version of the Othello game rule as shown in Appendix \ref{sec:QO}). The time to reach the maximum complexity
is determined by the nonlocality of the Hamiltonians.

Note that the complexity we calculated is the operator complexity of the time-evolution operators of 
diagonal Hamiltonians. 
Complexity is generically a basis-dependent quantity. Actually, although generic Hamiltonians are related to the diagonal Hamiltonian by a change of the basis of the Hilbert space, the complexity of a generic time-evolution operator is drastically different from that of the diagonal time-evolution operator. 
%Although we could argue that generic Hamiltonians are related to the diagonal Hamiltonian by a diagonalization \eqref{decomposition} with a unitary constant matrix $U_0$,
%this $U_0$ rotates the basis of the Hilbert space,  so the diagonalization drastically changes the value of the complexity.
%Note that we evaluated the rate of the complexity growth for generic reference states. Here, the eigenvalues
%of the Hamiltonian is assumed to be the major part of the complexity (see \eqref{decomposition}), and they 
%determine the rate. 
One may want to compare the rate with
that of the gravity dual \cite{Susskind:2014rva,Susskind:2014moa,Susskind:2016tae}.  
Unfortunately, the gravity dual description refers to 
a particular state (which is a thermal state with a given temperature),
while ours are in a micro-canonical ensemble and has no specification of the energy.\footnote{For the
study of variance of the eigenvalues and the speed of the complexity growth, see also
%\cite{Brown:2016wib,
\cite{Brown:2017jil}.} 
Therefore
one needs a further analysis which is a state-dependent argument, 
to compare the complexity growth rate of the gauge theories
and that of the black hole. 
It is also interesting to study for gauge theories the complexity of formation 
\cite{Chapman:2016hwi}, which is the complexity to create a TFD state of two CFTs from a disentangled state.

Given qubit distribution in lattice space, 
the nonlocality of the lattice gauge theory is rephrased as ``all-to-all'' couplings of qubits. 
Here all-to-all means that {\it independent} of the positions of the qubits, all pairs of the qubits interact.  
The Sachdev-Ye-Kitaev (SYK) model \cite{Sachdev:1992fk,Kitaev-talk-KITP} is a typical example of such all-to-all coupling models.
In fact, 
one can regard our qubit system as if there is a Majorana fermion at each plaquette.\footnote{Minor difference is that
% which allows only 2 level states - no fermion or one fermion and that's it due to exclusion principle. One minor difference is that 
such fermionic system has a different spin-statistics compared to our bosonic magnetic flux operator.} 
Then the $k$-local Hamiltonian of our study corresponds to the SYK model with $q=k$, whose Hamiltonian
is given by $H \sim \sum_{i_1,i_2,\cdots,i_q } j_{i_1 i_2\cdots} \chi_{i_1} \chi_{i_2} \cdots \chi_{i_q} $ 
where the number of the
Majorana fermion operator $\chi$ in each term is $q$, and $j$ is the random coupling.
The SYK model is known to show  \cite{Maldacena:2016hyu} a maximal Lyapunov exponent saturating 
the conjectured chaos bound \cite{Maldacena:2015waa}, 
with an appropriate scaling of the parameters.
The idea that the nonlocality triggers more chaos 
is natural since a many-body nonlocal interaction allows the system to reach the bound of 
phase space more quickly. 
Note that there is a difference between our model and the SYK model; the SYK model shows  
maximum chaos even with $q=4$, while, as we have seen, our model does not 
show maximum complexity ${\cal{O}}(e^S)$ for $k$-local if $k \ll L^2$. 
%In order to pursuing this non-locality idea furthermore, 
Given the similarity between the fastest time evolution of complexity and the fastest chaos development,  
it would be better even in Abelian gauge theories 
to understand the effects of locality/non-locality and connection between complexity and chaos furthermore.  
%not only  in Abelian gauge theories but also in non-Abelian gauge theories. 
%to construct possible models having a gravity dual, allow many-body nonlocal interactions such that the complexity growth becomes 
%faster and resultantly giving a stronger chaos. 
If we remove the restriction to diagonal Hamiltonians, we expect that locally interacting theories also have the maximal complexity with the order $e^S$. It is desired to develop new tractable quantities characterizing the complexity of states or operators, since the direct evaluation of the complexity for general Hamiltonians is difficult. 

We also comment on another approach by Nielsen \cite{Nielsen:2005} to define the complexity. 
In this approach, one defines the complexity of a unitary operator $U$ as a geodesic distance 
between $U$ and the identity operator with respect to a metric in the space of unitary operators. 
%The metric is defined so that an operator $U$ with a large complexity has a large distance form the identity operator. 
There is an ambiguity in the definition of the metric, which includes the ambiguity of the choice of the universal gate set. 
%This geometric approach has some advantages such that we do not need to 
%introduce a cut-off parameter $\epsilon$ as in \eqref{epdef}, 
%and the complexity takes continuum values. 
Since %Brown and Susskind 
\cite{Brown:2017jil} gives criteria of what class of metric we should consider for qubit systems, it may be a tractable question to see whether the geometric complexity behaves similarly to our result for $\mathbf{Z}_2$ gauge theories. 
%It is also interesting to consider the complexity geometry for general qudit systems, \textit{i.e.}, general $\mathbf{Z}_N$ gauge theories.  

Our analyses of the complexity of the classical/quantum $\mathbf{Z}_N$ gauge theories 
show that the time evolution of all the theories share the two stages:  the growth and the plateau. 
The difference among the theories or between the classical and the quantum theories
resides in the speed of the growth $v$ and the height of the plateau $\mathcal{C}_{\rm max}$.
They increase with more nonlocality. Only the maximally nonlocal quantum $\mathbf{Z}_N$ gauge theories
have $\mathcal{C}_{\rm max} =  {\cal{O}}(e^S)$, having the possibility of a dual gravity. 
We expect that for non-Abelian gauge groups, even local gauge theories may share the ${\cal{O}}(e^S)$ property
of the complexity. We plan to study if this is the case, in our future study. 

Finally, \cite{Heemskerk:2009pn, ElShowk:2011ag} discussed what kind of conditions are generically necessary in the boundary field theory  
to have a bulk Einstein gravity dual. Our analysis on the complexity and its growth speed might also shed new light on 
the condition. We hope to come back to such a deep question in the near future.

%%%%%%%%%%%%%%%%%%%%%%%%%%%%%%%%%%%%%%%%%%%%%%%%%%%%%%%%%%%%%%%%%%%%
\section*{Acknowledgement} 
%\hspace{0.51cm} 
It is our pleasure to thank Dan Kabat, Rob Myers, Kotaro Tamaoka, Tsuyoshi Yokoya, 
and Beni Yoshida for valuable discussions.
The work of K.H. was supported in part by JSPS KAKENHI Grant Numbers JP15H03658, JP15K13483, JP17H06462.  
The work of N.I. was supported in part by JSPS KAKENHI Grant Number JP25800143. 
S.S. is supported in part by the Grant-in-Aid for JSPS Research Fellow, Grant Number JP16J01004.  
%%%%%%%%%%%%%%%%%%%%%%%%%%%%%%%%%%%%%%%%%%%%%%%%%%%%%%%%%%%%%%%%%%%%
%%%%%%%%%%%%%%%%%%%%%%%%%%%%%%%%%%%%%%%%%%%%%%%%%%%%%%%%%%%%%%%%%%%%
%%%%%%%%%%%%%%%%%%%%%%%%%%%%%%%%%%%%%%%%%%%%%%%%%%%%%%%%%%%%%%%%%%%%

\appendix

%%%%%%%%%%%%%%%%%%%%%%%%%%%%%%%%%%%%%%%%%%%%%%%%%%%%%%%%%%%%%%%%%%%%
%%%%%%%%%%%%%%%%%%%%%%%%%%%%%%%%%%%%%%%%%%%%%%%%%%%%%%%%%%%%%%%%%%%%
\section{Review of $\mathbf{Z}_N$ Lattice Gauge Theory}
\label{AppA}
%%%%%%%%%%%%%%%%%%%%%%%%%%%%%%%%%%%%%%%%%%%%%%%%%%%%%%%%%%%%%%%%%%%%
%%%%%%%%%%%%%%%%%%%%%%%%%%%%%%%%%%%%%%%%%%%%%%%%%%%%%%%%%%%%%%%%%%%%
In this appendix, we review $\mathbf{Z}_N$ lattice gauge theory.

\subsection{Physical Hilbert space in lattice gauge theory}
In lattice gauge theories, the dynamical variables corresponding to the gauge fields live on links. 
For a gauge group $G$, a group element $L \in G$ is assigned to each link. 
Such an element is called a link variable. 
We represent by $L_{ij}$ the link variable at link $i$-$j$ where $i,j$ are labels for lattice vertices. 
Link variable $L_{ij}$ is essentially the exponential of the gauge field $A_{ij}$, 
$L_{ij}=e^{i a A_{ij}}$ where $a$ is a lattice spacing.  
All of the links are directed as Fig.~\ref{loopmain} and a link variable in the opposite direction gives the inverse element, i.e,   
\begin{align}
L_{ji} = (L_{ij})^{-1}. 
\label{app_inverse}
\end{align}
A gauge transformation at vertex $i$ changes the link variables on {\it all} the links connected to vertex $i$ as 
\begin{align}
L_{ij} \to g L_{ij}  \qquad ({}^\forall j  \text{ adjacent to } i), 
\end{align}
where $g$ is an arbitrary group element in $G$. 

We consider the physical Hilbert space $\mathcal{H}^\text{phys}$ in the temporal gauge $A_0=0$. \footnote{See, for example, \cite{Kogut:1974ag} and chapter 15 in \cite{Creutz:1984mg} for details of the Hamiltonian formalism in lattice gauge theories.} 
In terms of link variables, the temporal gauge fixes all of the time-directed link variables $L_{i0}$ to the unit element $1 \in G$, {\it i.e.}, $L_{i0}=1$.  
There are still residual time-independent gauge transformations which are compatible with the temporal gauge. 
The physical space $\mathcal{H}^\text{phys}$ is then given by the gauge-invariant space under the residual gauge transformations. 
Since we now have only time-independent gauge transformations, we do not need to consider the time-direction to define the physical space. 
Thus, in the following, we suppose that links are space-directed.   
Next we consider the case that the gauge group is $\mathbf{Z}_2$.

\subsection{$\mathbf{Z}_2$ gauge theory}
Group $\mathbf{Z}_2$ has only two elements $\pm 1$, or $e^{i n \pi}$ $(n=0,1)$. 
Thus, each link variable $L_{ij}$ in $\mathbf{Z}_2$ gauge theory takes the value $e^{i n \pi}$ $(n=0,1)$. 
In quantum mechanics, it means that each link has two independent states $\link{0}$ and $\link{1}$.\footnote{We leave usual ket-notation $\ket{\phantom{+}}$ for states on plaquettes.}  
Ignoring the gauge invariant condition, general states are spanned by $\otimes_\text{all links} \link{n_{ij}}_{ij}$ ($n_{ij}=0,1$). 
We represent this extended Hilbert space by $\mathcal{H}^{\bf{ext}}$.

A gauge transformation by the nontrivial element $-1$ in $\mathbf{Z}_2$ at vertex $i$ changes states on links connected to the vertex $i$ as $\link{n_{ij}}_{ij} \to \link{n_{ij} \oplus 1}_{ij}$ where $\oplus$ denotes addition modulo 2. 
Thus, the gauge transformation $g_i$ at vertex $i$ is represented by 
\begin{align}
g_i = \underset{j \text{ adjacent to } i}{\otimes} \sigma_1^{(ij)}, \quad 
\sigma_1^{(ij)}=
\begin{pmatrix}
0&1\\
1&0
\end{pmatrix} . 
\label{app_gauge_trsf_Z2}
\end{align}
The physical Hilbert space $\mathcal{H}^\text{phys}$ is a subspace of $\mathcal{H}^{\bf{ext}}$, 
which is invariant under $g_i$ for all vertices $i$, 
\begin{align}
\mathcal{H}^\text{phys} = \{ \ket\psi  \in \mathcal{H}^{\bf{ext}}\,\,  |\, g_{i} \ket{\psi} = \ket{\psi} \text{ for all vertices } i \}.  
\end{align} 
The constraint $g_{i} \ket{\psi} = \ket{\psi}$ is simply the Gauss's law.

\begin{figure}[htbp]
	\begin{center}
		\includegraphics[scale=0.25]{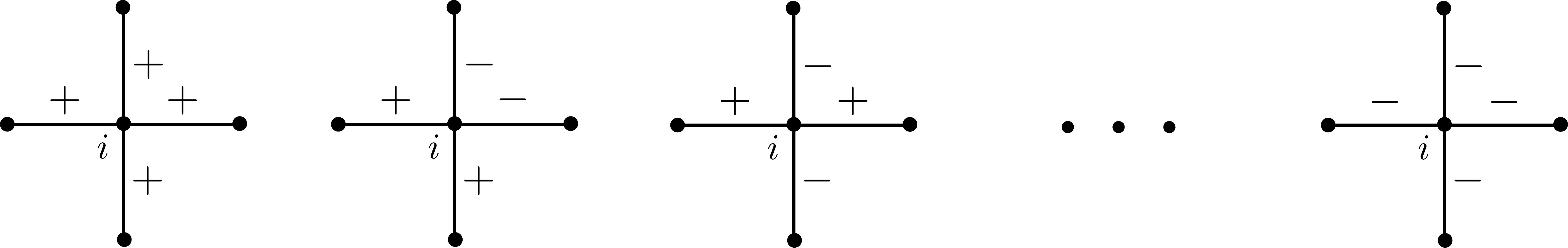}
		\caption{Gauss's law at vertex $i$ in $\mathbf{Z}_2$ gauge theory.   \label{Z2_Gauss_law}}
	\end{center}    
\end{figure}

In order to investigate the gauge invariance, it is more convenient to use the eigenvectors $\link{\pm}_{ij}$ of $\sigma_{1}^{(ij)}$, rather than $\link{n_{ij}}_{ij}$ (which are the eigenvectors of $\sigma_3$), such that   
\begin{align}
\sigma_1^{(ij)} \link{\pm}_{ij} = 
\pm \link{\pm}_{ij}\,.  
\label{app_s1_eigen}
\end{align}
With $\beta_{ij}=\pm$ as an eigenvalue of $\sigma_1^{(ij)}$, $\{\otimes_\text{all links} \link{\beta_{ij}}_{ij}\}$ is an orthogonal basis of the extended space $\mathcal{H}^{\bf{ext}}$. 
It is then clear that each orthogonal state $\otimes_\text{all links} \link{\beta_{ij}}_{ij}$ is an eigenstate of gauge transformation $g_{i}$ as 
\begin{align}
g_{i} \,(\otimes_\text{all links} \link{\beta_{ij}}_{ij}) = \left(\prod_{j \text{ adjacent to } i} \!\!\!\!\!\! \beta^{(ij)}\right)\, ( \otimes_\text{all links} \link{\beta_{ij}}_{ij})\,. 
\end{align} 
Therefore, states whose eigenvalues satisfy  
\begin{align}
\prod_{j \text{ adjacent to } i} \!\!\!\!\!\! \beta^{(ij)} =1 \quad \text{for all vertices $i$} 
\label{app_Gauss_law}
\end{align}
constitute a basis of the physical Hilbert space $\mathcal{H}^\text{phys}$. 
Eq.~\eqref{app_Gauss_law} means that there are even numbers of links whose states are $\link{-}_{ij}$ around each vertex $i$. 
For example, in two spatial dimensions, the gauge invariance at vertex $i$ allows states shown in Fig,~\ref{Z2_Gauss_law}. 
This is Gauss's law in pure gauge theory which states that an electric flux cannot terminate and it must form a closed loop. 
Actually $\sigma_1^{(ij)}$ is an operator measuring the electric flux penetrating the link $i$-$j$ because $\sigma_1^{(ij)}$ is the generator of the gauge transformation on the vertex $i$, which is the conjugate to the gauge field on the link $i$-$j$. 
This is a lattice analog of the standard fact in the continuum gauge theory; the physical condition in the temporal gauge leads to the Gauss constraint. 
 
Due to Gauss's law, a basis of the physical Hilbert space $\mathcal{H}^\text{phys}$ 
consists of states which have closed electric flux loops. 
One can further classify the flux loops topologically. 
For example, if we impose periodic boundary conditions on the spatial lattice such that space forms $T^2$, 
flux loops are classified according to their winding number on $T^2$. %whether they wind around spaces globally or not. 
However,  topologically different states belong to different superselection sectors, {\it i.e.}, they cannot mix by local operations. 
In this paper we consider only the topologically trivial sector.\footnote{Topologically nontrivial sectors are important in the context of quantum error correction such as the toric codes \cite{Kitaev:1997wr}.} 

In addition to electric flux operators $\sigma_1^{(ij)}$, there are magnetic loop operators that are defined on minimal plaquettes. 
On plaquette $p$, it is given by 
\begin{align}
\hat{f}_p \equiv \underset{(ij) \in p}{\otimes} \sigma_3^{(ij)},
\label{app_loop_op}
\end{align}
where the product is taken over all links belonging to the plaquette $p$. 
The operator $\hat{f}_p$ creates (or annihilates) a flux loop around the plaquette $p$. 
For example, if all link-states are $\link{+}$ around a plaquette, 
the loop operator at the plaquette changes each link-state to $\link{-}$. 
One can confirm that gauge invariant operators are 
only electric flux operators $\sigma_1^{(ij)}$ and magnetic loop operators $\hat{f}_p$.  
We often call the magnetic loop operators \textit{magnetic flux} operators.

Any orthogonal vectors $\{\otimes_\text{all links} \link{\beta_{ij}}_{ij}\}$ in the topologically trivial sector can be obtained by acting magnetic flux operators to the state $\otimes_\text{all links} \link{+}_{ij}$. 
Actually, in two spatial dimensions, any contractible flux loop $\ell$ can be created by a product of smaller loop operators $\prod_{p} \hat{f}_p$ where $p$ are plaquettes inside $\ell$. 
We now introduce a plaquette-basis notation which is a dual picture of the above link-basis.  
We first define $\otimes_\text{all plaquettes} \ket{0}_{p} \equiv \otimes_\text{all links} \link{+}_{ij}$, 
and then define $\ket{1}_{p} \equiv \hat{f}_p \ket{0}_{p}$. 
Since $\hat{f}_p^2=1$ from the definition \eqref{app_loop_op}, we also have $\ket{0}_{p} \equiv \hat{f}_p \ket{1}_{p}$. 
Furthermore, we have a nontrivial identification rule on the two spatial dimensions periodic lattice. 
On the periodic lattice, the operator $\prod_\text{all plaquettes} \hat{f}_p$ equals to the identity 
because $\sigma_3^{(ij)}$ appears twice for each link $(ij)$. 
It leads to the identification, $\otimes_\text{all plaquettes} \ket{0}_{p} = \otimes_\text{all plaquettes} \ket{1}_{p}$. Acting several $\hat{f}_p$ on the both sides, we obtain the global identification rule, 
\begin{align}
\otimes_\text{all plaquettes} \ket{m_p}_{p} = \otimes_\text{all plaquettes} \ket{m_p \oplus 1}_{p}\, \quad (m_p =0,1).
\label{app_identify_rule}
\end{align}
A basis in the physical Hilbert $\mathcal{H}^\text{phys}$ is therefore $\{\otimes_\text{all plaquettes} \ket{m_p}_{p}\}$ $(m_p =0,1)$ with the identification \eqref{app_identify_rule}. 
This space is the same as that of a qubit system with the identification under simultaneous flipping of all the qubits.

Magnetic flux operator $\hat{f}_p$ acts on the state $\otimes_{p'} \ket{m_{p'}}_{p'}$ as 
	\begin{align}
	\hat{f}_p \left(\otimes_{p'} \ket{m_{p'}}_{p'}\right) = \otimes_{p'} \ket{m_{p'}+\delta_{pp'}}_{p'}\,.
	\end{align}
Thus, in the plaquette basis $\{\otimes_\text{p} \ket{m_p}_{p}\}$, 
magnetic flux operator $\hat{f}_p$ is a non-diagonal operator.  
On the other hand, electric flux operator $\sigma_1^{(ij)}$ is a diagonal operator in this basis. Actually, it acts as a phase operator: 
\begin{align}
\sigma_1^{(ij)} \left(\otimes_{p} \ket{m_{p}}_{p}\right) = (-1)^{\sum_{p\ni (ij)} m_p} \left(\otimes_{p} \ket{m_{p}}_{p}\right)\,, 
\end{align} 
because $\sigma_1^{(ij)}$ anti-commutes with $\hat{f}_p$ if plaquette $p$ includes link $i$-$j$ and otherwise it commutes.

%%%%%%%%%%%%%%%%%%%%%%%%%%%%%%%%%%%%%%%%%%%%%%%%%%%%%%%%%%%%%%%%%%%%
%%%%%%%%%%%%%%%%%%%%%%%%%%%%%%%%%%%%%%%%%%%%%%%%%%%%%%%%%%%%%%%%%%%%
\subsection{$\mathbf{Z}_N$ gauge theory}
%%%%%%%%%%%%%%%%%%%%%%%%%%%%%%%%%%%%%%%%%%%%%%%%%%%%%%%%%%%%%%%%%%%%
%%%%%%%%%%%%%%%%%%%%%%%%%%%%%%%%%%%%%%%%%%%%%%%%%%%%%%%%%%%%%%%%%%%%
Generalizing $\mathbf{Z}_2$ to $\mathbf{Z}_N$ is straightforward. 
Group elements of $\mathbf{Z}_N$ are $\{e^{i \frac{2\pi n}{N}}|\, n=0,1,\cdots, N-1\}$. 
Thus link-states can be labeled by modulo $N$ integers $n=0,1,\cdots, N-1$ as $\link{n}_{ij}$. For $N\geq 3$, we should be careful about the orientation of links and note on the relation
$\link{n}_{ij}=\link{N-n}_{ji}$ due to eq.~\eqref{app_inverse}. 
As in the $\mathbf{Z}_2$ case, the extended Hilbert space $\mathcal{H}^{\bf{ext}}$ is spanned by $\{\otimes_\text{all links} \link{n_{ij}}_{ij}\}$. 

We now impose the gauge invariant condition on $\mathcal{H}^{\bf{ext}}$ to obtain the physical space $\mathcal{H}^\text{phys}$. 
Since group elements $e^{i \frac{2\pi n}{N}}$ $(n=2,\cdots, N-1)$ can be obtained as $n$-th power of $e^{i \frac{2\pi}{N}}$, 
it is enough to consider the invariance under the gauge transformation corresponding to $e^{i \frac{2\pi}{N}}$. 
The gauge transformation at vertex $i$ shifts states $\link{n_{ij}}_{ij}$ on links connected to vertex $i$ as 
\begin{align}
\link{n_{ij}}_{ij} \to \link{n_{ij}\oplus 1}_{ij}\,,
\end{align}
where $\oplus$ denotes addition modulo $N$.  
Thus the gauge transformation $g_i$ at vertex $i$ is represented by  $N$ by $N$ matrix $\tau_1^{(ij)}$\footnote{
	As in $\sigma_1^{(ij)}$ in $\mathbf{Z}_2$ theory, this $\tau_1^{(ij)}$ is the electric flux operator at link $i$-$j$.
	} 
\begin{align}
g_i = \underset{j \text{ adjacent to } i}{\otimes} \tau_1^{(ij)}, \quad 
\tau_1^{(ij)}\equiv
\begin{pmatrix}
0&0&\cdots&0&1\\
1&0&\cdots&0&0\\
0&1&\ddots&0&0\\ 
\vdots&&\ddots&&\vdots\\
0&0&&1&0
\end{pmatrix} \,, 
\label{app_gauge_trsf_ZN}
\end{align}
like eq.~\eqref{app_gauge_trsf_Z2} in the $\mathbf{Z}_2$ theory. 
It is convenient again to use eigenbasis of the permutation matrix $\tau_1^{(ij)}$. 
Since the eigenvalues of $\tau_1^{(ij)}$ are $e^{i \frac{2\pi}{N} \beta_{ij}}$ with  $\beta_{ij}=0,1,\cdots N-1$, we represent the corresponding eigenstates as $\link{\beta_{ij}}_{ij}$.  
Then a gauge transformation $g_i$ acts on a state $\otimes_\text{all links} \link{\beta_{ij}}_{ij}$ as 
\begin{align}
g_i \, \left( \otimes_\text{all links} \link{\beta_{ij}}_{ij} \right) 
=\left(\prod_{j \text{ adjacent to } i} \!\!\!\!\!\! e^{i \frac{2\pi}{N} \beta_{ij}} \right)\,\left( \otimes_\text{all links} \link{\beta_{ij}}_{ij} \right). 
\end{align}
Therefore, the physical Hilbert pace $\mathcal{H}^\text{phys}$ is spanned by $\{\otimes_\text{all links} \link{\beta_{ij}}_{ij}\}$ satisfying Gauss's law 
\begin{align}
\sum_{j \text{ adjacent to } i} \beta_{ij} = 0 \pmod{N} \quad \text{for all vertices $i$}. 
\end{align}
Note the orientation of links in the sum; they are all directed from vertex $i$ to its adjacent vertices $j$. 

Let us introduce a plaquette-basis as in the $\mathbf{Z}_2$ theory. 
In order to avoid confusion, we define the positive directions on the two spatial dimensional lattice; from left to right and from up to down as shown in Fig.~\ref{loopmain}.   %Fig.~\ref{app_loop}.  
%\begin{figure}[tbp]
%	\begin{center}
%		\includegraphics[height=5.0cm]{2dlattice_flux.pdf}
%		\caption{Loop operator $\hat{f}_p$ on a plaquette $p$. It creates a unit flux loop around $p$. \label{app_loop}}  
%	\end{center}
%\end{figure}
For a plaquette $p$ which is surrounded by links $i$-$j$, $j$-$k$, $k$-$l$, $l$-$i$, 
we define the loop operator $\hat{f}_p$ which creates the unit flux loop on $p$ by acting on link-states $\link{\beta_{ij}}_{ij}$$\link{\beta_{jk}}_{jk}$$\link{\beta_{lk}}_{lk}$$\link{\beta_{il}}_{il}$ as 
\begin{align}
\hat{f}_p\, \link{\beta_{ij}}_{ij}\link{\beta_{jk}}_{jk}\link{\beta_{lk}}_{lk}\link{\beta_{il}}_{il}
=\link{\beta_{ij}+1}_{ij}\link{\beta_{jk}+1}_{jk}\link{\beta_{lk}-1}_{lk}\link{\beta_{il}-1}_{il}\,. 
\end{align}
The magnetic flux operator satisfies $(\hat{f}_p)^N=1$. 
As in the $\mathbf{Z}_2$ theory, we define the state $\otimes_\text{all links} \link{\beta_{ij}=0}_{ij}$ as $\otimes_\text{all plaquettes} \ket{0}_{p}$. 
Then, the plaquette basis $\{\otimes_\text{all plaquettes} \ket{m_p}_{p}\, |\,m_p =0,1,\ldots N-1 \pmod{N} \}$ is obtained by acting magnetic flux operators $\hat{f}_p$ on $\otimes_\text{all plaquettes} \ket{0}_{p}$ as 
\begin{align}
\hat{f}_p \ket{m_p}_{p}= \ket{m_p\oplus 1}_{p}\,.
\end{align}

In the case where the lattice space forms periodic boundary conditions, namely, no boundary, if we act the same strength fluxes on all the plaquettes, its effects on each link always cancel.  Therefore, only in such case, we have an additional identification rule $\otimes_\text{all plaquettes} \, \hat{f}_p=1$, which implies 
\begin{align}
\otimes_\text{all plaquettes} \ket{m_p}_{p} = \otimes_\text{all plaquettes} \ket{m_p\oplus 1}_{p} \,.
\label{app_identify_ZN}
\end{align}
On the plaquette basis, the topologically trivial sector of $\mathcal{H}^\text{phys}$ 
is $\{\otimes_\text{all plaquettes} \ket{m_p}_{p}\}$ with the identification \eqref{app_identify_ZN}. 
As in $\mathbf{Z}_2$ gauge theory, magnetic flux $\hat{f}_p$ is a non-diagonal operator and electric flux $\tau_1^{(ij)}$ is a diagonal phase operator in the plaquette basis $\{\otimes_p \ket{m_p}_{p}\}$.

%%%%%%%%%%%%%%%%%%%%%%%%%%%%%%%%%%%%%%%%%%%%%%%%%%%%%%%%%%%%%%%%%%%%
\section{Quantum Othello}
\label{sec:QO}
%%%%%%%%%%%%%%%%%%%%%%%%

%\noindent
%{\it Othello rule as a noncommutativity}

\setlength{\fboxsep}{0mm}
%
%The $\mathbf{Z}_2$ gauge theory systems which we studied so far are constructed by just the commuting
%operations $M$, the magnetic flux operator. The simplicity of the system comes from the commutativity.
%Generic quantum systems have operators which are not mutually commutative. 
%Here we introduce a new term in the Hamiltonian which is not commutative with the original
%Hamiltonian. In fact, it defined a quantum Othello game.

In this appendix we show that the Othello rule is equivalent to the CNOT gate once it is treated 
quantum mechanically.

The main rule of the Othello game is to flip over all disks which are surrounded by your colored disks. 
This rule can be implemented at least for $1\times 3$-plaquette system, 
\fbox{$\hspace{4mm}\mid\hspace{3mm}\mid\hspace{4mm}$}. The white or black color of the disk
can be specified by the state $|m_1\rangle\otimes|m_2\rangle\otimes|m_3\rangle$
where $m_1,m_2,m_3=0,1$.
The white disk means $|0\rangle$, while the black disk means $|1\rangle$. In this notation, the
Othello procedure of flipping the disks would correspond to a map
\begin{align}
|1\rangle\otimes|0\rangle\otimes|1\rangle \quad \to \quad
|1\rangle\otimes|1\rangle\otimes|1\rangle \, . 
\end{align}
The disk at the center is flipped.

Now, notice that 
any quantum system preserves probability and thus needs to be unitary. The operation above 
can be unitary once it is supplemented by another rule
\begin{align}
|1\rangle\otimes|1\rangle\otimes|1\rangle \quad \to \quad
|1\rangle\otimes|0\rangle\otimes|1\rangle \,
\end{align}
at the same time. 

We can express these rules as a unitary matrix acting on the states in the Hilbert space. 
Due to the global gauge symmetry which flips all the plaquettes at the same time, 
we find that the Hilbert space is four-dimensional, spanned by the following states
\begin{align}
|1\rangle\otimes|0\rangle\otimes|0\rangle \, (= |0\rangle\otimes|1\rangle\otimes|1\rangle) \, , 
\quad
|1\rangle\otimes|1\rangle\otimes|0\rangle \, (= |0\rangle\otimes|0\rangle\otimes|1\rangle) \, , 
\nonumber \\
|1\rangle\otimes|1\rangle\otimes|1\rangle \, (= |0\rangle\otimes|0\rangle\otimes|0\rangle) \, , 
\quad
|1\rangle\otimes|0\rangle\otimes|1\rangle \, (= |0\rangle\otimes|1\rangle\otimes|0\rangle) \, .
\nonumber
\end{align}
With this basis, any state is given by a four-vector whose components are complex constants. Then the
Othello rule defined above is expressed by a matrix ${\cal R}$
\begin{align}
{\cal R} \equiv
\left(
\begin{array}{cccc}
1& &  & \\
& 1 & & \\
& & &1 \\
& &1 & 
\end{array}
\right)\,.
\end{align}
Immediately one notices that this matrix is the same as that for the CNOT gate, \eqref{CNOTdef}.

We demonstrated here that the Othello rule in the $1\times 3$ site model is equivalent to the CNOT,
but more general cases can be treated in the same manner. In general,
multiple products of the CNOT gates provide longer rules of the Othello.
The essence of the Othello rule is, as emphasized in Sec.~3, to introduce a nonlocal interaction.
Since the CNOT gate is an entangling gate, the successful
products of the CNOT gates produce the nonlocality.

%%%%%%%%%%%%%%%%%%%

%%%%%%%%%%%%%%%%%%%%%%%%%%%%%%%%%%%%%%%%%%%%%%%%%%%%%%%%%%%%%%%%%%%%
%%%%%%%%%%%%%%%%%%%%%%%%%%%%%%%%%%%%%%%%%%%%%%%%%%%%%%%%%%%%%%%%%%%%
%%%%%%%%%%%%%%%%%%%%%%%%%%%%%%%%%%%%%%%%%%%%%%%%%%%%%%%%%%%%%%%%%%%%

\end{document}